\documentclass[12pt]{article}

\usepackage{graphicx}
\usepackage{amssymb}
\usepackage{epstopdf}
\usepackage{times}
\usepackage[dvips]{color}

\newcommand{\caQ}{\mathcal{Q}}
\newcommand{\caW}{\mathcal{W}}
\newcommand{\caP}{\mathcal{P}}
\newcommand{\caF}{\mathcal{F}}
\newcommand{\caE}{\mathcal{E}}
\newcommand{\caL}{\mathcal{L}}
\newcommand{\caM}{\mathcal{M}}
\newcommand{\caI}{\mathcal{I}}
\newcommand{\caJ}{\mathcal{J}}
\newcommand{\bfT}{\mathbf{T}}
\newcommand{\bfx}{\mathbf{x}}
\newcommand{\bfq}{\mathbf{q}}

\newcommand{\bfzeta}{\mbox{\boldmath $\bf\zeta$}}
\newcommand{\bfeta}{\mbox{\boldmath \small$\bf\eta$}}
\newcommand{\bfmu}{\mbox{\boldmath \small$\bf\mu$}}
\newcommand{\NESSW}{physical NESS work $\,\!$}
\newcommand{\NESSw}{physical NESS work}

\newcommand{\NESSWs}{physical NESS works $\,\!$}

\newcommand{\spaEq}{\hspace{2 em}}
\newcommand{\transP}[5]{
   {#1}\!\left(\begin{array}{c} \!\! #2 \!\! \\ 
   \!\! #3 \!\! \end{array}\right| \left. \begin{array}{c}
   \!\! #4 \!\! \\ \!\! #5 \!\! \end{array}\right) }
\newcommand{\pathaveA}[1]{
   \left\langle\!\!\left\langle \;#1\;
   \right\rangle\!\!\right\rangle_{t} }

\newcommand{\vspfigA}{\vspace{0cm}}  
\newcommand{\vspfigB}{\vspace{0cm}} 
\newcommand{\vspfigC}{\vspace{0.6cm}}
\newcommand{\hightfigA}{0.22\textheight}
\newcommand{\hightfigB}{0.46\textheight}

\begin{document}

%\baselineskip 5 ex

%%%%%%%%%%%%%%%%%%%%%%%%%%%%%%%%%%%%%%%%%%%%%%%%%%%%%%%%%%%%%%%%%%%%%%

\begin{center}

\textbf{\LARGE \hspace{-5cm}
Nonequilibrium Steady State Thermodynamics 
\hspace{-5cm}}

\vspace{-0.3cm}
\textbf{\LARGE %\hspace{-5cm}
and Fluctuations for Stochastic Systems %\hspace{-5cm}
}

\vspace{0.3cm} 
{\bf Tooru Taniguchi and E. G. D. Cohen}

\end{center} 
 
\vspace{-0.8cm} 
\begin{quote}  
\begin{center} 
 
{\small \it The Rockefeller University, 1230 York 
Avenue, New York, NY 10021, USA.} 

\vspace{-0.1cm} (\today)

\end{center} 
\vspace{0.2cm}

   We use the work done on and the heat removed from a system 
to maintain it in a nonequilibrium steady state for 
a thermodynamic-like description of such a system
as well as of its fluctuations. 
   Based on a generalized Onsager-Machlup theory 
for nonequilibrium steady states
we indicate two ambiguities, not present in an equilibrium state, 
in defining such work and heat: 
one due to a non-uniqueness of time-reversal procedures 
and another due to multiple possibilities to separate heat
into work and an energy difference in nonequilibrium steady states.    
   As a consequence, for such systems, the work and heat 
satisfy multiple versions of the first and second laws 
of thermodynamics as well as of their fluctuation theorems. 
%   In addition, with the work so defined, we obtain nonequilibrium 
%detailed balance relations, which lead to work fluctuation theorems.
    Unique laws and relations appear only to be obtainable for 
concretely defined systems, using physical arguments to choose 
the relevant physical quantities. 
   This is illustrated on a number of systems, including 
a Brownian particle in an electric field, a driven torsion pendulum, electric circuits and an energy transfer driven
by a temperature difference. 

\end{quote} 

\vspace{0.5cm} 
%%%%%%%%%%%%%%%%%%%%%%%%%%%%%%%%%%%%%%%%%%%%%%%%%%%%%%%%%%%%%%%%%%%%%%
\section{Introduction}
\label{Intro}

   Of all steady states of systems, the equilibrium state is 
by far the most studied.
   First, a thermodynamic description has been developed involving 
the work done by or on the system or the heat produced by or 
removed from the system. 
   This leads to the first law of thermodynamics, i.e. the law 
of energy conservation, while the introduction of entropy leads 
to the second law of thermodynamics, i.e. that entropy
changes in a closed system have to be non-negative.

   A generalization to systems in nonequilibrium steady states 
(NESS) has been made as a special case of the general theory of thermodynamics of irreversible processes
(irreversible thermodynamics) for systems in local, i.e. near equilibrium \cite{Gro84}. 
   This theory has in turn been enlarged to an extended 
irreversible thermodynamics \cite{Jou01}, where in addition 
to the usual local quantities (local mass density, local velocity
and local energy density) also the corresponding irreversible thermodynamic currents of mass, momentum and energy (or heat) 
are taken into account for a description of the system.  
   Also in that context NESSs can be considered. 
   NESSs, as well as their fluctuations, have also be 
considered in hydrodynamics \cite{ZS06}. 

   The major difference between all these thermodynamic theories 
of NESSs and the attempt proposed here to describe NESSs, 
is that all these theories are ultimately based on a direct 
generalization of equilibrium thermodynamics and, in particular, the 
use of the same concepts of work and heat as in equilibrium.
   To the contrary, the theory developed here introduces 
fundamentally different definitions of work and heat, 
associated with a NESS, rather than those used in the 
above equilibrium thermodynamic based theories.
   In fact, we propose a thermodynamic-like description of systems 
in NESSs by defining the work associated with such a system as 
the work that has to be done on the system to maintain it 
in its NESS and prevent it from decaying to an equilibrium state.    
   Similarly we define the heat associated with such a system 
as the heat that has to be removed from the system to eliminate 
the heat produced by the irreversible (nonequilibrium) processes 
which take place in such systems.

   We develop this theory for NESS using a generalization of the classical path integral theory of Onsager and Machlup \cite{OM53} 
for fluctuations in the equilibrium state to NESSs, 
used already by us in two 
previous papers for a specific model \cite{TC07a,TC07b}. 
   An introduction to this theory, relevant for this paper, 
can be found in the first paper \cite{TC07a}. 
   The present paper attempts to present the general structure, 
applied to a variety of models, for discussing the NESS 
as a generalization of the equilibrium state and to exhibit 
the conceptual differences between these two steady states 
of a system and their fluctuations.

   In fact, contrary to the NESS, the equilibrium state is an 
absolutely stable state, which maintains itself without 
the necessity of any work to be done on it, nor of the removal 
of any spontaneously produced heat, since this heat vanishes 
on average in an equilibrium state. 
   For that reason the Onsager-Machlup theory of fluctuations 
of the equilibrium state, does not consider any work done on 
the system and only considers the entropy production rate 
associated with the fluctuations in the equilibrium state.
   The absence of any work allows the formulation of a theory 
of fluctuations in the equilibrium state, based on the entropy production associated with these fluctuations, alone. 
   Therefore the Onsager-Machlup theory does not contain 
any equilibrium thermodynamic feature.

   This is completely different for a NESS, where the presence 
of external ``forces'' (characterized by appropriate 
nonequilibrium parameters), keeps the system permanently out of 
equilibrium, requiring ``actions'' involving work and heat 
to maintain this system in a NESS and prevent it from decaying 
to the absolutely stable equilibrium state. 
   Then a generalization of the two laws of equilibrium
thermodynamics is possible, which require, however, NESS 
adapted definitions of work and heat, which are, together with 
the internal energy, the ingredients of a
thermodynamic-like formulation of the NESS.

%   To treat, in addition, the fluctuations of these quantities 
%in a NESS, 
   The generalization of the Onsager-Machlup theory for fluctuations 
in the equilibrium state to one for NESS,   
turned out to be non-trivial. 
   This,  since such a generalization 
%from the equilibrium state to the NESS leads 
involves in general 
ambiguities, i.e. multiple a priori possible choices 
for the work, heat and energy and their fluctuations in a NESS.  
   The Onsager-Machlup theory for the equilibrium state
provides us though with a starting point to deal with these 
problems and to obtain a physically unique description 
of the thermodynamic laws and the fluctuations of a
NESS, at least for the specific models considered in this paper.

   Implementation of the above outlined program is based on 
Onsager and Machlup's path integral method. 
   This involves, not only the above mentioned new definitions 
of the NESS adapted thermodynamic-like quantities of work and heat, 
but also new definitions of forward and corresponding backward 
paths in time because of the presence of external nonequilibrium parameters in the NESS. 
   The main difficulty and the origin of these ambiguities arising 
then is that an appropriate choice of a backward path for a 
given forward path is not unique and depends on the nature of
the system in the NESS. 
   In addition, there is an intrinsic ambiguity 
because work and energy differences can only 
be defined up to a common quantity. 
   It appears at present that these ambiguities can only 
be resolved on physical grounds for specific concrete models.

   The contents of this paper are organized 
in three parts as follows.     
   After the introduction in this section \ref{Intro}, 
we introduce in Sec. \ref{NESSmodel} 
the class of systems in NESSs which we will 
consider in this paper. 
   We studied three classes. 
   (A) Systems under a constant force, such as an electrically 
charged Brownian particle in a fluid subject to an external 
electric field $\caE$ [cf. Fig. \ref{fig1modelA}(a)];
   (B) Systems coupled to an oscillator, as, e.g. a Brownian 
particle confined by a harmonic oscillator, which 
is dragged through the fluid with a constant velocity 
$v$ by an outside force  [cf. Fig. \ref{fig2modelB}(a)];
   (C) Systems with two random noise sources. 
   An example is two independent heat reservoirs at different 
temperatures, each containing a Brownian particle, which
are coupled to each other harmonically, allowing an energy current 
from one reservoir to the other [cf. Fig. \ref{fig3modelC}(a)]. 
   For each class we introduce not only Brownian particle
models but also corresponding electric circuit models. 
   In total we consider eight models: two of Class A, four of 
Class B and two of Class C. 
%some of which have been studied in the literature [--].
   Each of these models can be described by a Langevin equation, 
which is given explicitly in a common form 
in the next section  \ref{OMtheory} by Eq. (\ref{LangeEquat1}).  
   
   In the second part of the paper, in Sec. \ref{OMtheory}, 
we discuss the generalized Onsager-Machlup theory for a NESS   
and show how to obtain  
appropriate definitions of the heat and work 
from our general point of view. 
   We first introduce the path integral method 
%to be used in this paper, 
for the Langevin dynamics (\ref{LangeEquat1}),  
based on a Lagrangian to give a probability functional of paths. 
   Then, following Onsager and Machlup, 
we write this Lagrangian for the NESS as a sum of 
two dissipation functions and an entropy production rate, 
where the latter allows us a definition of the heat in the NESS 
by integrating over time and multiplying by the temperature 
of a heat reservoir, connected to the system.% 
\footnote{   
   Note that in this paper we consider 
the entropy \emph{production} in NESSs, rather than a 
NESS entropy itself \cite{GC06}.
}
  By minimizing this Lagrangian we obtain the average path, 
which then leads to the non-negativity of the 
entropy production for the average path, i.e. the validity 
of the second law of thermodynamics for the average path.  
%{\underline{independent}} of the sign $(\pm)$ of the external
%nonequilibrium parameter $\mu$, i.e. valid 
%for $\mu=+$ {\it{and}} -. 
   Finally, using the energy conservation law 
(the first law of thermodynamics), the work is obtained 
as a sum of the heat and the internal energy difference.  
   This work consists of four parts: 
(i) work given by the partial time-derivative of 
the internal energy, 
(ii) work done by an external driving force, 
(iii) work caused by a time-irreversible force, 
and (iv) work by a temperature difference between reservoirs.

   In this part of the paper we also discuss in detail 
the role of time-reversal for our NESS Onsager-Machlup theory. 
   We point out in detail the difficulties associated 
with the ambiguity 
of defining an appropriate backward path associated with a given 
forward path due to the presence of external nonequilibrium 
parameters $\bfmu$, e.g. a dragging velocity $v$ or an electric 
field $\caE$, which specify the NESS forces or currents. 
%, where we restrict ourselves here 
%to the case that the ${\bf{\mu}}$, e.g. a drag-velocity $v$
%or an external electric field $E$, can only be either + or -, 
%indicating a $+v$ or $+E$ and a $-v$ or $-E$, respectively.
%   We indicate the ambiguities inherent in the appropriate choice 
%of corresponding forward and backward paths  
%   We point out the difficulties associated with the ambiguity 
%of defining an appropriate backward path associated with a given 
%forward path 
%where we restrict ourselves here 
%to the case that the components of $\bfmu$ can only take 
%two values of opposite sign, i.e. ${\bf{\pm}}$.
%
   To formulate this ambiguity,  
a time-reversal operator $\hat{I}_{\pm}$ is introduced 
which reverses (indicated by a hat) 
the direction of the (internal) motion 
(the velocity) of the system, 
as compared with that on the forward path, 
\emph{as well as} a possible, 
but not necessary, reversal of the sign of the external 
nonequilibrium parameter $\bfmu$ 
(indicated by $\pm$ in $\hat{I}_{\pm}$).%
\footnote{ 
   A time-reversal procedure involving a change of sign 
of a nonequilibrium parameter was already 
used before in shear flow systems \cite{EM90,TM02}.
}  
   Two possible definitions can therefore 
be given for the heat, corresponding to a $+$ or $-$ sign 
in $\hat{I}_\pm$, respectively, as well as for the work 
and the internal energy, leading to two possible expressions 
for the energy conservation law, or the first law as well as 
for the second law of (NESS) thermodynamics 
for \emph{each} (not only the average) path.  
%   Another ambiguity, in which the work and 
%the energy can each be defined up to a common quantity only,  
%is also discussed in this part of the paper.  
%   These ambiguities are present 
%in the level of the general theory.  
%   These definitions are, however, not unique and give two 
%possible choices for the heat, work and energy as well as for the 
%two laws of thermodynamics for NESS.  
%   The same is true for the fluctuation relations. 

    In Sec. \ref{WorkFluctTheor} we discuss the nonequilibrium 
detailed balance relations and the transient fluctuation theorems 
\cite{ES94}
for work, which hold for both $\hat{I}_{+}$ and $\hat{I}_{-}$. 
%   Similarly, the transient fluctuation relation also
%holds for both $\mu = \pm$. 
   All the above laws and relations are therefore unaffected 
by the ambiguities mentioned above, i.e. they are valid relations 
for the NESS, \emph{in}dependent of the above ambiguities.
   For the transient fluctuation theorems this must be 
due to the fact that they are mathematical identities \cite{CG99}.  
%[independent of the nature of the quantities occurring in them.]  
   This means here that one obtains two formal identities, 
without the need to identify the appropriate thermodynamic work 
on physical grounds, as is necessary for a physical discussion 
of particular systems. 
   This ``universal'' validity of the transient fluctuation 
theorems could disappear for asymptotic fluctuation theorems 
\cite{GC95} for NESSs as was indeed shown
in a previous paper for the case of a dragged Brownian particle 
model \cite{TC07a}.

   In the third part of the paper, Sec. \ref{WorkModel}, 
we will illustrate how the above mentioned ambiguities can be eliminated  
and lead to unique choices of 
the heat and work to maintain a NESS
 %work, heat and energy 
%and implement the results of the second part 
on a variety of models introduced in Sec. \ref{NESSmodel}. 
%which provide examples to test the general procedure described 
%in the second part of this paper. 
%   In the third part of this paper, Sec. \ref{WorkModel}, 
%we illustrate on the above mentioned models 
%   Especially, we discuss 
%how the above ambiguities to determine thermodynamic quantities 
%can be eliminated on physical grounds 
%%by a proper physical choice 
%and this then leads to class (i.e. model) dependent 
%unique definitions 
%of the heat and work to maintain a NESS, satisfying 
%the first and second laws of NESS thermodynamics. 
%how the above ambiguities 
%associated with defining an appropriate backward path can be 
%eliminated by a proper physical choice and lead to class - i.e. 
%model - dependent unique definitions of what we will call 
%the \NESSh, work and energy, which satisfy 
%the first and second laws of NESS thermodynamics. 
%
%   We also consider a frame dependence of our theory 
%using a dragged Brownian particle model and indicate 
%an inertial effect in the term of d'Alembert type force 
%in a comoving frame description of this model. 
%
   Although these models are all linear we do not expect the
nature of our considerations to be qualitatively changed, 
if non-linearities in the potentials, occurring in these models, 
are introduced. 
   However, the dependence of, in particular, 
fluctuations on the properties of the stochastic 
noise is much less clear \cite{BC04,TcC07}.

%%%%%%%%%%%%%%%%%%%%%%%%%%%%%%%%%%%%%%%%%%%%%%%%%%%%%%%%%%%%%%%%%%%%%%
\section{NESS Models}
\label{NESSmodel}

   Before discussing our generalized 
Onsager-Machlup theory for NESSs, 
we introduce some typical NESS models all 
described by Langevin equations. 
   Using these models, we give concrete examples 
of external nonequilibrium parameters 
which specify the system in a NESS (so are zero at equilibrium)
and change their signs with a reversal of the steady state 
force or current. 
   These parameters play a crucial role in this paper  
and co-determine the choice of the proper 
time reversal procedure to calculate relevant work and heat 
to associate with a system, as will be discussed later. 
   The internal energies for these models are also given 
in this section and will be used to determine the work to maintain 
a NESS in the following sections. 
%   We also emphasize that NESS Brownian 
%models can have the corresponding electric circuit models 
%described as the same type of stochastic processes. 
   As mentioned in Sec. \ref{Intro}, 
we discuss these NESS models by separating them into three classes: 
Class A for models driven by a constant external force, 
Class B for systems coupled to an oscillator, 
and Class C for models with two random noises. 
%   Each class consists of Brownian particle models as well as 
%electric circuit models.   

\begin{figure}[!t]
\vspfigA
\begin{center}
\resizebox{!}{\hightfigA}{\includegraphics{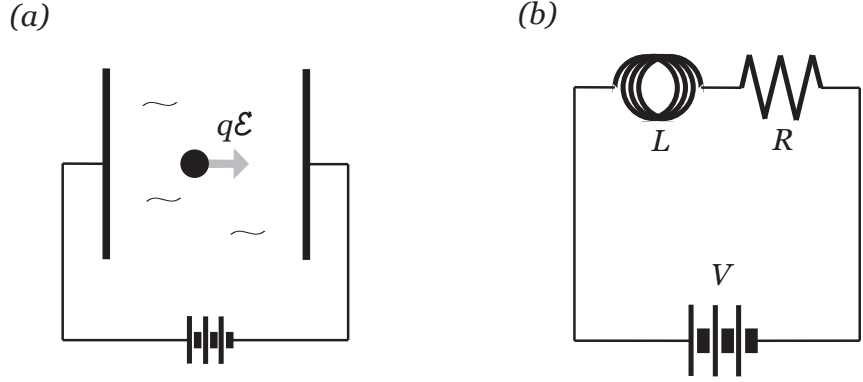}}
\vspfigB
\caption{NESS models of Class A: 
   (a) a charged particle driven by a constant electric field  
   and (b) an electric circuit consisting of an inductor 
   and resistor in series. 
      For an explanation of the symbols, see the text.}
\label{fig1modelA}
\end{center}
\vspfigC
\end{figure}

%%%%%%%%%%%%%%%%%%%%%%%%%%%%%%%%%%%%%%%%%%%%%%%%%%%%%%%%%%%%%%%%%%%%%%
\subsection{Class A: Systems under a Constant Force} 
\label{ClassA}

a)   The first (and possibly simplest) example is an electrically 
charged Brownian particle in a fluid in a uniform electric field. 
   The Langevin equation for this system is given by    
\begin{eqnarray}
   m\ddot{x}_{s} 
   = q\caE -\alpha \dot{x}_{s} 
   + \zeta_{s} 
\label{LangeEquatA1}
\end{eqnarray}
for the particle position $x_{s}$ at the time $s$, 
where $m$ is the mass, 
$q$ the electric charge of the particle, 
$\caE$ a constant external electric field,  
$\alpha$ the friction coefficient of the particle in the fluid, 
$\ddot{x}_{s}\equiv d^{2} x_{s}/ds^{2}$ and 
$\dot{x}_{s}\equiv d x_{s}/ds$. %
%\footnote{
%In this paper, we use the notations $\dot{X}_{s}$ and 
%$\ddot{X}_{s}$ for the first and second time-derivative 
%for any time-dependent quantity $X_{s}$.
%} 
   Here, $\zeta_{s}$ is a Gaussian-white random force 
whose first two auto-correlations are given by   
$\langle \zeta_{t} \rangle = 0 $ and 
$\langle \zeta_{t_{1}}\zeta_{t_{2}} \rangle 
 = (2\alpha/\beta)\delta(t_{1}-t_{2})$, respectively, 
with $\beta$ the inverse temperature of the heat reservoir   
and the notation $\langle \cdots\rangle$ 
for an ensemble average.%
\footnote{
Note the coefficient $2\alpha/\beta$ 
in $\langle \zeta_{t_{1}}\zeta_{t_{2}} \rangle$ is due to 
the fluctuation dissipation theorem, 
which is, strictly speaking, justified 
around equilibrium. 
   In this report, we \emph{assume} that it is still correct 
for our NESS models. 
}   
   The Brownian particle is driven by a constant 
force $q\caE$ via the external field $\caE$ which 
plays the role of the external nonequilibrium parameter 
in this model and vanishes at equilibrium. 
   A schematic illustration for this system is given in 
Fig. \ref{fig1modelA}(a). 
   The internal energy $E$ of this system 
is given by 
\begin{eqnarray}
   E(\dot{x}_{s}) = 
      \frac{1}{2}m \dot{x}_{s}^{2} .
\label{EnergA1}
\end{eqnarray}
%
%with $\dot{x}_{s}\equiv dx_{s}/ds$. 
   It is important to note that here we regards 
$q\caE$ as an ``external'' driving force 
and its corresponding potential energy    
is not included in the ``internal'' energy $E$.  
   In this system, the Brownian particle achieves 
a constant average velocity $\overline{v} =q\caE/\alpha$ 
in a NESS. %
%\footnote{
%   This model can be for electrons  
%in a direct current circuit with voltage $|\caE|$ 
%and electric current density $\rho \overline{v}$ 
%($\rho$: the density of electrons).} 
   Note that a nonequilibrium state driven by a constant 
force can be realized in variety of other ways, 
for example, in a Brownian particle under a constant 
gravitational force. 
% by replacing $q\caE$ with $mg\mathbf{i}$ 
% with the acceleration of gravity $g$ and an unit vector 
% $\mathbf{i}$ pointing to the direction of the gravitational force.  

%---------------------------------------------------------------------
%--------------------------------------------------------------------- 

b)   As a second example we consider an electric circuit 
consisting of an inductor (with self-inductance $L$) 
and a resistor (with resistance $R$) in series \cite{07BMN}, 
as shown in Fig. \ref{fig1modelA}(b). 
   In this circuit, the voltage $V$ of the battery is equal 
to $L \dot{I}_{s} + R I_{s} + \delta V_{s}$ with 
the electric current $I_{s}$ through the resistor 
and a voltage random noise 
$\delta V_{s}$ in the resistor. 
   Combining this with $I_{s}=\dot{q}_{s}$,  
where $q_{s}$ is the charge in the resistor, % at time $s$, 
we obtain the Langevin equation   
\begin{eqnarray}
   L \ddot{q}_{s} 
   = V - R \dot{q}_{s} - \delta V_{s} .
\label{LangeEquatA2}
\end{eqnarray}
%, 
%with $\ddot{q}_{s}\equiv d^{2} q_{s}/ds^{2}$, and 
%$\dot{q}_{s}\equiv d q_{s}/ds$. 
   Here, we assume that $\delta V_{s}$ is a Gaussian-white 
random noise whose first two auto-correlations are given by   
$\langle \delta V_{s} \rangle = 0$ and 
$\langle \delta V_{s}\delta V_{s'} \rangle 
      = (2R/\beta)\delta(s-s')$
by the Johnson-Nyquist theorem \cite{J28,N28}. 
   The external nonequilibrium parameter in this model 
is given by the voltage $V$ of the battery. 
   We note that the two Langevin equations 
(\ref{LangeEquatA1}) and (\ref{LangeEquatA2}) have the same form. 
   We summarized correspondences of the quantities 
in these two equations in Table \ref{CorreModel1}. 
   Noting these correspondences, the energy 
$E$ of this electric circuit model is given by Eq. (\ref{EnergA1}) 
with a replacement of $m$ and $x_{s}$ by $L$ and $q_{s}$, 
respectively, i.e. by $E(\dot{q}_{s})=(1/2)L\dot{q}_{s}^{2}$. 

%---------------------------------------------------------------------
\newcommand{\widthtableA}{3.5cm} 
\newcommand{\widthtableB}{0.9cm} 
\newcommand{\widthtableC}{1.3cm}
\newcommand{\widthtableD}{2cm}
\newcommand{\widthtableE}{1cm}
\begin{table}[!t]
\vspfigA
\begin{center}
\begin{tabular}{|c|cccc|c|c|}
\hline 
    \makebox[\widthtableA]{Model} & 
    \makebox[\widthtableB]{} & 
    \makebox[\widthtableB]{} & 
    \makebox[\widthtableB]{} & 
    \makebox[\widthtableB]{} & 
    \makebox[\widthtableC]{Class A} & 
    \begin{tabular}{c}
       \makebox[\widthtableD]{Class B} 
    \end{tabular} \ \\
 \hline 
 \hline 
    Brownian particle & 
    $x_{s}$ & $m$ & $\alpha$ &$\zeta_{s}$ & $q\caE$&
    \begin{tabular}{cc}
       \makebox[\widthtableE]{$\kappa$} 
       & \makebox[\widthtableD]{$v$}  
    \end{tabular} \\
 \hline 
    Electric circuit &
    $q_{s}$ & $L$ & $R$  &$-\delta V_{s}$ & V &
    \begin{tabular}{cc}
       \makebox[\widthtableE]{$1/C$} 
       & \begin{tabular}{c}
          \makebox[\widthtableD]{$I$ (parallel)} \\
          $CA$ (serial)
        \end{tabular} 
    \end{tabular} \\
 \hline 
    Torsion pendulum &
    $\theta_{s}$ & $\caI$ & $\nu$ &$\zeta_{s}$ &  &
    \begin{tabular}{cc}
       \makebox[\widthtableE]{$\sigma$} 
       & \makebox[\widthtableD]{$\xi/\sigma$}  
    \end{tabular} \\
 \hline 
\end{tabular}
\end{center}
\vspfigB
\caption{
      Correspondences of quantities in various NESS models. 
      The external nonequilibrium parameters characterizing the  
   deviations of the systems from an equilibrium state 
   are $\caE$ and $V$ in Class A, 
   and $v$, $I$, $A$ and $\xi$ in Class B, respectively. 
      Explanation of symbols is in the text.     
}
\label{CorreModel1}
\vspfigC
\end{table}%
%---------------------------------------------------------------------

%%%%%%%%%%%%%%%%%%%%%%%%%%%%%%%%%%%%%%%%%%%%%%%%%%%%%%%%%%%%%%%%%%%%%%
\subsection{Class B: Systems coupled to a Harmonic Oscillator}
\label{ClassB} 

\begin{figure}[!t]
\vspfigA
\begin{center}
\resizebox{!}{\hightfigB}{\includegraphics{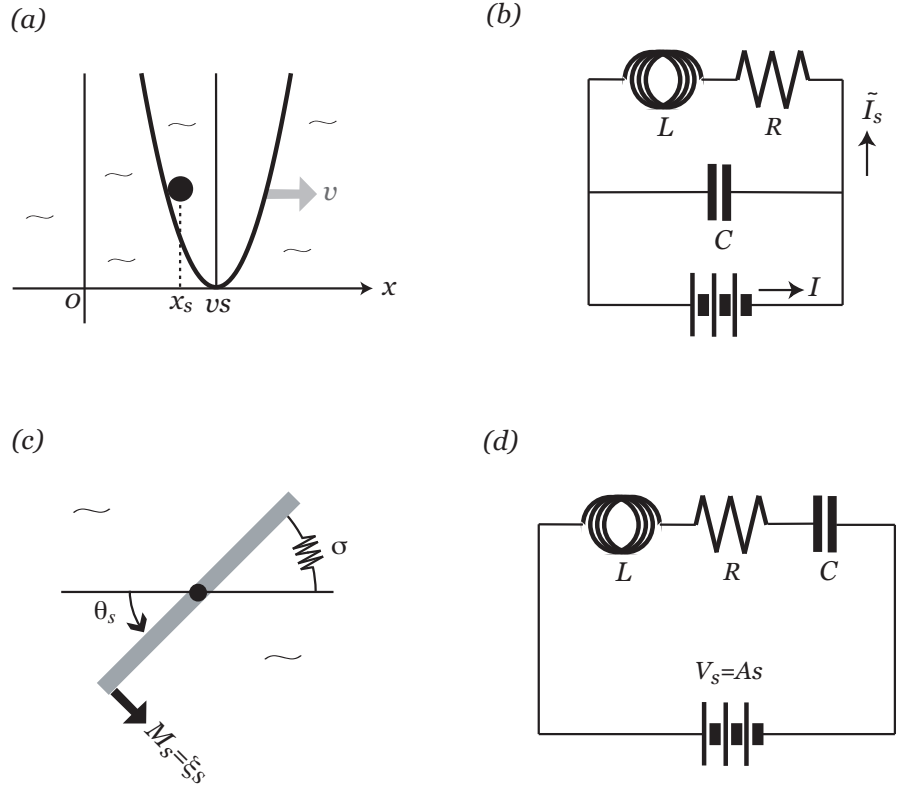}}
\vspfigB
\caption{NESS models of Class B: 
   (a) a particle dragged by a harmonic potential 
   with a constant velocity,  
   (b) an electric circuit with a serial inductor-resistor coupled 
      to a capacitor in parallel, 
   (c) a torsion pendulum confined by a spring with 
      an external torque, and 
   (d)  an electric circuit with an inductor, resistor 
      and capacitor in series. 
   Symbols are defined in the test.}
\label{fig2modelB}
\end{center}
\vspfigC
\end{figure}

   As the second class of NESS models, 
we consider a system under an oscillating force. 

a)   The first example in this class 
is a Brownian particle confined by a harmonic potential 
which is dragged by a constant velocity $v$ in a fluid 
 \cite{TC07a,TC07b,ZC03b,S98}. 
   The Langevin equation for this system is given by 
\begin{eqnarray}
   m\ddot{x}_{s} 
   = -\kappa (x_{s}-v s) -\alpha \dot{x}_{s} 
   + \zeta_{s} 
\label{LangeEquatB1}
\end{eqnarray}
for the particle position $x_{s}$ % at time $s$,  
with the oscillator spring constant $\kappa$ and 
the Gaussian-white random force $\zeta_{s}$. 
   In this model, the dragging velocity $v$ plays the role 
of the external nonequilibrium parameter 
which is zero at equilibrium. 
   A schematic illustration of this model is 
given in Fig. \ref{fig2modelB}(a). 
%   This model may be simply regarded as a Brownian particle 
%dragged by a spring with a constant velocity \cite{S98}, 
%and can be realized as a Brownian particle 
%captured in an optical trap which moves with a constant velocity 
%through a fluid \cite{ZC03b,WSM02}. 
%   This model was also considered using the NESS  
%Onsager-Machlup theory in our previous papers \cite{TC07a,TC07b}. 
   In this model the internal energy $E$  
of the particle is given by 
\begin{eqnarray}
   E(\dot{x}_{s},x_{s}) = 
      \frac{1}{2}m \dot{x}_{s}^{2} 
      +\frac{1}{2}\kappa (x_{s}-vs)^{2} 
\label{EnergB1}
\end{eqnarray}
where the second term on its right-hand side 
is the potential energy of the particle 
in the harmonic oscillator.    
%$(1/2)\kappa (x_{s}-vs)^{2}$. 

   Although the model described by Eq. (\ref{LangeEquatB1}) 
may be regarded as a Brownian particle 
model producing a constant average velocity of the particle, 
like the electric field driven model described 
by Eq. (\ref{LangeEquatA1}), 
we should notice there are clear differences between these two models.  
   In this dragged Brownian particle model (Class B), 
the particle moves with a constant 
velocity even if there is no friction, 
since the average velocity of the particle is 
independent of the friction constant $\alpha$. 
   On the other hand, in the electric field driven model (Class A), 
the average velocity of the particle will depend on the 
friction constant and if there is no friction 
then the particle accelerates indefinitely. 
   We also note that there is no 
explicit time-dependent parameter in the force  
in the electric field driven model, 
while in a Brownian particle dragged 
by a constant velocity there is an explicit time-dependence 
in the force via $v s$ in Eq. (\ref{LangeEquatB1}). 
   These ``simple'' differences will manifest themselves 
in \emph{different} definitions of work and heat in NESSs, 
as will be discussed in Secs. \ref{WorkModelA} 
and \ref{WorkModelB}.

%---------------------------------------------------------------------
%--------------------------------------------------------------------- 

b)   The same form as the Langevin equation (\ref{LangeEquatB1}) 
appears for an electric circuit in which an inductor and resistor  
in series are coupled with a capacitor in parallel \cite{ZCC04}, 
as shown in Fig. \ref{fig2modelB}(b). 
   We first derive a Langevin equation for this system. 
   Denoting the electric current  
through the resistor by $\tilde{I}_{s}=\dot{q}_{s}$ 
with $q_{s}$ the charge through the resistance,  
as shown in Fig. \ref{fig2modelB}(b), 
the voltage difference $V_{1s}$ between the ends 
of the inductor and resistor in series with 
the Johnson-Nyquist voltage 
fluctuation $\delta V_{s}$ as a Gaussian-white noise is given by 
$V_{1s}= L \dot{\tilde{I}\;}_{s} + R \tilde{I}_{s} + \delta V_{s}$. 
   On the other hand, the voltage $V_{2s}$ applied to the capacitor 
(with the capacitance $C$) is given by $V_{2s}=(Is-q_{s})/C$ 
where $I$ is the constant electric current from the battery, 
as shown in Fig. \ref{fig2modelB}(b).  
   Here, we used that the charge of the capacitor is given 
by $Is-q_{s}$, i.e. the charge $Is$ received from the battery  
minus the charge $q_{s}$ taken to the resistor. 
   Using that $V_{1s}=V_{2s}$ and $\tilde{I}_{s}=\dot{q}_{s}$ 
we obtain 
\begin{eqnarray}
   L \ddot{q}_{s} 
   = -\frac{1}{C}(q_{s}-Is) - R \dot{q}_{s} - \delta V_{s} ,
\label{LangeEquatB2}
\end{eqnarray}
which is the Langevin equation for the charge $q_{s}$. 
%in the resistor. %at time $s$. 
   The external nonequilibrium parameter of this system is 
given by $I$. 
   Note that, different from the previous electric circuit 
model in Class A [cf. Eq. (\ref{LangeEquatA2})]  
in which the voltage of the battery is constant, 
in this electric circuit model, described by Eq. (\ref{LangeEquatB2}),  
the electric current $I$ from the battery is assumed to be constant.  
%   This model is considered in Ref. \cite{ZCC04} 
%to discuss the extended heat fluctuation theorem.  
   The energy of this system is given by 
Eq. (\ref{EnergB1}) with a replacement of $m$, $x_{s}$, $\kappa$, 
and $v$ by $L$, $q_{s}$, $1/C$ and $I$, respectively 
(cf. Table \ref{CorreModel1}).

c)  The third example of Class B is a torsion pendulum 
under an external torque in a fluid \cite{DJG06}. 
   A schematic illustration of this model is given in Fig. 
\ref{fig2modelB}(c) as a rod 
with the total moment of inertia $\caI$, 
rotating around its center 
with a spring functioning as a torsion. 
   The time-derivative $\caI \ddot{\theta}_{s}$ 
of the angular momentum $\caI \dot{\theta}_{s}$ 
for an angular displacement $\theta_{s}$ 
must be equal to the torque applied to the rod, 
so that the equation of motion for $\theta_{s}$  %at time $s$ 
is given by the Langevin equation  
\begin{eqnarray}
   \caI \ddot{\theta}_{s} 
   = 
   -\sigma\theta_{s} + \caM_{s} 
   - \nu \dot{\theta}_{s} + \zeta_{s}
\label{LangeEquatB3}
\end{eqnarray}
where $\nu$ is the viscous damping, $\sigma$ the elastic torsional 
stiffness of the pendulum, and $\caM_{s}$ the external torque.  
%$\ddot{\theta}_{s}\equiv d^{2}\theta_{s}/ds^{2}$ and 
%$\dot{\theta}_{s}\equiv d \theta_{s}/ds$. 
   For this model, we consider the case of a linear external 
torque of   
\begin{eqnarray} 
   \caM_{s} = \xi s
\label{LineaForciB3}
\end{eqnarray}
with a force constant $\xi$. 
   Since the pendulum is driven externally by 
the torque (\ref{LineaForciB3}), its coefficient 
$\xi$ plays the role of the external nonequilibrium parameter. 
   %
%   Noting that the total mechanical torque is represedted by  
%$-\sigma\theta_{s} + \caM_{s} = -\sigma [\theta_{s}-(\xi/\sigma)s]$, 
%the Langevin equation (\ref{LangeEquatI2}) 
%can also be regarded as a pendulum which is rotated 
%by dragging it via spring with the spring constant $\sigma$ 
%by a constant angular velocity $\xi/\sigma$.  
   In this model the internal energy $E$  
of the particle is given by 
\begin{eqnarray}
   E(\dot{\theta}_{s},\theta_{s}) = 
      \frac{1}{2} \caI  \dot{\theta}_{s}^{2} 
      +\frac{1}{2}\sigma \theta_{s}^{2} .
\label{EnergB2}
\end{eqnarray}
as the sum of the kinetic energy and the torsion energy. 

   It is important to note that although the Lagrangian equation 
(\ref{LangeEquatB3}) with the torque (\ref{LineaForciB3}) 
has the same form as Eq. (\ref{LangeEquatB1}) 
(cf. Table \ref{CorreModel1}), the energy (\ref{EnergB2}) 
in this driven torsion pendulum model 
does not have the same form as the energy (\ref{EnergB1}) 
for the dragged Brownian particle model. 
   As shown later in this paper, this difference 
of energy also appears as a difference 
in the definition of the work. 
   Another difference between these two models 
is that for the driven torsion pendulum model 
the average internal energy increases with time in a NESS, leading to 
a time-proportional work rate as will be discussed in 
Sec. \ref{WorkModelB}, while in the dragged Brownian particle 
model the average internal energy is independent of time in a NESS  
with a constant average work rate.

d)   As the last example in Class B, we consider 
an electric circuit consisting of an inductor, 
resistor and capacitor in series 
with a time-dependent applied voltage 
$V_{s}=As$ with a constant $A$ \cite{ZCC04}. 
   [See Fig. \ref{fig2modelB}(d) for a schematic illustration 
of this model.]
   The Langevin equation for the charge $q_{s}$ %at time $s$ 
is given by 
\begin{eqnarray}
   L \ddot{q}_{s}  
   = -\frac{q_{s}}{C} + A s - R \dot{q}_{s} - \delta V_{s} .
\label{LangeEquatB4}
\end{eqnarray}
   In this model, the equilibrium state is realized 
when $A=0$, so that $A$ is the external nonequilibrium parameter. 
   Note that the Langevin equation (\ref{LangeEquatB4}) 
has the same form as Eqs. (\ref{LangeEquatB1}), 
(\ref{LangeEquatB2}) and (\ref{LangeEquatB3}) 
(cf. Table \ref{CorreModel1}).
   The energy of this system is given by 
Eq. (\ref{EnergB2}) with the replacements of $m$, $x_{s}$ and 
$\kappa$ by $L$, $q_{s}$ and $1/C$, respectively 
(cf. Table \ref{CorreModel1}).

%%%%%%%%%%%%%%%%%%%%%%%%%%%%%%%%%%%%%%%%%%%%%%%%%%%%%%%%%%%%%%%%%%%%%%
\subsection{Class C: Systems with Two Random Noises}
\label{ClassC}

   As the last category of NESS models discussed in this paper, 
we introduce stochastic models with two random noises. 
%   Different from the previous two classes of nonequilibrium models, 
%an energy transfer from one reservoir to another can occur and  
%there is no average motion of a Brownian particle to realize a NESS. 
%   Another reason to discuss this class of models is that 
%the work and heat to sustain a NESS 
%is not trivial in models of this class, 
%so that we can demonstrate the usefulness 
%of our generalized Onsager-Machlup theory by discussing them. 

%---------------------------------------------------------------------
\begin{figure}[!t]
\vspfigA
\begin{center}
\resizebox{!}{\hightfigA}{\includegraphics{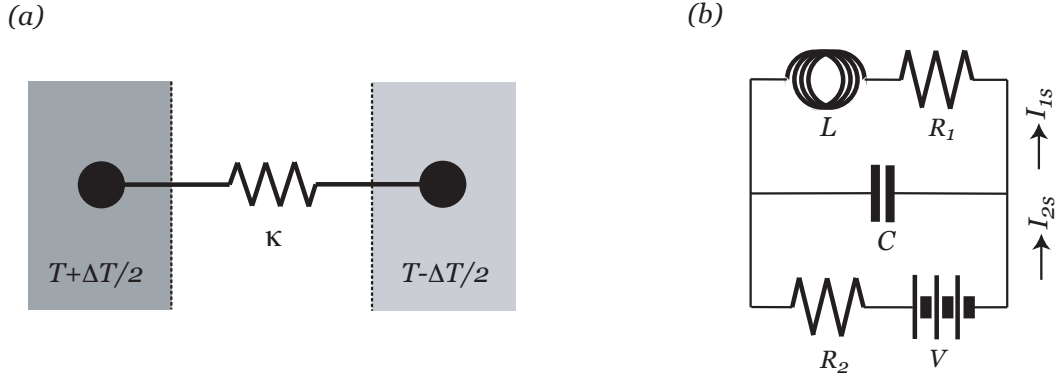}}
\vspfigB
\caption{
      NESS models of Class C: 
   (a) two harmonically coupled particles connected 
   to two heat reservoirs 
      with different temperatures, and 
   (b) an electric circuit with two resistors.}
\label{fig3modelC}
\end{center}
\vspfigC
\end{figure}  
%--------------------------------------------------------------------- 

a)   The first example in this category consists of two Brownian 
particles coupled by a spring, where each particle is confined 
to a heat reservoir at a different temperature 
(cf. Refs. \cite{S98,GS06}).  
   We give a schematic illustration of 
this model in Fig. \ref{fig3modelC}(a). 
   The Langevin equation for the positions $x_{1s}$ and 
$x_{2s}$ of the first and second particle, respectively, 
is given by 
\begin{eqnarray}
   m \ddot{x}_{js} 
   = -\kappa \left(x_{js} - x_{ks}\right) 
   -\alpha \dot{x}_{js} +\zeta_{js} 
\label{LangeEquatC1}
\end{eqnarray}
with $j\neq k$, $j=1,2$ and $k=1,2$. 
%the mass $m_{j}$, and 
%$\ddot{x}_{js}\equiv d^{2}x_{js}/ds^{2}$
%and $\dot{x}_{js}\equiv dx_{js}/ds$. 
   Here, $\zeta_{1s}$ and $\zeta_{2s}$ are two 
independent Gaussian-white random forces at different temperatures 
$T+\Delta T/2$ and $T-\Delta T/2$, respectively, 
so that $\langle \zeta_{js}\rangle = 0 $ and 
$\langle \zeta_{js}\zeta_{j's'}\rangle 
= (2\alpha/\beta_{j}) \delta_{jj'} \delta(s-s')$ 
with the inverse temperatures 
$\beta_{j}\equiv \{k_{B}[T+(-1)^{j+1}\Delta T/2]\}^{-1}$. 
   For simplicity we assumed identical masses 
and friction coefficients for the two particles. 
   In this system, a NESS is sustained 
with an energy transfer between the two reservoirs 
due to the temperature difference $\Delta T$ as 
an external nonequilibrium parameter.   
   The internal energy $E$  
of this system is given by 
\begin{eqnarray}
   E(\dot{\bfx}_{s},\bfx_{s}) = 
     \sum_{j=1}^{2} \frac{1}{2}m \dot{x}_{js}^{2}
      + \frac{1}{2} \kappa \left(x_{1s} - x_{2s} 
      \right)^{2}
\label{EnergC1}
\end{eqnarray}
with $\bfx_{s}\equiv (x_{1s},x_{2s})$.  
% $\dot{\bfx}_{s}\equiv (\dot{x}_{1s},\dot{x}_{2s})$. 

b)   We can consider a similar stochastic system with 
two random noises in an electric circuit as in 
Fig. \ref{fig3modelC}(b), 
which is like that in Fig. \ref{fig2modelB}(b),   
except for an additional resistance next to the battery. 
   We denote by $I_{1s}=\dot{q}_{1s}$ 
the electric current associated with charge $q_{1s}$ 
through the resistor $1$ 
(with resistance $R_{1}$) next to the coil  
and by $I_{2s}=\dot{q}_{2s}$ the electric current 
associated with charge $q_{2s}$ through the resistor $2$ 
(with resistance $R_{2}$) next to the battery. 
   The charge on the capacitor is given by 
$q_{2s} - q_{1s}$, so that the voltage drop 
in the capacitor is $(q_{2s} - q_{1s})/C$. 
   This voltage drop is equal to 
the voltage difference  
$L \dot{I}_{1s} +R_{1} I_{1s} +\delta V_{1s}$ 
between the ends of the inductor and the resistor $1$ in series, 
and also to the one $V -R_{2} I_{2s} - \delta V_{2s}$
between the ends of the battery and the resistor $2$.  
   Here, $\delta V_{1s}$ and $\delta V_{2s}$ are 
independent Gaussian-white random noises in  
the resistors $1$ and $2$, respectively,  
with $\langle \delta V_{js}\rangle = 0 $ and 
$\langle \delta V_{js} \delta V_{j's'}\rangle 
= (2R_{j}/\beta) \delta_{jj'} \delta(s-s')$, 
$j=1,2$ and $j'=1,2$. 
   Using these voltages, 
the Langevin equation for the charges 
$q_{js}$ is given by 
\begin{eqnarray}
   L \ddot{q}_{1s} 
   +R_{1} \dot{q}_{1s} +\delta V_{1s} 
   = V -R_{2} \dot{q}_{2s} - \delta V_{2s}
   = \frac{q_{2s} - q_{1s}}{C} .
\label{LangeEquatC2}
\end{eqnarray}
%
%with $\ddot{q}_{js}\equiv d^{2}q_{js}/ds^{2}$
%and $\dot{q}_{js}\equiv dq_{js}/ds$. 
   In this system the electric current is driven 
by the voltage $V$, which is the external nonequilibrium 
parameter. 
   The energy $E$ of this system is given by 
\begin{eqnarray}
   E(\dot{\bfq}_{s},\bfq_{s}) = 
      \frac{1}{2}L \dot{q}_{1s}^{2}
      + \frac{\left(q_{1s} - q_{2s} 
      \right)^{2}}{2C} 
\label{EnergC2}
\end{eqnarray}
with $\bfq_{s}\equiv (q_{1s},q_{2s})$.  

   Although we categorize the above two models as a single class C, 
the corresponding Langevin equations (\ref{LangeEquatC1}) and 
(\ref{LangeEquatC2}) do not have exactly the same form, 
different from the models in Class A or B. 
   However, we can introduce a single   
%   To discuss these two equations simultaneously, we consider the 
Langevin equation, which reduces to Eqs. (\ref{LangeEquatC1}) and  
(\ref{LangeEquatC2}) as special cases:  
\begin{eqnarray}
   m_{j} \ddot{x}_{js} 
      &=& \Gamma \delta_{j2} -\kappa \left(x_{js} - x_{ks}\right) 
      -\alpha_{j} \dot{x}_{js} +\zeta_{js}, 
      \label{LangeEquatC3} 
\end{eqnarray}
with $j\neq k$, $j=1,2$ and $k=1,2$, a constant $\Gamma$ and two   
independent Gaussian-white random noises $\zeta_{1s}$ 
and $\zeta_{2s}$ at different temperatures 
$T+\Delta T/2$ and $T-\Delta T/2$, respectively, 
so that $\langle \zeta_{js}\rangle = 0$ and 
$\langle \zeta_{js}\zeta_{j's'}\rangle 
= (2\alpha_{j}/\beta_{j}) \delta_{jj'} \delta(s-s')$. 
   We take $\bfmu=(\Delta T,\Gamma)$ with two  
external nonequilibrium parameters 
$\Delta T$ and $\Gamma$ in this model. 
   The internal energy $E$ of this system is given by  
\begin{eqnarray}
   E(\dot{x}_{s},x_{s}) 
   = \sum_{j=1}^{2} \frac{1}{2}m_{j} \dot{x}_{js}^{2} 
   + \frac{1}{2}\kappa \left(x_{1s} - x_{2s}\right)^{2}
\label{EnergC3}.
\end{eqnarray}
   Eqs. (\ref{LangeEquatC3}) and (\ref{EnergC3}) become 
Eqs. (\ref{LangeEquatC1}) and  (\ref{EnergC1}), 
respectively, in the case of $m_{1}=m_{2}=m$, 
$\alpha_{1}=\alpha_{2}=\alpha$ and $\Gamma =0$,  
while they reduce Eq. (\ref{LangeEquatC2}) and 
(\ref{EnergC2}), respectively, 
in the case of $x_{js}=q_{js}$, $m_{1}=L$, $m_{2}=0$, 
$\Gamma =V$
$\alpha_{j}=R_{j}$, $\kappa=1/C$, $\zeta_{js}=-\delta V_{js}$ 
and $\beta_{1}=\beta_{2}=\beta$ (i.e. $\Delta T=0$). 

   After having introduced here the NESS models 
which we will use in this paper,
we now discuss the theory which we will apply to them.

%%%%%%%%%%%%%%%%%%%%%%%%%%%%%%%%%%%%%%%%%%%%%%%%%%%%%%%%%%%%%%%%%%%%%%
\section{NESS Onsager-Machlup Theory}
\label{OMtheory}

   In this section we discuss a generalized Onsager-Machlup 
theory for NESSs, or simply the NESS Onsager-Machlup theory, 
for linear stochastic models, including those introduced 
in the previous section.

\subsection{Path Integral Approach to Stochastic Dynamics}

   We can write the Langevin equations for all the models 
in the previous section \ref{NESSmodel}, in the form: 
\begin{eqnarray}
   m_{j} \ddot{x}_{js} = F_{j}(\bfx_{s},s;\bfeta) 
   - \alpha_{j} \dot{x}_{js} +\zeta_{js} , 
\label{LangeEquat1}
\end{eqnarray}
$j=1,2,\cdots,N$, for a system with $N$ degrees of freedom  
described by 
$\bfx_{s}\equiv(x_{1s},x_{2s},\cdots x_{Ns})$. %at time $s$.  
   Here, $x_{js}$ is a position (charge), 
$m_{j}$ a mass (self-inductance), 
$\alpha_{j}$ a friction coefficient (resistance), and 
$F_{j}$ a mechanical force (voltage) 
including the external nonequilibrium parameter $\bfeta$ 
in Brownian (electric circuit) models, respectively. 
   Furthermore, $\zeta_{js}$ incorporates 
a Gaussian-white random noise, 
so that $\langle \zeta_{js}\rangle = 0 $ and 
$\langle \zeta_{js}\zeta_{j's'}\rangle 
= (2\alpha_{j}/\beta_{j}) \delta_{jj'} \delta(s-s')$ 
with the inverse temperatures $\beta_{j}$.

   Note that systems with $N$ degrees of freedom have already been 
considered in Onsager and Machlup's original theory   
using $N$ independent variables $\alpha_{1}$, $\alpha_{2}$, $\cdots$, 
$\alpha_{N}$  for any integer number $N$ \cite{OM53}. 
   (Also see the introduction of Ref. \cite{TC07a}.) 
   However, we emphasize that for our models introduced 
in Sec. \ref{NESSmodel} it is enough to consider 
$N=1$ or $2$, i.e. $N=1$ for Classes A and B 
or $N=2$ for Class C, 
although our general theory developed 
in Secs. \ref{OMtheory} and \ref{WorkFluctTheor} 
is formally correct for any $N$.

   Although the mechanical force $F_{j}(\bfx_{s},s;\bfeta)$ 
could be a nonlinear function of $\bfx_{s}$ 
in the Langevin equation (\ref{LangeEquat1}), 
%in some parts of our arguments below, 
we restrict ourselves in this paper 
to functions linear in $x_{js}$, 
consistent with the linear Langevin equations used in 
the previous section \ref{NESSmodel}, 
which is sufficient for the purposes of this paper. 
   Thus we will impose the condition 
\begin{eqnarray}
   \langle F_{j}(\bfx_{s},s;\bfeta)\rangle 
   = F_{j}(\langle\bfx_{s}\rangle,s;\bfeta)
\label{LineaForce1}
\end{eqnarray}
for the ensemble average of the force 
$F_{j}(\bfx_{s},s;\bfeta)$ which is linear   
with respect to $\bfx_{s}$.  
   We will use this condition (\ref{LineaForce1}) 
to discuss the second law of thermodynamics 
in Sec. \ref{SeconLawTherm} of this paper.  
   Otherwise, the linearity of the force $F_{j}$ 
is not used in the general theory developed in this paper. 
%  An investigation of concrete nonlinear models 
%is needed to see whether the NESS theory 
%presented here has any validity   
%if non-linearity enters into the formal considerations given here. 
 
   Similarly, we assume for the models of Class C that 
the temperature difference between the two heat reservoirs is small, 
so that  
\begin{eqnarray}
   \left|\frac{\Delta T_{j}}{\overline{T}}\right| <\!< 1 
\label{SmallTempe1}
\end{eqnarray}
where $\overline{T}$ is the average temperature 
$\overline{T} \equiv (1/N)\sum_{j=1}^{N} T_{j}$ 
and $\Delta T_{j}$ is the deviation 
$\Delta T_{j} \equiv  T_{j} - \overline{T}$ 
of the temperature $T_{j}\equiv 1/(k_{B}\beta_{j})$ 
of the $j$-th reservoir from $\overline{T}$ 
with the Boltzmann constant $k_{B}$. 
   We will calculate quantities like work and heat 
up to the lowest non-vanishing order 
in $|\Delta T_{j}/\overline{T}|$.    
%(the first order, or the second order if the first order is zero),  
%although a formal generalization of these results  
%to higher orders of $|\Delta T_{j}/\overline{T}|$ is straightforward.  
   Here, $\Delta \bfT \equiv 
(\Delta T_{1}, \Delta T_{2},\cdots,\Delta T_{N})$ 
plays a role of the \emph{thermal} nonequilibrium 
parameter and combining it with the \emph{mechanical} nonequilibrium 
parameter $\bfeta$ we obtain the total external 
nonequilibrium parameter 
$\bfmu = (\bfeta, \Delta \bfT )$. 

   For later use, we now implement the stochastic dynamics, 
given by the Langevin equation (\ref{LangeEquat1}),  
using a path integral approach. 
   Thereto, we note that 
the probability functional  
$\caP_{\zeta}(\{\bfzeta_{s}\})$ of 
the Gaussian-white random noise 
$\bfzeta_{s}\equiv (\zeta_{1s},\zeta_{2s},\cdot,\zeta_{Ns})$ 
is given by 
$\caP_{\zeta}(\{\bfzeta_{s}\}) = C_{\zeta} $ $\exp\{ 
$ $- \sum_{j=1}^{N}[\beta_{j}/(4\alpha_{j})] 
\int_{t_{0}}^{t}ds \; \zeta_{js}^{2}\}$
with the normalization constant $C_{\zeta}$. 
   By inserting $\bfzeta_{s}$ from the Langevin equation 
(\ref{LangeEquat1}) 
into this functional $\caP_{\zeta}(\{\bfzeta_{s}\})$ and 
interpreting $\caP_{\zeta}(\{\bfzeta_{s}\})$ then as the 
probability functional $\caP_{x}(\{\bfx_{s}\})$ 
for the path $\{\bfx_{s}\}_{s\in[t_{i},t_{f}]}$, 
we obtain \cite{TC07b},  
apart from a normalization constant, 
\begin{eqnarray}
   \caP_{x}(\{\bfx_{s}\}) = 
   C_{x}\exp\left[\int_{t_{0}}^{t}ds\; 
   \caL\!\left(\ddot{\bfx}_{s},\dot{\bfx}_{s},
   \bfx_{s},s;\bfmu\right)\right]
\label{ProbaFunctX1}
\end{eqnarray}
where the function 
$\caL(\ddot{\bfx}_{s},\dot{\bfx}_{s},\bfx_{s},s;\bfmu)$ 
of $\ddot{\bfx}_{s}$, $\dot{\bfx}_{s}$, $\bfx_{s}$ and $s$ 
is a Lagrangian given by \cite{TC07b} 
\begin{eqnarray}
   \caL\!\left(\ddot{\bfx}_{s},\dot{\bfx}_{s},
   \bfx_{s},s;\bfmu \right) = 
   -\sum_{j=1}^{N}\frac{\alpha_{j}\beta_{j}}{4}
   \left[\dot{x}_{js} 
   - \frac{1}{\alpha_{j}} F_{j}(\bfx_{s},s;\bfeta)
   +\frac{m_{j}}{\alpha_{j}}\ddot{x}_{js} \right]^{2} 
\label{LagraOM1}
\end{eqnarray}
with the normalization constant $C_{x}$. 
   [See also, for example, Ref. \cite{R89} for a derivation 
of the probability functional (\ref{ProbaFunctX1})  
via the Fokker-Planck equation corresponding 
to the Langevin equation (\ref{LangeEquat1}).]

%%%%%%%%%%%%%%%%%%%%%%%%%%%%%%%%%%%%%%%%%%%%%%%%%%%%%%%%%%%%%%%%%%%%%
\subsection{Time-Reversal in NESS} 
\label{TimeReverProce}

   Time-reversal plays a crucial role   
in nonequilibrium thermodynamics.  
%   First of all, the time-irreversible feature of thermodynamic 
%systems is represented as the second law of thermodynamics. 
   For example, the Onsager-Casimir symmetry relations between 
two linear transport coefficients have a different sign, 
depending on the behavior of thermodynamic 
variables under time reversal  \cite{O31a,C45}.      
   Moreover, in the Onsager-Machlup fluctuation theory 
around equilibrium \cite{OM53}, the entropy production rate 
is directly related to the difference of a Lagrangian 
for a forward path and the corresponding Lagrangian 
for a time-reversed (or backward) path. 
   In the next subsection we will discuss a generalization of 
this argument for the entropy production around equilibrium states 
to NESSs.  
   But first, before such a discussion, we must clarify 
an essential difference of a time-reversal procedure in NESSs  
from that in equilibrium states. 

   In equilibrium states, the time-reversal of the dynamics  
is unique and is simply given by a change of sign of the particle 
velocity $\dot{\bfx}_{s}$. 
   On the other hand, the time-reversal in NESSs is not unique.  
   This is due to the two independent kinds of motions in 
such states:  
an \emph{internal} intrinsic particle motion given by 
% their position change 
$\dot{\bfx}_{s}$, %(or a charge change $\dot{q}_{s}$) 
but, in addition, by an \emph{externally} induced motion 
characterized by the external 
nonequilibrium parameter $\bfmu$. %\cite{EM90}. 
   Therefore, in NESSs, we have two choices for a time-reversal 
procedure of the dynamics: 
either a change of sign of $\dot{\bfx}_{s}$ only, which has always 
to be done to obtain a time-reversed path, 
or a change of the the signs of both $\dot{\bfx}_{s}$ and $\bfmu$. 
   To discuss these two time-reversal procedures explicitly, 
we introduce the time-reversal operator $\hat{I}_{\pm}$ 
for NESSs by   
\begin{eqnarray}
   \hat{I}_{\pm}X(\{\bfx_{s}\};\bfmu) 
      = X(\{\bfx_{t+t_{0}-s}\};\pm\bfmu) 
\label{TimeRever1}
\end{eqnarray}
for any functional $X(\{\bfx_{s}\};\bfmu)$ of the path 
$\{\bfx_{s}\}_{s\in[t_{0},t]}$ and the external nonequilibrium 
parameter $\bfmu$.% 
\footnote{
Here and in the rest of the paper we adopt the convention 
that any equation containing the symbols $\pm$ 
on the left- and right-hand sides, denote  
\emph{two} equations, one with only  
the upper symbol ($+$ in $\pm$) 
and the other with only the lower symbol ($-$ in $\pm$). 
} 
   Under this time-reversal operation, the direction of 
motion of the particle on the forward path 
$\{\bfx_{s}\}_{s\in[t_{0},t]}$ in the functional 
$X(\{\bfx_{s}\};\bfmu)$ is transformed into 
$\{\bfx_{t+t_{0}-s}\}_{s\in[t_{0},t]}$ with 
the same geometry of the path but with  
the initial and final positions  
on the forward path (on the time-reversed path) 
given by $\bfx_{t_{0}}$ and $\bfx_{t}$ ($\bfx_{t}$ 
and $\bfx_{t_{0}}$), respectively. 
   This time-reversal operation for the internal motion  
represented by the particle position $\bfx_{s}$ is 
indicated by the hat $\hat{ }$ on the operator $\hat{I}_{\pm}$.  
   On the other hand, the other time-reversal procedure associated 
with the external nonequilibrium parameter $\bfmu$ 
as well is referred to by adding the subscripts $\pm$ 
in the operator $\hat{I}_{\pm}$, 
so that $\hat{I}_{-}$ ($\hat{I}_{+}$) does change (does not change) 
the sign of the nonequilibrium parameter $\bfmu$
under this time-reversal operation. 
%   The time-reversal operator in Eq. (\ref{TimeRever1}) 
%depends on the time-interval $[t_{0},t]$, i.e. on the  
%choice of its initial time $t_{0}$ and its final time $t$, 
%but we suppress such dependences of the operator $\hat{I}_{\pm}$ 
%in order to simplify the notation. 

   So far, we have chosen the initial time $t_{0}$ and the final 
time $t$ independently, to make clear their roles.  
   However, for convenience, in the remaining part of this paper, 
we choose, without loss of generality, the origin of the time 
in the middle of the initial time $t_{0}$ and the final time $t$, 
so that $t_{0}=-t$. 
   By taking this origin of the time, the 
length of the time interval for $s\in[t_{0},t]$ is given by 
$t-t_{0}=2t$. 

   We now discuss some properties of the time-reversal 
operator $\hat{I}_{\pm}$ useful for later.  
   First, by Eq. (\ref{TimeRever1}) the time-reversal operator 
$\hat{I}_{\pm}$ satisfies the relation 
\begin{eqnarray}
   \hat{I}_{\pm}{}^{2} = 1. 
\label{ProjeI1}
\end{eqnarray}
   Second, it can also be shown for this time-reversal operator 
that 
\begin{eqnarray}
   \hat{I}_{\pm}\int_{-t}^{t} ds\; 
      Y\!\left(\ddot{\bfx}_{s},\dot{\bfx}_{s},\bfx_{s},s;\bfmu\right) 
      &=& \int_{-t}^{t} ds\; Y
      \!\left(\frac{d^{2}\bfx_{-s}}{ds^{2}},
      \frac{d\bfx_{-s}}{ds},\bfx_{-s},s;\pm\bfmu\right) 
      \\
   &=& \int_{-t}^{t} ds\; 
      Y\!\left(\ddot{\bfx}_{s},-\dot{\bfx}_{s},\bfx_{s},
      -s;\pm\bfmu\right), 
\label{TimeRever2} 
\end{eqnarray}
%
%with $t+t_{0}=0$, 
for any function $Y(\ddot{\bfx}_{s},\dot{\bfx}_{s},
\bfx_{s},s;\bfmu)$ of $\ddot{\bfx}_{s}$, $\dot{\bfx}_{s}$, 
$\bfx_{s}$, $s$ and $\bfmu$. 
   Eq. (\ref{TimeRever2}) means that the effect 
of the time-reversal operator $\hat{I}_{\pm}$ on  
a functional of the form $\int_{-t}^{t} ds\; 
Y(\ddot{\bfx}_{s},\dot{\bfx}_{s},\bfx_{s},s;\bfmu)$ 
is expressed not only by the change 
$\dot{\bfx}_{s}\rightarrow-\dot{\bfx}_{s}$ 
of the (internal) particle velocity, but also by the changes 
$s\rightarrow -s$ 
of the explicit time-dependence,  
and by $\bfmu\rightarrow\pm\bfmu$ 
for the external nonequilibrium parameter in the function 
$Y(\ddot{\bfx}_{s},\dot{\bfx}_{s},\bfx_{s},s;\bfmu)$.

%%%%%%%%%%%%%%%%%%%%%%%%%%%%%%%%%%%%%%%%%%%%%%%%%%%%%%%%%%%%%%%%%%%%%
\subsection{Dissipation Functions, Entropy Production and the Second Law of Thermodynamics} 
\label{SeconLawTherm}

   In this and the next subsections, 
using the time-reversal procedure introduced   
in the previous subsection \ref{TimeReverProce}, we 
formulate a generalized Onsager-Machlup theory  
for NESSs in three steps: 
[i] calculation of the entropy production rate 
as the time-irreversible part of the Lagrangian, 
leading to the second law of thermodynamics 
(Sec. \ref{SeconLawTherm}),   
[ii] introduction of the heat via the entropy production rate  
(Sec. \ref{HeatWork}), and 
[iii] introduction of the work from the heat and 
an internal energy difference using the energy conservation 
or the first law of thermodynamics 
(Sec. \ref{HeatWork}). 

   To discuss the first step 
for the NESS Onsager-Machlup theory, 
we separate the Lagrangian $\caL$ 
into a time-reversal invariant (even) part and 
a time-irreversible (odd) part as  
\begin{eqnarray}
   \caL\!\left(\ddot{\bfx}_{s},\dot{\bfx}_{s},\bfx_{s},s;\bfmu\right) 
      = -\frac{1}{2 k_{B}}
      \left[\Phi_{\pm}(\ddot{\bfx}_{s},\bfx_{s},s;\bfmu) 
      -\dot{S}_{\pm}(\ddot{\bfx}_{s},\dot{\bfx}_{s},
      \bfx_{s},s;\bfmu)\right] ,
\label{LagraOM3}
\end{eqnarray}
where $\Phi_{\pm}$ and $\dot{S}_{\pm}$ are defined by 
\begin{eqnarray}
   \Phi_{\pm}(\ddot{\bfx}_{s},\dot{\bfx}_{s},\bfx_{s},s;\bfmu)
      &\equiv& - k_{B}
      \left[\caL\!\left(\ddot{\bfx}_{s},\dot{\bfx}_{s},
      \bfx_{s},s;\bfmu\right) 
      + \caL\!\left(\ddot{\bfx}_{s},-\dot{\bfx}_{s},
      \bfx_{s},-s;\pm\bfmu\right)\right] ,
      \label{DissiFunct1}
   \\
   \dot{S}_{\pm}(\ddot{\bfx}_{s},\dot{\bfx}_{s},\bfx_{s},s;\bfmu) 
   &\equiv& k_{B}\left[\caL\!\left(\ddot{\bfx}_{s},\dot{\bfx}_{s},
      \bfx_{s},s;\bfmu\right) 
      - \caL\!\left(\ddot{\bfx}_{s},-\dot{\bfx}_{s},
      \bfx_{s},-s;\pm\bfmu\right)\right] . 
\label{EntroProdu1}
\end{eqnarray}
   From Eqs. (\ref{DissiFunct1}) and (\ref{EntroProdu1}), 
using the property (\ref{TimeRever2}) for the time-reversal operator  
$\hat{I}_{\pm}$, we obtain  
\begin{eqnarray}
   \hat{I}_{\pm}\int_{-t}^{t}ds\; 
       \Phi_{\pm}(\ddot{\bfx}_{s},\dot{\bfx}_{s},
          \bfx_{s},s;\bfmu) &=& 
       \int_{-t}^{t}ds\; 
       \Phi_{\pm}(\ddot{\bfx}_{s},\dot{\bfx}_{s},\bfx_{s},s;\bfmu) , 
       \\
   \hat{I}_{\pm} \int_{-t}^{t}ds\; 
      \dot{S}_{\pm}\!(\ddot{\bfx}_{s},\dot{\bfx}_{s},
         \bfx_{s},s;\bfmu) 
      &=& -\int_{-t}^{t}ds\; 
      \dot{S}_{\pm}\!(\ddot{\bfx}_{s},\dot{\bfx}_{s},
      \bfx_{s},s;\bfmu) 
      \label{ReverEntro1}
\end{eqnarray}
for the time-reversal part $\Phi_{\pm}$ 
and the time-irreversible part $\dot{S}_{\pm}$ 
of the Lagrangian $\caL$, respectively. 

   A major point in the Onsager-Machlup theory  
is then that the odd part of the Lagrangian is 
identified with the entropy production rate \cite{OM53}.  
%while the time-reversal part $\Phi_{\pm}$ is connected 
%to the so-called dissipation functions (defined later). 
%   \cor{To discuss this point for NESSs, we restrict ourself 
%in the cases of (i) both $\dot{S}_{+}$ ($\Phi_{+}$) 
%and $\dot{S}_{-}$  ($\Phi_{-}$) 
%for $\Delta T_{j}=0$ or (ii) only $\dot{S}_{+}$ ($\Phi_{+}$) 
%for $\Delta T_{j}\neq 0$, hereafter 
%in Secs \ref{OMtheory} and \ref{WorkFluctTheor}.}%
%\footnote{
%\cor{
%   As will be shown in Sec. \ref{WorkModel},  
%these two cases (i) and (ii) are enough to discuss the physical 
%work and heat to maintain a NESS for all models introduced 
%in Sec. \ref{NESSmodel}.
%}}
   To discuss the physical interpretation for  
the odd part $\dot{S}_{\pm}$ 
of the Lagrangian $\caL$ for NESSs, 
we first have to give more explicit forms for $\dot{S}_{\pm}$ 
and $\Phi_{\pm}$. 
%   For these cases, we first give more explicit forms 
%for $\dot{S}_{\pm}$ and $\Phi_{\pm}$. 
   By inserting the Lagrangian (\ref{LagraOM1}) into 
Eqs. (\ref{DissiFunct1}) and (\ref{EntroProdu1}), we obtain  
\begin{eqnarray}
   \Phi_{\pm}(\ddot{\bfx}_{s},\bfx_{s},s;\bfmu)
      &=& \Phi_{\pm}^{(1)}(\dot{\bfx}_{s},\bfx_{s},s;\bfmu) 
      +\Phi_{\pm}^{(2)}(\dot{\bfx}_{s},\bfx_{s},s;\bfmu) 
      +\Phi_{\pm}^{(3)}(\dot{\bfx}_{s},\bfx_{s},s;\bfmu),
      \label{DissiFunct2}
   \\
   \dot{S}_{\pm}(\ddot{\bfx}_{s},\dot{\bfx}_{s},\bfx_{s},s;\bfmu) 
   &=&  \sum_{j=1}^{N}\frac{1}{T_{j}}
      \left[F_{j\pm}^{(e)}(\bfx_{s},s;\bfeta)-m_{j}\ddot{x}_{js}
      \right]\left[\dot{x}_{js} 
      - \frac{1}{\alpha_{j}}F_{j\pm}^{(o)}(\bfx_{s},s;\bfeta)\right] 
      \nonumber \\
   &&\spaEq 
      +\Phi_{\pm}^{(3)}(\dot{\bfx}_{s},\bfx_{s},s;\bfmu).
\label{EntroProdu2}
\end{eqnarray}
where $\Phi_{\pm}^{(k)}$, $k=1,2,3$ are defined by 
\begin{eqnarray}
   \Phi_{\pm}^{(1)}(\dot{\bfx}_{s},\bfx_{s},s;\bfmu) &\equiv& 
      \sum_{j=1}^{N}\frac{\alpha_{j}}{2T_{j}}\left[\dot{x}_{js} 
      -\frac{1}{\alpha_{j}}
      F_{j\pm}^{(o)}(\bfx_{s},s;\bfeta)\right]^{2} ,
      \label{DissiFunct3a}\\
   \Phi_{\pm}^{(2)}(\ddot{\bfx}_{s},\bfx_{s},s;\bfmu) &\equiv& 
      \sum_{j=1}^{N}\frac{1}{2\alpha_{j}T_{j}}
      \left[F_{j\pm}^{(e)}(\bfx_{s},s;\bfeta)
      -m_{j}\ddot{x}_{js} \right]^{2}
      \label{DissiFunct3b} \\
%   \Phi_{+}^{(3)}(\dot{\bfx}_{s},\bfx_{s},s;\bfmu) &\equiv& 0
%      \label{UpsilonFunct1} \\
   \Phi_{\pm}^{(3)}(\dot{\bfx}_{s},\bfx_{s},s;\bfmu) &\equiv& 
      \left[1+(-1)^{\pm 1}\right]
      \sum_{j=1}^{N} \frac{\alpha_{j}}{2} 
      \frac{\Delta T_{j}}{\overline{T}^{2}- \Delta T_{j}^{2}} 
      \nonumber \\
   &&\spaEq \times
      \left[\dot{x}_{js} 
      + \frac{1}{\alpha_{j}} F_{j}(\bfx_{s},-s;\pm\bfeta)
      -\frac{m_{j}}{\alpha_{j}}\ddot{x}_{js} \right]^{2}
      \label{UpsilonFunct2}
\end{eqnarray}
and $F_{j\pm}^{(e)}$ and $F_{j\pm}^{(o)}$ are defined by 
\begin{eqnarray}
   F_{j\pm}^{(e)}(\bfx_{s},s;\bfeta) &\equiv& 
      \frac{1}{2}\left[F_{j}(\bfx_{s},s;\bfeta)
      +F_{j}(\bfx_{s},-s;\pm\bfeta)\right] , 
      \label{ForceEven1} \\ 
   F_{j\pm}^{(o)}(\bfx_{s},s;\bfeta) &\equiv& 
      \frac{1}{2}\left[F_{j}(\bfx_{s},s;\bfeta)
      -F_{j}(\bfx_{s},-s;\pm\bfeta)\right]  
      \label{ForceOdd1}
\end{eqnarray} 
as the even (e) part and the odd (o) part 
of the force $F_{j}= F_{j\pm}^{(e)}+ F_{j\pm}^{(o)}$, respectively, 
under the time-reversal procedures $s\rightarrow -s$ 
and either $\bfmu \rightarrow +\bfmu$ or $\bfmu \rightarrow -\bfmu$.   
%in a similar way to Eqs. (\ref{DissiFunct1}) and (\ref{EntroProdu1}) 
%for the even part $\Phi_{\pm}$ and the odd part $\dot{S}_{\pm}$ 
%of the Lagrangian $\caL$. 
%$F_{j}(\bfx_{s},s;\bfeta) 
%= F_{j\pm}^{(e)}(\bfx_{s},s;\bfeta) 
%+ F_{j\pm}^{(o)}(\bfx_{s},s;\bfeta)$, respectively. 
   Here, $\Phi_{\pm}^{(1)}$ and $\Phi_{\pm}^{(2)}$  
correspond to the dissipation 
functions in the Onsager-Machlup theory \cite{TC07a}  
and $\Phi_{+}^{(3)}=0$. 
   Using Eqs. (\ref{LagraOM3}) and (\ref{DissiFunct2}) the Lagrangian 
can be represented as the sum of the dissipation functions, 
$\Phi_{+}^{(3)}$ and 
the minus entropy production rate, i.e.  
$\caL=-[1/(2k_{B})][\Phi_{\pm}^{(1)}+\Phi_{\pm}^{(2)} 
+ \Phi_{\pm}^{(3)}-\dot{S}_{\pm}]$ 
for NESSs, 
in a similar way as in the Onsager-Machlup theory 
for equilibrium states. 

   Eqs. (\ref{DissiFunct3a}) and (\ref{DissiFunct3b}) 
show that the dissipation functions $\Phi_{\pm}^{(1)}$ 
and $\Phi_{\pm}^{(2)}$ are always non-negative 
and time-reversal invariant, i.e. $\Phi_{\pm}^{(k)} \geq 0$ and 
$\hat{I}_{\pm} \int_{-t}^{t}ds\; \Phi_{\pm}^{(k)} = 
\int_{-t}^{t}ds\; \Phi_{\pm}^{(k)}$, $k=1,2$.  
   This non-negativity of the dissipation functions is directly 
related to the non-negativity of the \emph{average} entropy production 
rate, namely the second law of thermodynamics, in the linear regime. 
   To show this, we note that in our NESS Onsager-Machlup 
theory, the \emph{average} path 
$\{\langle\bfx_{s}\rangle\}_{s\in[-t,t]}$ 
is given by the variational principle 
\begin{eqnarray}
    \caL\!\left(\ddot{\bfx}_{s},\dot{\bfx}_{s},
   \bfx_{s},s;\bfmu\right) = \mbox{minimum} \;\;\;\; 
   \mbox{for}\;\; \bfx_{s} = \langle\bfx_{s}\rangle ,
\label{AveraX1}
\end{eqnarray}
leading to the average Langevin equation  
$m\langle\ddot{x}_{js}\rangle 
= F_{j}(\langle\bfx_{s}\rangle,s;\bfeta) 
-\alpha_{j} \langle\dot{x}_{js}\rangle$ using 
Eqs. (\ref{LineaForce1}) and (\ref{LagraOM1}), i.e. 
\begin{eqnarray}
   \langle\dot{x}_{js}\rangle 
   -\frac{1}{\alpha_{j}}
   F_{j\pm}^{(o)}(\langle\bfx_{s}\rangle,s;\bfeta)
   =
   \frac{1}{\alpha_{j}}\left[ 
   F_{j\pm}^{(e)}(\langle\bfx_{s}\rangle,s;\bfeta) 
   - m\langle\ddot{x}_{js}\rangle\right]  
\label{LangeAvera1}
\end{eqnarray}
with Eqs. (\ref{ForceEven1})  and (\ref{ForceOdd1}). 
   Using Eq. (\ref{LangeAvera1}), we see that 
2 times the dissipation functions 
(\ref{DissiFunct3a}) and (\ref{DissiFunct3b}) 
and the entropy production rate (\ref{EntroProdu2}) 
minus $\Phi_{\pm}^{(3)}$ 
coincide with each other for the average path, i.e.: 
\begin{eqnarray}
    && \dot{S}_{\pm}\!(\langle\ddot{\bfx}_{s}\rangle,
      \langle\dot{\bfx}_{s}\rangle,\langle\bfx_{s}\rangle,s;\bfmu) 
      - \Phi_{\pm}^{(3)}(\langle\dot{\bfx}_{s}\rangle,
      \langle\bfx_{s}\rangle,s;\bfmu) 
      \nonumber \\
   &&\spaEq = 
   2\Phi_{\pm}^{(1)}(\langle\dot{\bfx}_{s}\rangle,
      \langle\bfx_{s}\rangle,s;\bfmu) 
   = 2\Phi_{\pm}^{(2)}(\langle\ddot{\bfx}_{s}\rangle,
      \langle\bfx_{s}\rangle,s;\bfmu) .
\label{AveraEntro1}
\end{eqnarray}
   Combining Eq. (\ref{AveraEntro1}) with the non-negativity 
 of the dissipation functions $\Phi_{\pm}^{(1)}$ 
and $\Phi_{\pm}^{(2)}$ we obtain 
\begin{eqnarray}
   \dot{S}_{\pm}\!(\langle\ddot{\bfx}_{s}\rangle,
   \langle\dot{\bfx}_{s}\rangle,\langle\bfx_{s}\rangle,s;\bfmu) 
      - \Phi_{\pm}^{(3)}(\langle\dot{\bfx}_{s}\rangle,
      \langle\bfx_{s}\rangle,s;\bfmu) 
    \geq 0. 
\label{SeconLaw1}
\end{eqnarray}
   This is a statement of the second law of thermodynamics 
in our generalized Onsager-Machlup theory for NESSs.%
\footnote{As will be shown in Sec. \ref{WorkModel},  
the term $\Phi_{\pm}^{(3)}$ disappears 
for the physical entropy production rate 
to maintain a NESS for all models introduced 
in Sec. \ref{NESSmodel}. 
} 
%justifying that $\dot{S}_{\pm}(\ddot{\bfx}_{s},\dot{\bfx}_{s},
%\bfx_{s},s;\bfmu)$ is regarded as the entropy production rate. 
   We note that Eq. (\ref{SeconLaw1}) expresses a   
non-negativity of $\dot{S}_{\pm} - \Phi_{\pm}^{(3)}$ for 
the average path $\{\langle\bfx_{s}\rangle\}_{s\in [-t,t]}$,  
but the quantity $\dot{S}_{\pm} - \Phi_{\pm}^{(3)}$ itself 
can be negative for other paths, 
in contrast to the dissipation functions  $\Phi_{\pm}^{(1)}$ 
and $\Phi_{\pm}^{(2)}$,  
which are always non-negative for any path.

%%%%%%%%%%%%%%%%%%%%%%%%%%%%%%%%%%%%%%%%%%%%%%%%%%%%%%%%%%%%%%%%%%%%%
\subsection{Heat, Energy, Work and the First Law of Thermodynamics}
\label{HeatWork}

   We now discuss, as the main results of this paper, 
the appropriate heat and work to  
maintain the NESSs from the entropy production rate 
$\dot{S}_{\pm}$ discussed in the previous subsection 
\ref{SeconLawTherm}. 

   Using the entropy production rate $\dot{S}_{\pm}$, 
we introduce the heat $\caQ_{\pm}$ produced in the system 
on the trajectory $\{\bfx_{s}\}_{s\in [t_{0},t]}$ by  
\begin{eqnarray}
   \caQ_{\pm}(\{\bfx_{s}\};\bfmu) &\equiv& \overline{T} 
      \int_{-t}^{t}ds\; 
      \dot{S}_{\pm}\!(\ddot{\bfx}_{s},\dot{\bfx}_{s},
      \bfx_{s},s;\bfmu) .   
      \label{Heat1} 
\end{eqnarray}
  Inserting Eq. (\ref{EntroProdu2}) into (\ref{Heat1}) 
and using $T_{j}^{-1}
%=\overline{T}^{-1}-\Delta T_{j}\;\overline{T}^{-2}
=\overline{T}^{-1}[1-\Delta T_{j}/\overline{T}] 
+\mathcal{O}(|\Delta T_{j}/\overline{T}|^{2})$ 
and the condition (\ref{SmallTempe1}) we obtain 
\begin{eqnarray}
   && \caQ_{\pm}(\{\bfx_{s}\};\bfmu)  
      \nonumber \\
   &&\spaEq=   \sum_{j=1}^{N}  
      \left(1-\frac{\Delta T_{j}}{\overline{T}}\right)
      \int_{-t}^{t}ds\;
      \left[F_{j\pm}^{(e)}(\bfx_{s},s;\bfeta)-m_{j}\ddot{x}_{js}
      \right]
%      \nonumber \\
%   &&\spaEq\spaEq \times
      \left[\dot{x}_{js} 
      - \frac{1}{\alpha_{j}}F_{j\pm}^{(o)}(\bfx_{s},s;\bfeta)\right] 
      \nonumber \\
   &&\spaEq\spaEq +  \left[1+(-1)^{\pm 1}\right]
      \sum_{j=1}^{N}  \frac{\alpha_{j}}{2} 
      \frac{\Delta T_{j}}{\overline{T}} \int_{-t}^{t}ds\;  
      \left[\dot{x}_{js} 
      + \frac{1}{\alpha_{j}} F_{j}(\bfx_{s},-s;\pm\bfeta)
      -\frac{m_{j}}{\alpha_{j}}\ddot{x}_{js} \right]^{2}
      \nonumber \\
   &&\spaEq\spaEq
      +\mathcal{O}\left(\left|\frac{\Delta T_{j}}{\overline{T}}
      \right|^{2}\right)  
\label{Heat2}
\end{eqnarray}
as a concrete form of the heat $\caQ_{\pm}$
up to the first order in $\Delta T_{j}/\overline{T}$. 
   The term involving $\Delta T_{j}$ 
on the right-hand side of Eq. 
(\ref{Heat2}) gives the heat produced by the system  
with temperature differences between reservoirs.  
%   It should be noted that $\caQ_{\pm}$ is the heat 
%produced from the system, and is not the heat 
%current from a high temperature region 
%to a low temperature region. 

   We now discuss properties of the heat $\caQ_{\pm}$ 
from its definition (\ref{Heat1}). 
   We first note that due to Eqs. (\ref{ReverEntro1}) 
and (\ref{Heat1}) the heat $\caQ_{\pm}$ is anti-symmetric 
under time-reversal, i.e.  
\begin{eqnarray}
   \hat{I}_{\pm} 
      \caQ_{\pm}(\{\bfx_{s}\};\bfmu)
      = - \caQ_{\pm}(\{\bfx_{s}\};\bfmu) .
\label{ReverHeat1}
\end{eqnarray}
   Using Eqs. (\ref{ProbaFunctX1}), (\ref{EntroProdu1}) 
and (\ref{Heat1}) we can also show that  
\begin{eqnarray}
   \caQ_{\pm}(\{\bfx_{s}\};\bfmu) &=& 
      \overline{\beta}^{-1}\int_{-t}^{t}ds\;\left[ 
      \caL\!\left(\ddot{\bfx}_{s},\dot{\bfx}_{s},
         \bfx_{s},s;\bfmu\right) 
      - \caL\!\left(\ddot{\bfx}_{s},-\dot{\bfx}_{s},
         \bfx_{s},-s;\pm\bfmu\right)\right]
      \label{Heat3}\\
   &=& \overline{\beta}^{-1}\ln\frac{\caP_{x}(\{\bfx_{s}\};\bfmu)}
      {\hat{I}_{\pm}\caP_{x}(\{\bfx_{s}\};\bfmu)} .
      \label{Heat4}
\end{eqnarray}
with $\overline{\beta}\equiv 1/(k_{B}\overline{T})$. 
%   [It may be noted that Eq. (\ref{ReverHeat1}) can also be 
%derived directly from Eq. (\ref{Heat4}) using 
%Eq.(\ref{ProjeI1}).] %$\hat{I}_{\pm}^{2}=1$.]
   Eq. (\ref{Heat3}) connects directly the heat $\caQ_{\pm}$  
with a time-irreversible part of the Lagrangian $\caL$.  
   Eq. (\ref{Heat4}) implies that the behavior of the dynamics 
under time-reversal %irreversibility of the dynamics  
makes the probability functional $\caP_{x}(\{\bfx_{s}\};\bfmu)$ 
for the forward path not equal to the corresponding  
probability functional $\hat{I}_{\pm}\caP_{x}(\{\bfx_{s}\};\bfmu)$ 
for the backward path and a nonzero heat is due to this 
non-equality. 

   We now proceed to introduce the internal energy and then 
the work using the energy conservation law   
as a relation among the heat, the internal energy and the work. 
%   First, we introduce the internal energy in our general theory. 
%   We first note that in physical processes 
%the heat is caused by two sources: a change of the internal energy 
%stored in the system and an external work. 
%   So, we now discuss the separation of the heat  
%into the internal energy change and the external work 
%in our general formulation. 
%   To discuss this separation of the heat into the internal 
%energy change and the external work for NESSs, 
   To introduce the internal energy,  
we first separate the even part 
$F_{j\pm}^{(e)}$ of the force $F_{j}$ 
into a force due to the internal potential $U_{\pm}$ and 
a force $f_{j\pm}$ due to the external driving force 
(e.g. an external electric force on a charged particle): 
\begin{eqnarray}
   F_{j\pm}^{(e)}(\bfx_{s},s;\bfeta) 
   = - \frac{\partial U_{\pm}(\bfx_{s},s;\bfeta)}
   {\partial \bfx_{s}} + f_{j\pm}(\bfx_{s},s;\bfeta) .
\label{Poten1}
\end{eqnarray}
%
%   Usually, $f_{j\pm}$ is chosen as a force 
%to cause a nonequilibrium particle current, 
%e.g. an external electric force 
%for a charged particle, etc..  
%and does not contribute to the internal potential energy 
%because this force 
%is regarded as the one to do the work only to 
%drive the system in a NESS without 
%injecting aenergy into the system.   
   The separation of the force into a potential 
force and an external driving force, like in Eq. (\ref{Poten1}),  
has already been used before, cf. Ref. \cite{K98}.  
   Using the potential $U_{\pm}$, we next introduce 
the internal energy $E_{\pm}$ by 
\begin{eqnarray}
   E_{\pm}(\dot{\bfx}_{s},\bfx_{s},s;\bfeta)  
   \equiv \sum_{j=1}^{N} \frac{1}{2} m_{j} 
   |\dot{x_{j}}_{s}|^{2} + U_{\pm}(\bfx_{s},s;\bfeta)
\label{Energ1}
\end{eqnarray}
as the sum of the kinetic energy 
$\sum_{j=1}^{N} (1/2) m_{j} |\dot{x_{j}}_{s}|^{2}$ 
and the potential energy $U_{\pm}$. 
   Using this energy, we introduce  
the energy difference $\Delta E_{\pm}$ by 
\begin{eqnarray}
   \Delta E_{\pm} \equiv E_{\pm}(\dot{\bfx}_{t},\bfx_{t},t;\bfeta) 
   - E_{\pm}(\dot{\bfx}_{-t},\bfx_{-t},-t;\bfeta)  
\label{EnergDiffe1}
\end{eqnarray}
as the difference of the internal energy at the final time $t$ 
and the initial time $t_{0}=-t$. 
   
   Now we will introduce the work. 
   We first note that in physical processes 
the external work is transformed into heat 
and a change of the internal energy,  
%the heat is caused by two sources: a change of the internal energy 
%stored in the system and an external work, 
as expressed in the energy conservation law.  
   From this, the work $\caW_{\pm}$ done along the trajectory 
$\{\bfx_{s}\}_{s\in[-t,t]}$ is then defined by  
\begin{eqnarray}
   \caW_{\pm}(\{\bfx_{s}\};\bfmu) 
   = \caQ_{\pm}(\{\bfx_{s}\};\bfmu) + \Delta E_{\pm} . 
\label{Work1}
\end{eqnarray}
   Eq. (\ref{Work1}) leads to an expression of the first law 
of thermodynamics for NESSs, by taking its functional average. 

   As a remark about the internal energy and the work, 
in the above argument we first introduced the heat 
$\caQ_{\pm}$ by Eq. (\ref{Heat1}), 
then separated it into the energy difference 
$\Delta E_{\pm}$ and the work $\caW_{\pm}$ via 
$\caQ_{\pm} = \caW_{\pm} - \Delta E_{\pm}$, i.e. Eq. (\ref{Work1}). 
   However, this separation of the heat into the work 
and the energy difference is not unique, due to the non-uniqueness 
of the separation (\ref{Poten1}) of the force $F_{j\pm}^{(e)}$ 
into an external force $f_{j\pm}$ and 
a potential force $- \partial U_{\pm}/\partial \bfx_{s}$, 
which introduces the potential $U_{\pm}$. 
   This non-uniqueness of the potential actually 
happens in the models discussed in Sec. \ref{ClassB} 
where the dragged Brownian particle model 
and the driven torsion pendulum model are described by 
the same Langevin equations, but have 
different internal energies.  
   This non-uniqueness, or ambiguity, 
in the introduction of a potential, 
by a separation of the force on the particles 
into a force due to an internal potential 
and an external driving force, 
can only be resolved for specific models, it seems, 
on physical grounds, rather than by mathematical arguments alone. 
  
   A related remark about the above argument 
to introduce the work and the energy in a NESS  
is that we \emph{assumed} that 
there is no contribution from 
the odd part $F_{j\pm}^{(o)}$ of the force $F_{j}$  
to the internal potential $U_{\pm}$, 
therefore to the energy $E_{\pm}$. 
   This implies that the internal energy $E_{\pm}$ 
must be time-reversal invariant, i.e.  
\begin{eqnarray}
   E_{\pm}(-\dot{\bfx}_{s},\bfx_{s},-s;\pm\bfeta)
   = E_{\pm}(\dot{\bfx}_{s},\bfx_{s},s;\bfeta) .
\label{ReverEnerg1}
\end{eqnarray}
   Using Eq. (\ref{ReverEnerg1}) we obtain 
\begin{eqnarray}
   \hat{I}_{\pm}\Delta E_{\pm}  
   = \hat{I}_{\pm}\int_{-t}^{t}ds\; 
   \frac{d E_{\pm}(\dot{\bfx}_{s},\bfx_{s},s;\bfeta)}{ds} 
   = -\Delta E_{\pm} , 
\label{ReverEnerg2}
\end{eqnarray} 
i.e. the energy difference $\Delta E_{\pm}$ is 
anti-symmetric under time-reversal.  
   Eqs. (\ref{ReverHeat1}), (\ref{Work1}) and (\ref{ReverEnerg2}) 
lead to 
\begin{eqnarray}
   \hat{I}_{\pm}\caW_{\pm}(\{\bfx_{s}\};\bfmu) 
   = -\caW_{\pm}(\{\bfx_{s}\};\bfmu) , 
\label{ReverWork1}
\end{eqnarray}
so that the work done on a backward path has 
the same magnitude but the opposite sign 
of the work done on the corresponding forward path.  
   Eq. (\ref{ReverWork1}) will play an important role to derive 
the correct work fluctuation theorem in Sec. \ref{FluctRelatFED}. 

   To discuss the physical meaning of the work 
$\caW_{\pm}$ defined formally by Eq. (\ref{Work1}), 
we now give it in more explicit form. 
   Inserting Eqs. (\ref{Heat2}) and (\ref{EnergDiffe1}) into Eq. 
(\ref{Work1}) we obtain 
\begin{eqnarray}
    \caW_{\pm}(\{\bfx_{s}\};\bfmu) 
    &=& \caW_{\pm}^{(e)}(\{\bfx_{s}\};\bfeta) 
    + \caW_{\pm}^{(f)}(\{\bfx_{s}\};\bfeta)
    +  \caW_{\pm}^{(o)}(\{\bfx_{s}\};\bfeta)  
      \nonumber \\
   &&\spaEq
    + \caW_{\pm}^{(t)}(\{\bfx_{s}\};\bfmu) 
      +\mathcal{O}\left(\left|\frac{\Delta T_{j}}{\overline{T}}
      \right|^{2}\right) 
\label{Work2}
\end{eqnarray}
%
%\pagebreak
where $\caW_{\pm}^{(e)}$, $\caW_{\pm}^{(f)}$,  
$ \caW_{\pm}^{(o)}$ and $\caW_{\pm}^{(t)}$ are defined by 
\begin{eqnarray}
    \caW_{\pm}^{(e)}(\{\bfx_{s}\};\bfeta) 
       &\equiv& \int_{-t}^{t}ds\;\frac{\partial 
       E_{\pm}(\dot{\bfx}_{s},\bfx_{s},s;\bfeta)}{\partial s}, 
      \label{Work3a} \\
   \caW_{\pm}^{(f)}(\{\bfx_{s}\};\bfeta)
      &\equiv&  \sum_{j=1}^{N} \int_{-t}^{t}ds\;
      f_{j\pm}(\bfx_{s},s;\bfeta)\dot{x}_{js} ,
      \label{Work3b} \\ 
   \caW_{\pm}^{(o)}(\{\bfx_{s}\};\bfeta)
      &\equiv& - \sum_{j=1}^{N}\frac{1}{\alpha_{j}} \int_{-t}^{t}ds\;
      \left[F_{j\pm}^{(e)}(\bfx_{s},s;\bfeta)
      -m_{j}\ddot{x}_{js} \right]
      F_{j\pm}^{(o)}(\bfx_{s},s;\bfeta) ,  
      \label{Work3c} \\ 
   \caW_{\pm}^{(t)}(\{\bfx_{s}\};\bfmu) 
      &\equiv& - \sum_{j=1}^{N} \frac{\Delta T_{j}}{\overline{T}}
      \int_{-t}^{t}ds\; 
      \left[F_{j\pm}^{(e)}(\bfx_{s},s;\bfeta)-m_{j}\ddot{x}_{js}
      \right]\left[\dot{x}_{js} 
      - \frac{1}{\alpha_{j}}F_{j\pm}^{(o)}(\bfx_{s},s;\bfeta)\right] 
      \nonumber \\
   && \spaEq + \left[1+(-1)^{\pm 1}\right]
      \sum_{j=1}^{N}  \frac{\alpha_{j}}{2} 
      \frac{\Delta T_{j}}{\overline{T}} 
      \nonumber \\
   && \spaEq\spaEq \times
      \int_{-t}^{t}ds\;  
      \left[\dot{x}_{js} 
      + \frac{1}{\alpha_{j}} F_{j}(\bfx_{s},-s;\pm\bfeta)
      -\frac{m_{j}}{\alpha_{j}}\ddot{x}_{js} \right]^{2}.
%     \nonumber \\
     \label{Work3d}
\end{eqnarray}
[See Appendix \ref{WorkOMtheory} for a derivation of Eq. 
(\ref{Work2}).]
   In Eq. (\ref{Work2}) the total work $\caW_{\pm}$ has been 
separated into the four parts: $\caW_{\pm}^{(e)}$, $\caW_{\pm}^{(f)}$,   
$\caW_{\pm}^{(o)}$ and $\caW_{\pm}^{(t)}$. 
   The first part $\caW_{\pm}^{(e)}$ 
comes from the partial time-derivative of the energy (e) $E_{\pm}$.  
%as a contribution of a part of the even ($e$) part  
%$F_{j\pm}^{(e)}$ of the force $F_{j\pm}$.   
   This is the work used in Refs. \cite{J97,C98}. 
   The second part $\caW_{\pm}^{(f)}$ 
is the work done by the external driving force ($f$), $f_{\pm}$,  
while the third part $ \caW_{\pm}^{(o)}$ 
of the work is due to the odd ($o$) part 
$F_{j\pm}^{(o)}$ of the force. 
   We remark that this third part of the work  
includes a d'Alembert type force $-m\ddot{\bfx}_{s}$
as noted by Onsager and Machlup \cite{OM53} 
and also by the authors \cite{TC07a,TC07b}. 
   The last part $\caW_{\pm}^{(t)}$ is due to 
the temperature ($t$) differences $\Delta T_{j}$ among reservoirs. 
   In Sec. \ref{WorkModel}, we will consider   
concrete examples for the four different kinds of works 
$\caW_{\pm}^{(e)}$, $\caW_{\pm}^{(f)}$, $\caW_{\pm}^{(o)}$ 
and $\caW_{\pm}^{(t)}$ using the NESS models  
discussed in Sec. \ref{NESSmodel}.  
%appearing on the right-hand side of Eq. (\ref{Work2}), 
%respectively. 
%
   From the explicit form (\ref{Work2}) of the work, together  
with Eqs. (\ref{Work3a}), (\ref{Work3b}), (\ref{Work3c}) and 
(\ref{Work3d}), we can show that the work $\caW_{\pm}$ 
is a purely nonequilibrium quantity and vanishes at equilibrium,  
$\caW_{\pm}=0$,   
because the energy $E_{\pm}$ does not have an explicit 
time-dependence (so that $\caW_{\pm}^{(e)}=0$), 
$f_{j\pm}=0$ (so that $\caW_{\pm}^{(f)}=0$), 
$F_{j\pm}^{(o)}=0$ (so that $\caW_{\pm}^{(o)}=0$) 
and $\Delta T_{j}=0$ (so that $\caW_{\pm}^{(t)}=0$). 
   At equilibrium, where Onsager and Machlup formulated  
their fluctuation theory \cite{OM53}, the energy balance equation is 
simply given by $\caQ_{\pm} = - \Delta E_{\pm}$ 
from Eq. (\ref{Work1}) using $\caW_{\pm}=0$.

   Finally, we want to make some remarks about the ambiguities 
caused by the non-uniqueness of the time-reversal operator 
$\hat{I}_{\pm}$ in NESSs. 
   In this section we have discussed 
how to introduce thermodynamic quantities like the entropy 
production rate, the heat, the internal energy and the work, etc., 
suitable for a NESS.  
   However, so far we could only specify for each of them  
two possibilities, e.g. $\dot{S}_{+}$ or $\dot{S}_{-}$ 
for the entropy production rate, $\caW_{+}$ or $\caW_{-}$ 
for the work, and so on, due to the two possibilities 
($+$ or $-$) contained in the time-reversal operator 
$\hat{I}_{\pm}$ for NESSs. 
   The above arguments do not answer the question, 
which of these two actually represents the physical work, etc. 
to maintain a NESS.  
%   Note that this ambiguity in choice of the 
%NESS quantities is a purely nonequilibrium effect 
%related to the external nonequilibrium parameter $\bfmu$  
%and does not exist at equilibrium.    
   Both works $\caW_{+}$ or $\caW_{-}$ 
and heats $\caQ_{+}$ or $\caQ_{-}$ satisfy 
the energy conservation (first) law (\ref{Work1})  
and both entropy production rates  
$\dot{S}_{+}$ and $\dot{S}_{-}$ satisfy 
the second law of thermodynamics (\ref{SeconLaw1}). 
%in the linear case of Eq. (\ref{LineaForce1}). 
%
   One of the possible criteria to choose 
the physical work, etc. to maintain a NESS, is that we choose 
one of the two time-reversal operators $\hat{I}_{+}$ 
and $\hat{I}_{-}$ in such a way that the internal energy $E$ 
must have the time-reversal symmetry (\ref{ReverEnerg1}).  
   In this way, for example, as will be discussed  
in Sec. \ref{WorkModelB}, 
we can choose the correct time-reversal operator 
$\hat{I}_{-}$ for the dragged Brownian particle model 
in the laboratory frame, noting that the energy 
(\ref{EnergB1}) has the time-reversal symmetry 
for the time-reversal operator $\hat{I}_{-}$ with $\bfmu=v$. 
   Therefore, the quantities with the suffix ``${}_{-}$'',  
like $\caW_{-}$, $\dot{S}_{-}$ and $\caQ_{-}$, 
give the physical quantities to maintain a NESS  
for the dragged Brownian particle model. 
   Another criterion to choose the physical 
quantities to maintain a NESS  
is that the average of the work and the heat must be 
strictly positive, because in the NESS we always need to do 
positive average work (as well as remove positive average heat) 
to sustain the system in a NESS. 
%and such positive work is partly transformed 
%into a non-zero positive entropy production.  
   This condition also leads to a strictly positive 
entropy production in NESSs. 
   As we will see in  Sec.\ref{WorkModel}, 
for some NESS models, 
only \emph{one} of the averages of $\dot{S}_{+}$ and $\dot{S}_{-}$ 
is strictly positive while the other vanishes in 
NESSs. 
   In such a case, we can regard the strictly positive 
entropy production as the physical one to maintain a NESS    
and we can choose the heat, energy and work, etc., 
corresponding to this physical entropy production.  
%   (Note that the second law of thermodynamics 
%(\ref{SeconLaw1}) still allow the zero average 
%of the entropy production rate.) 
%
   In Sec. \ref{WorkModel}, we will illustrate the resolution 
of the above ambiguities in the  
choice for the physical heat and work, etc, using 
the systems discussed in Sec. \ref{NESSmodel}. 
   To discuss this point, hereafter we will use 
the terminology of the ``physical NESS'' heat and work 
for the heat and work 
to sustain a NESS, respectively.   
%   We also discuss a frame-dependence of work and heat. 
%   Besides, the separation of the even part 
%$F_{j\pm}^{(e)}$ as  
%the first and the second term of the right-hand side of 
%Eq. (\ref{Poten1}) is not unique mathematically. 

%%%%%%%%%%%%%%%%%%%%%%%%%%%%%%%%%%%%%%%%%%%%%%%%%%%%%%%%%%%%%%%%%%%%%%
\section{Fluctuation Theorems} 
\label{WorkFluctTheor}

   So far, we have discussed the definition of 
the entropy production, the heat, the internal energy 
and the work in NESSs. 
   These quantities are introduced as functions of a path, 
so that they include information not only 
of their ensemble averages (over all paths) 
but also of their fluctuations.  
   In this section %, as the last topic of this paper, 
we discuss their fluctuating properties in terms of    
fluctuation theorems using our NESS Onsager-Machlup theory. 
   We restrict our arguments to the   
NESS detailed balance condition  
and the corresponding transient 
fluctuation theorem where the initial state is an equilibrium state. 
   For other fluctuation theorems, like 
the asymptotic fluctuation theorem for any initial state,  
we refer to our previous papers \cite{TC07a,TC07b} 
for a dragged Brownian particle model.

%%%%%%%%%%%%%%%%%%%%%%%%%%%%%%%%%%%%%%%%%%%%%%%%%%%%%%%%%%%%%%%%%%%%%%
\subsection{Nonequilibrium Detailed Balance Relation}

   Up until now we have emphasized the role of the work $\caW_{\pm}$ 
to distinguish NESSs from equilibrium states.  
   However, this work also plays an important role in 
a detailed balance condition for NESSs, which 
we call the \emph{nonequilibrium detailed balance relation}.  
   This relation has been already discussed in Ref. \cite{TC07a} 
for the dragged Brownian particle model and was used there  
to derive transient fluctuation theorems for work. 
   In this section we generalize this to the classes 
of NESS models discussed in Sec. \ref{NESSmodel}. 
%it for a wider class of NESS models, 
%based on our general theory discussed in Sec. \ref{OMtheory}. 

   The nonequilibrium detailed balance relation including 
the work $\caW_{\pm}$ can be derived from 
Eqs. (\ref{Heat4}), (\ref{EnergDiffe1}) and (\ref{Work1}) as 
\begin{eqnarray}
    &&
    e^{-\overline{\beta}\left[\caW_{\pm}(\{\bfx_{s}\};\bfmu)
    - \Delta\caF_{\pm} \right]} \; 
    \caP_{x}(\{\bfx_{s}\};\bfmu) \; 
    \varrho_{\pm}(\dot{\bfx}_{-t},\bfx_{-t},-t;\bfeta) 
      \nonumber \\
   &&\spaEq 
    =       
    \varrho_{\pm}(\dot{\bfx}_{t},\bfx_{t},t;\bfeta) \; 
    \hat{I}_{\pm}\caP_{x}(\{\bfx_{s}\};\bfmu) 
\label{DetaiBalan1}
\end{eqnarray}
where $\varrho_{\pm}(\dot{\bfx}_{s},\bfx_{s},s;\bfeta)$ 
is a canonical-like distribution function defined by 
\begin{eqnarray}
   \varrho_{\pm}(\dot{\bfx}_{s},\bfx_{s},s;\bfeta) 
   \equiv \exp\left\{\overline{\beta}\left[\caF_{\pm}(s;\bfeta) 
   -E_{\pm}(\dot{\bfx}_{s},\bfx_{s},s;\bfeta) 
   \right]\right\}
\label{CanonDistr1}
\end{eqnarray}
with $\caF_{\pm}$ given by 
\begin{eqnarray}
   \caF_{\pm}(s;\bfeta) \equiv -\frac{1}{\overline{\beta}} \ln
   \int d\dot{\bfx}_{s}\int d\bfx_{s}\; 
   \exp\left[-\overline{\beta} 
   E_{\pm}(\dot{\bfx}_{s},\bfx_{s},s;\bfeta) \right] . 
\label{FreeEnerg1}
\end{eqnarray}
to normalize the canonical-like distribution function  
$\varrho_{\pm}(\dot{\bfx}_{s},\bfx_{s},s;\bfeta)$. 
   Here, in Eq. (\ref{DetaiBalan1}), 
$\Delta\caF_{\pm}$ is defined by 
\begin{eqnarray}
   \Delta\caF_{\pm} \equiv  \caF_{\pm}(t;\bfeta) 
   - \caF_{\pm}(-t;\bfeta) ,
%\label{}
\end{eqnarray}
i.e. the difference of $\caF_{\pm}$ between the initial time 
$t_{0}=-t$ and the final time $t$. 
   
   We will now show that 
the nonequilibrium detailed balance relation 
(\ref{DetaiBalan1})
reduces to the well-known  
equilibrium detailed balance condition. 
% \cite{TC07a}.  
   We first note that at equilibrium  
$\Delta\caF_{\pm}=0$, $\caW_{\pm}=0$     
and that there is then also no distinction between 
the two time-reversal operators,  
i.e. $\hat{I}_{+}=\hat{I}_{-}$,  
since the external nonequilibrium parameter is zero: 
$\bfmu=\mathbf{0}$.  
   Thus, at equilibrium we can write 
$\varrho_{\pm}=f^{[eq]}(\dot{\bfx}_{s},\bfx_{s})$ using an  
equilibrium canonical distribution function 
$f^{[eq]}(\dot{\bfx}_{s},\bfx_{s})$. 
%   Inserting these equations $\Delta\caF_{\pm}=0$, $\caW_{\pm}=0$,  
%$\hat{I}_{\pm} \equiv \hat{I}$ and 
%$\varrho_{\pm}=f^{[eq]}$ into Eq. (\ref{DetaiBalan1}), 
%we obtain the relation $\caP_{x}(\{\bfx_{s}\};\bfmu) \; 
%f^{[eq]}(\dot{\bfx}_{-t},\bfx_{-t}) 
%= f^{[eq]}(\dot{\bfx}_{t},\bfx_{t}) \;
%\hat{I}\caP_{x}(\{\bfx_{s}\};\bfmu)$ at equilibrium.  
%
   Secondly, we introduce the transition probability 
$\transP{P}{\bfx_{f},\dot{\bfx}_{f}}{t}
{\bfx_{i},\dot{\bfx}_{i}}{t_{0}}$
from an initial (i) point $(\bfx_{i},\dot{\bfx}_{i})$ at time $t_{0}$ 
to a final (f) point $(\bfx_{f},\dot{\bfx}_{f})$ at time $t$,  
which is given in terms of the probability functional 
$\caP_{x}(\{\bfx_{s}\})$ for paths $\{\bfx_{s}\}_{s\in[t_{0},t]}$ by 
$\transP{P}{\bfx_{f},\dot{\bfx}_{f}}{t}
   {\bfx_{i},\dot{\bfx}_{i}}{t_{0}}
   $  $= \int_{(\bfx_{t_{0}},\dot{\bfx}_{t_{0}})
   =(\bfx_{i},\dot{\bfx}_{i})}
   ^{(\bfx_{t},\dot{\bfx}_{t})=(\bfx_{f},\dot{\bfx}_{f})} 
   \mathcal{D}\bfx_{s}\;$ $\caP_{x}(\{\bfx_{s}\})$, 
i.e. taking the path integral 
(denoted by $\mathcal{D}\bfx_{s}$) of $\caP_{x}(\{\bfx_{s}\})$ 
over all paths $\{\bfx_{s}\}_{s\in[t_{0},t]}$ under the conditions 
$(\bfx_{t_{0}},\dot{\bfx}_{t_{0}})=(\bfx_{i},\dot{\bfx}_{i})$
and $(\bfx_{t},\dot{\bfx}_{t})=(\bfx_{f},\dot{\bfx}_{f})$. 
%   This transition probability satisfies the Fokker-Planck 
% equation corresponding to the Langevin equation (\ref{LangeEquat1}).  
%
   Using this and inserting the above equations 
$\Delta\caF_{\pm}=0$, $\caW_{\pm}=0$  %$\hat{I}_{\pm} \equiv \hat{I}$ 
and $\varrho_{\pm}=f^{[eq]}$ for equilibrium states 
into Eq. (\ref{DetaiBalan1}), 
we obtain the equilibrium detailed 
balance condition $\transP{P}{\bfx_{t},\dot{\bfx}_{t}}{t}{\bfx_{-t},
\dot{\bfx}_{-t}}{-t} $ $f^{[eq]}(\dot{\bfx}_{-t},\bfx_{-t}) 
$ $= f^{[eq]}(\dot{\bfx}_{t},\bfx_{t})
$ $\transP{P}{\bfx_{-t},\dot{\bfx}_{-t}}{t}{\bfx_{t},
\dot{\bfx}_{t}}{-t} $. 
%for 
%the transition probability 
%$\transP{P}{\bfx_{f},\dot{\bfx}_{f}}{t}
%{\bfx_{i},\dot{\bfx}_{i}}{-t}$ and 
%the equilibrium canonical distribution function 
%$f^{[eq]}(\dot{\bfx}_{s},\bfx_{s})$. 
%
%   Namely, the nonequilibrium detailed balance relation 
% (\ref{DetaiBalan1}) includes 
%the equilibrium detailed balance condition as 
%a special case. 

%%%%%%%%%%%%%%%%%%%%%%%%%%%%%%%%%%%%%%%%%%%%%%%%%%%%%%%%%%%%%%%%%%%%%%
\subsection{Transient Fluctuation Theorems for Work}
\label{FluctRelatFED}

   The nonequilibrium detailed balance relation (\ref{DetaiBalan1}) 
imposes a special relation on the probability distribution function 
for work, which is called the transient fluctuation theorems 
\cite{ES94}. 
   To discuss this theorem, we introduce the 
probability distribution function $P_{w\pm}(W,t;\bfmu)$  
for the work $\caW_{\pm}$ as 
\begin{eqnarray}
   P_{w\pm}(W,t;\bfmu) = \pathaveA{\delta\!\left(
   W-\caW_{\pm}(\{\bfx_{s}\};\bfmu)\right)} .
\label{WorkDistr1}
\end{eqnarray}
   Here, $\pathaveA{\cdots}$ 
indicates the functional average, which is defined by  
\begin{eqnarray}
   \pathaveA{X(\{\bfx_{s}\})} 
      &\equiv& \int d\bfx_{f}\int d\dot{\bfx}_{f}
      \int_{(\bfx_{-t},\dot{\bfx}_{-t})
         =(\bfx_{i},\dot{\bfx}_{i})}%
      ^{(\bfx_{t},\dot{\bfx}_{t})=(\bfx_{f},\dot{\bfx}_{f})} 
      \mathcal{D}\bfx_{s} \int d\bfx_{i}\int d\dot{\bfx}_{i}\; 
      \nonumber \\
   &&\spaEq \times 
      X(\{\bfx_{s}\})\caP_{x}(\{\bfx_{s}\}) 
      f(\dot{\bfx}_{i},\bfx_{i},-t)  
\label{PathAvera1}
\end{eqnarray}
for any functional $X(\{\bfx_{s}\})$ of the path 
$\{\bfx_{s}\}_{s\in[-t,t]}$, where $f(\dot{\bfx}_{i},\bfx_{i},-t)$ 
is the initial distribution function 
of the initial position $\bfx_{i}$ 
and the initial velocity $\dot{\bfx}_{i}$. % 
%\footnote{
%%   It may be noted that t
%   Using this initial distribution function  
%$f(\dot{\bfx}_{i},\bfx_{i},-t)$ and the transition probability 
%$\transP{P}{\bfx_{f},\dot{\bfx}_{f}}{t}
%   {\bfx_{i},\dot{\bfx}_{i}}{-t}$, the probability 
%$f(\bfx_{t},\dot{\bfx}_{t},t)$ of the position $\bfx_{t}$ 
%and the velocity $\dot{\bfx}_{t}$ at time $t$ 
%is given by $f(\bfx_{t},\dot{\bfx}_{t},t) 
%=  \int d\bfx_{i}\int d\dot{\bfx}_{i} 
%\transP{P}{\bfx_{t},\dot{\bfx}_{t}}{t}{\bfx_{i},\dot{\bfx}_{i}}{-t} 
%f(\dot{\bfx}_{i},\bfx_{i},-t) $. 
%}
%   Using this functional average, the normalization condition 
%for the probability functional $\caP_{x}(\{\bfx_{s}\})$ is 
%expressed as $\pathaveA{1}=1$.
%
   The work distribution function (\ref{WorkDistr1}) 
can be rewritten in the form (cf. Ref. \cite{TC07a})
\begin{eqnarray}
   P_{w\pm}(W,t;\bfmu) 
   = \frac{1}{2\pi}\int_{-\infty}^{+\infty} 
   d\sigma\; e^{i\sigma W} \caE_{w\pm}(i\sigma,t;\bfmu)
\label{WorkDistr2}
\end{eqnarray}
with $\caE_{\pm}(\sigma,t;\bfmu)$ defined by 
\begin{eqnarray}
   \caE_{w\pm}(\sigma,t;\bfmu) 
   \equiv \pathaveA{e^{-\sigma  
   \caW_{\pm}(\{\bfx_{s}\};\bfmu)}} .
\label{Efunct1}
\end{eqnarray}
   The function $\caE_{w\pm}(i\sigma,t;\bfmu)$ can be 
regarded as a Fourier transformation of the work distribution 
function $P_{w\pm}(W,t;\bfmu)$, as well as a generation function 
for the work $\caW_{\pm}(\{\bfx_{s}\};\bfmu)$. 
   The work distribution function $P_{w\pm}(W,t;\bfmu)$ was 
obtained explicitly for all $W$ and $t$ 
by carrying out path integrals for dragged Brownian 
particle models as in Refs. \cite{TC07a,TC07b}. 

   In the remaining of this section, we assume 
that the initial distribution function 
$f(\bfx_{-t},\dot{\bfx}_{-t},-t)$ at the initial time $t_{0}=-t$ 
is given by the canonical-like distribution function 
(\ref{CanonDistr1}), i.e. 
\begin{eqnarray}
   f(\bfx_{-t},\dot{\bfx}_{-t},-t) 
   = \varrho_{\pm}(\dot{\bfx}_{-t},\bfx_{-t},-t;\bfeta) . 
\label{InitiCondi1}
\end{eqnarray}
   In that case, we obtain  
\begin{eqnarray}
   \caE_{w\pm}(\overline{\beta}-\sigma,t;\bfmu) 
   =  e^{-\overline{\beta} \Delta\caF_{\pm}}
   \hat{I}_{\pm} \caE_{w\pm}(\sigma,t;\bfmu) .
\label{FluctTrans1}
\end{eqnarray}
   Eq. (\ref{FluctTrans1}) leads then 
to two generalized transient fluctuation theorems:  
\begin{eqnarray}
   \frac{P_{w\pm}(W,t;\bfmu)}
   {\hat{I}_{\pm} P_{w\pm}(-W,t;\bfmu)} = 
   e^{\overline{\beta} (W-\Delta\caF_{\pm})}
\label{FluctTrans2}
\end{eqnarray}
for the work distribution function $P_{w\pm}(W,t;\bfmu)$. 
   [See Appendix \ref{TransFluctAp} for a derivation  
of Eqs. (\ref{FluctTrans1}) and (\ref{FluctTrans2}).]
   We emphasize that Eq. (\ref{FluctTrans2}) 
is a relation for work fluctuations 
described by the distribution function $P_{w\pm}(W,t;\bfmu)$,  
for the special initial condition (\ref{InitiCondi1}) 
[cf. Eq. (\ref{CanonDistr1})]. 

   The two generalized transient fluctuation theorems 
(\ref{FluctTrans2}) reduce both to 
the usual transient fluctuation theorem  \cite{ES94} 
%   The usual transient fluctuation theorem holds 
when the energy $E_{\pm}$ is the equilibrium energy 
independent of the external nonequilibrium parameter $\bfmu$   
[so that then the initial distribution function  (\ref{InitiCondi1}) 
is an equilibrium canonical distribution function 
$f^{[eq]}(\dot{\bfx}_{s},\bfx_{s})$] and also 
the condition $\Delta\caF_{\pm}=0$
is satisfied.%
\footnote{ 
   The energy $E$ is independent of the 
external nonequilibrium parameter $\bfmu$ 
 for all the models discussed 
in Sec. \ref{NESSmodel} of this paper, except 
for the dragged Brownian particle model 
in the comoving frame and its corresponding electric circuit model, 
when $\varrho_{\pm}(\dot{\bfx}_{s},\bfx_{s},s;\bfeta)$ 
is an equilibrium canonical distribution function. 
   In addition, the condition $\Delta\caF_{\pm}=0$ 
is satisfied for all the models discussed in this paper. 
}  
%   If all these conditions are fulfilled, the 
%two generalized transient fluctuation theorems 
%(\ref{FluctTrans2}) reduce both to 
%the usual transient fluctuation theorem. 
   Note that Eq. (\ref{FluctTrans2}) implies 
two transient fluctuation theorems; one for $P_{w+}(W,t;\bfmu)$ 
and the other for $P_{w-}(W,t;\bfmu)$, as mathematical identities 
satisfied at any time for a canonical-like initial distribution. 
%especially for both the works $\caW_{+}$ and $\caW_{-}$ 
%calculated concretely in Sec. \ref{WorkModel} for NESS models. 
   Both these transient fluctuation theorems 
could be checked experimentally. 
   These two different transient fluctuation theorems, 
for example, for the \NESSW and for a work to overcome the friction 
in the dragged Brownian particle model, have already been 
discussed in Ref. \cite{TC07a}. 

   Finally, one may notice that the transient fluctuation theorems  
(\ref{FluctTrans2}) have a similar form as the so-called 
Crooks theorem \cite{C99}, 
which is a type of transient fluctuation theorem involving  
a free energy difference. %, as a mathematical identity. 
   However, Eq. (\ref{FluctTrans2}) is not exactly the same 
as the Crooks theorem.  
   First, in the Crooks theorem the work is always given by  
a time-integral of the partial 
time-derivative of the energy, like in Eq. (\ref{Work3a}), 
while in Eq. (\ref{FluctTrans2}) the work $\caW_{\pm}$ 
does not have such a simple form,  
noting that the work (\ref{Work2}) can contain, in general, 
other contributions, as given by Eqs. (\ref{Work3b}), 
(\ref{Work3c}) and (\ref{Work3d}). 
   Second, the energy $E_{\pm}$ is, in general, not the equilibrium 
energy because it can include the external nonequilibrium 
parameter $\bfmu$, so that the quantity $\caF_{\pm}$ defined 
by Eq. (\ref{FreeEnerg1}) does not have  
to be an equilibrium free energy as appearing in the Crooks theorem.

%%%%%%%%%%%%%%%%%%%%%%%%%%%%%%%%%%%%%%%%%%%%%%%%%%%%%%%%%%%%%%%%%%%%%%
\section{Illustration of the Forgoing Theory on Specific NESS Systems} 
\label{WorkModel} 

   In Sec. \ref{OMtheory}, we obtained  
the work and the heat for NESS systems 
described by a linear Langevin equation. 
   In this section, first, using the simple models 
introduced in Secs. \ref{ClassA} and \ref{ClassB} 
as Classes A and B, we check 
that our formal expressions for the work and the heat 
in Sec. \ref{OMtheory} indeed yield the  
\NESSW and heat to maintain a NESS. 
   For these discussions, we mainly use 
the dragged Brownian particle and pendulum models 
and omit arguments for the corresponding electric circuit models 
since their results can be obtained by the correspondences 
in Table \ref{CorreModel1} between the Brownian particle 
(and pendulum) models 
and the electric circuit models.  
   After that, we will also discuss the \NESSW 
for more complicated NESS  
models introduced in Sec. \ref{ClassC} as Class C.

%%%%%%%%%%%%%%%%%%%%%%%%%%%%%%%%%%%%%%%%%%%%%%%%%%%%%%%%%%%%%%%%%%%%%%
\subsection{Class A}
\label{WorkModelA}

   First, we apply our NESS Onsager-Machlup theory 
to the models discussed in Sec. 
\ref{ClassA}, categorized as Class A.

   For the electric field driven model in Class A, 
using the force $F_{1}=q\caE$  
the Lagrangian (\ref{LagraOM1}) is represented by 
$\caL =  - (\alpha\beta/4) [\dot{x}_{s} - q\caE/\alpha 
+ (m/\alpha)\ddot{x}_{s} ]^{2} $. 
%
%\begin{eqnarray}
%   \caL\!\left(\ddot{x}_{s},\dot{x}_{s},
%   x_{s},s;\caE \right) 
%   =  - \frac{\alpha\beta}{4}
%         \left(\dot{x}_{s} - \frac{q\caE}{\alpha} 
%         + \frac{m}{\alpha}\ddot{x}_{s} \right)^{2} 
%\label{LagraOMA1}
%\end{eqnarray} 
%
%   Here, we expressed the Lagrangian $\caL$ using the notations 
%for the Brownian particle models described 
%by Eqs. (\ref{LangeEquatA1}) and (\ref{LangeEquatB1}), 
%but the Lagrangians for other models like the driven torsion 
%pendulum model and electric circuit models in the classes 
%A and B can be obtained immediately from Eq. (\ref{LagraOMAB1}) 
%with the correspondences shown in Table \ref{CorreModel1}. 
%
   Using this Lagrangian %(\ref{LagraOMA1}), 
we can calculate 
the entropy production rate $\dot{S}_{\pm}$ 
by Eq. (\ref{EntroProdu1}), 
then the heat $\caQ_{\pm}$ by Eq. (\ref{Heat1})  
and the work $\caW_{\pm}$ by Eq. (\ref{Work1}) 
using the energy of Eq. (\ref{EnergA1}). 
   Alternatively, using the force $F_{1}=q\caE$ 
with the external nonequilibrium parameter $\bfmu=\caE$,  
the work can also be calculated directly by 
Eq. (\ref{Work2}) with Eqs. (\ref{Work3a}), 
(\ref{Work3b}), (\ref{Work3c}) and (\ref{Work3d}). 
   These calculations are straightforward, 
so we omit their details 
and only show their results in Table \ref{EnergWorkModel1}. 
   In this Table we exhibited especially the work 
$\caW_{\pm}$, since the heat $\caQ_{\pm}$ 
is given by $\caQ_{\pm} = \caW_{\pm} - \Delta E_{\pm}$ 
[i.e. Eq. (\ref{Work1})] in terms of the work $\caW_{\pm}$ 
and the energy difference $\Delta E_{\pm}$ 
of Eq. (\ref{EnergDiffe1}). 
%   [As references, in this Table we also showed 
%the Langevin equations (L.E.), the energy 
%$E(\dot{x}_{s},x_{s},s;\bfmu)$, and the external nonequilibrium 
%parameter (N.P.), and the necessary time reversal 
%operator (T.R.) to derive these works.] 
%   The results in Table \ref{CorreModel1} are presented 
%using the notations of the Browian particle models 
%and the driven torsion pendulum model, 
%but results for the corresponding electric circuit models 
%of Classes A and B are given using this Table and 
%the correspondences of quantities in Table \ref{CorreModel1}. 

   In this model, the \NESSW should be given by 
the work done by the external electric force $q\caE$. 
   This work indeed appears as $\caW_{+}=\int_{-t}^{t}ds \; 
q\caE \dot{x}_{s}$  
in the above calculation for the work (\ref{Work2}).  
   This \NESSW is zero at equilibrium $\caE=0$. 
   Using the average velocity 
$\langle \dot{x}_{s}\rangle=\overline{v}=q\caE/\alpha$ 
for this model in the NESS, the average work rate  
corresponding to this \NESSW $\caW_{+}$ 
is given by $(q\caE)^{2}/\alpha$ in the NESS, which is 
proportional to the square of the external nonequilibrium 
parameter $\caE$. 
   Therefore, the average work, as well as the average 
heat and entropy production rate, 
are strictly positive in NESSs, 
as we required in the end of Sec. \ref{HeatWork}. 
%as shown in Table \ref{EnergWorkModel1}. 

%   The time-reversal operator $\hat{I}_{\pm}$ to derive 
%the \NESSW are different in Class A 
%and Class B: the operators $\hat{I}_{+}$ 
%for the \NESSW $\caW_{+}$ for Class A 
%and the operators $\hat{I}_{-}$ for the \NESSW 
%$\caW_{-}$ for Class B. 
   Although it is not the \NESSw, 
we can still calculate the work $\caW_{-}$ 
for the electric field driven model. 
   It is given by $\int_{-t}^{t} m\ddot{x}_{s}\overline{v}$, 
i.e. the work to maintain a motion of the particle   
with an average velocity 
$\overline{v} (=q\caE/\alpha)$ by the total force 
$m\ddot{x}_{s} (= q\caE -\alpha \dot{x}_{s} + \zeta_{s})$. 
   In the NESS, the average of this work $\caW_{-}$ is zero 
since then $\langle \ddot{x}_{s}\rangle=0$, 
so that following our requirement  
for the average \NESSW to be strictly positive 
in the NESS, this work $\caW_{-}$ cannot be the \NESSw. 
%   In the electric field model the average of the heat should 
%give an Ohmic heat in the case of electric current. 

%---------------------------------------------------------------------
%\def\arraystretch{1.5}
%\def\Topspc{\rule{0pt}{12pt}}
%\def\Btmspc{\rule[-10pt]{0pt}{0pt}}
\newcommand{\widthTableA}{0.1\textwidth}
\newcommand{\widthTableB}{0.15\textwidth}
\newcommand{\widthTableC}{0.2\textwidth}
\newcommand{\widthTableD}{0.21\textwidth}
\newcommand{\widthTableE}{0.15\textwidth}
\newcommand{\widthHeightA}{1.5cm}
\newcommand{\widthHeightB}{2.5cm}
\newcommand{\widthHeightC}{1.1cm}
\newcommand{\widthHeightD}{2.2cm}
\newcommand{\widthHeightE}{1.3cm}
\begin{table}[t!]
\vspfigA
\begin{center}
\renewcommand{\arraystretch}{2}
\begin{tabular}{|c||c|c|c|c|}
\hline 
   Class & Class A & \multicolumn{3}{|c|}{Class B}\\
\hline 
   \parbox[c][\widthHeightA]{\widthTableA}{
    \begin{minipage}{\widthTableA} \begin{center}
      \baselineskip 2 ex
      Model  
      \end{center}\end{minipage} }
   & \begin{minipage}{\widthTableB} \begin{center}
      \baselineskip 0pt
      {\small Electric field}\\ {\small driven model} 
      \end{center}\end{minipage} 
   & \begin{minipage}{\widthTableC} \begin{center}
      \baselineskip 0pt
      {\small Dragged Brownian}\\ {\small particle in the }
      \\ {\small laboratory frame} 
      \end{center}\end{minipage} 
   & \begin{minipage}{\widthTableD} \begin{center}
      \baselineskip 0pt
      {\small Dragged Brownian }\\ {\small particle in the }
      \\ {\small comoving frame} 
      \end{center}\end{minipage} 
   & \begin{minipage}{\widthTableE} \begin{center}
      \baselineskip 0pt 
      {\small Driven torsion }\\ {\small pendulum} 
      \end{center}\end{minipage} \\
\hline 
\hline 
   \parbox[c][\widthHeightD]{\widthTableA}{
    \begin{minipage}{\widthTableA} \begin{center}
      \baselineskip 2 ex
      Langevin \\ equation, \\ Figure, \\ Section
      \end{center}\end{minipage} }
   & \begin{minipage}{\widthTableB} \begin{center}
      \baselineskip 2.5 ex
      Eq. (\ref{LangeEquatA1})[(\ref{LangeEquatA2})],  
      \\ Fig. \ref{fig1modelA}(a)[(b)],    
      \\ Secs. \ref{ClassA} 
      \\ and \ref{WorkModelA}
      \end{center}\end{minipage}
   & \begin{minipage}{\widthTableB} \begin{center}
      \baselineskip 2.5 ex
      Eq. (\ref{LangeEquatB1})[(\ref{LangeEquatB2})], 
      \\  Fig. \ref{fig2modelB}(a)[(b)], 
      \\  Secs. \ref{ClassB} 
      \\ and \ref{WorkModelB}
      \end{center}\end{minipage}
   & \begin{minipage}{\widthTableB} \begin{center}
      \baselineskip 2.5 ex
      Eq. (\ref{LangeEquatB5}) 
      \\ in Sec. \ref{ComovModel} 
      \end{center}\end{minipage}
   & \begin{minipage}{\widthTableB} \begin{center}
      \baselineskip 2.5 ex
      Eq. (\ref{LangeEquatB3})[(\ref{LangeEquatB4})],  
      \\  Fig. \ref{fig2modelB}(c)[(d)], 
      \\ Secs. \ref{ClassB} 
      \\ and \ref{WorkModelC}
      \end{center}\end{minipage}\\
\hline 
   \parbox[c][\widthHeightC]{\widthTableA}{
   \begin{minipage}{\widthTableA} \begin{center}
      \baselineskip 0 ex
      Nonequi- \\ librium \\ parameter
      \end{center}\end{minipage}} 
   & $\caE$ 
   & \multicolumn{2}{|c|}{$v$} 
   & $\xi$ \\
\hline 
   \parbox[c][\widthHeightC]{\widthTableA}{
   \begin{minipage}{\widthTableA} \begin{center}
      \baselineskip 2 ex
      Internal \\ energy 
      \end{center}\end{minipage}}
   & $\frac{1}{2}m \dot{x}_{s}^{2}$ 
   & $\frac{1}{2}m \dot{x}_{s}^{2} 
         \!+\!\frac{1}{2}\kappa (x_{s}\!-\!vs)^{2}$  
   & $\frac{1}{2} m \dot{y}_{s}^{2} 
         \!+\!\frac{1}{2}\kappa y_{s}^{2}$
   & $\frac{1}{2}\caI\dot{\theta}_{s}^{2} 
         \!+\!\frac{1}{2} \sigma \theta_{s}^{2}$  \\
\hline 
   \parbox[c][\widthHeightC]{\widthTableA}{
   \begin{minipage}{\widthTableA} \begin{center}
      \baselineskip 0 ex
      Time-reversal \\ operator
      \end{center}\end{minipage}}
   & $\hat{I}_{+}$  
   & \multicolumn{3}{|c|}{$\hat{I}_{-}$} \\
\hline 
   \parbox[c][\widthHeightB]{\widthTableA}{
      $\;$ Forces}
   & \begin{minipage}{\widthTableB} \begin{center}
      $F_{1+}^{(e)} \!=\! q\caE $, \\
      $F_{1+}^{(o)} \!=\! 0$,  \\ 
      $f_{1+}\!=\! q\caE $
      \end{center}\end{minipage}
   & \begin{minipage}{\widthTableC} \begin{center}
      $F_{1-}^{(e)} \!=\! -\! \kappa (x_{s}\!-\!vs) $, \\
      $F_{1-}^{(o)} \!=\! 0$,  \\ 
      $f_{1-}\!=\! 0$
      \end{center}\end{minipage}
   & \begin{minipage}{\widthTableD} \begin{center}
      $F_{1-}^{(e)} \!=\! -\! \kappa y_{s} $, \\
      $F_{1-}^{(o)} \!=\! -\!\alpha v$,  \\ 
      $f_{1-}\!=\! 0$
      \end{center}\end{minipage}
   &  \begin{minipage}{\widthTableB} \begin{center}
      $F_{1-}^{(e)} \!=\!   -\sigma\theta_{s} + \caM_{s} $, \\
      $F_{1-}^{(o)} \!=\! 0$,  \\ 
      $f_{1-}\!=\! \caM_{s} $
      \end{center}\end{minipage}\\
\hline 
   \parbox[c][\widthHeightA]{\widthTableA}{
      \begin{minipage}{\widthTableA} \begin{center}
      \baselineskip 0 ex
      NESS \\ work
      \end{center}\end{minipage}}
   & \begin{minipage}{\widthTableB} \begin{center}
      $\caW_{+}\!=\!\caW_{+}^{(f)}$ \\
      $\!=\!\int_{t_{0}}^{t} \! ds  \; q\caE  \dot{x}_{s}$ 
      \end{center}\end{minipage}
   &\begin{minipage}{\widthTableC} \begin{center}
      $\caW_{-}\!=\!\caW_{-}^{(e)}\!=\!$ \\
      $\!-\!\int_{t_{0}}^{t} \! ds \; \kappa (x_{s}\!-\!vs) v$
      \end{center}\end{minipage}
   & \begin{minipage}{\widthTableD} \begin{center}
      $\caW_{-}\!=\!\caW_{-}^{(o)}\!=\!$\\
      $\!-\!\int_{t_{0}}^{t} \! ds  
        \; (\kappa y_{s}\!+\!m\ddot{y}_{s}) v$ 
      \end{center}\end{minipage} 
   &\begin{minipage}{\widthTableE} \begin{center}
      $\caW_{-}\!=\!\caW_{-}^{(f)}$ \\
      $\!=\!\int_{t_{0}}^{t} \! ds \; \caM_{s} \dot{\theta}_{s}$
      \end{center}\end{minipage}\\ 
\hline 
   \parbox[c][\widthHeightE]{\widthTableA}{
   \begin{minipage}{\widthTableA} \begin{center}
      \baselineskip 0 ex
      Average \\ NESS  %\\ thermo-\\dynamic 
      \\ work rate  
      \end{center}\end{minipage}} 
   & $\frac{(q\caE)^{2}}{\alpha} > 0$ 
   & \multicolumn{2}{|c|}{$\alpha v^{2} >0$}
   & $\frac{\xi^{2}}{\sigma} t > 0$ \\
\hline 
   \parbox[c][\widthHeightB]{\widthTableA}{
   \begin{minipage}{\widthTableA} \begin{center}
      \baselineskip 2 ex
      Non-NESS \\  work 
      \end{center}\end{minipage}}
   &\begin{minipage}{\widthTableB} \begin{center}
      $\caW_{-} \!=\! \caW_{-}^{(o)}$ $\!=\!\int_{t_{0}}^{t} \! ds  
        \; m\ddot{x}_{s} \overline{v}$ 
        {\small (Zero average in NESS)}
      \end{center}\end{minipage} 
   & \begin{minipage}{\widthTableC} \begin{center}
      No $\caW_{+}$ \\ 
      $[$The energy does not satisfy  
      Eq. (\ref{ReverEnerg1}) for $\hat{I}_{+}$.$]$ 
      \end{center}\end{minipage} 
   & \begin{minipage}{\widthTableD} \begin{center}
      $\caW_{+} \!=\! \caW_{+}^{(f)}$ \\$
      \!=\!-\!\int_{t_{0}}^{t} \! ds  
      \; \alpha\dot{y}_{s} v$ 
      {\small (Zero average in NESS)}
      \end{center}\end{minipage} 
   & \begin{minipage}{\widthTableE} \begin{center}
      $\caW_{+} \!=\! \caW_{+}^{(o)}$ $
      \!=\frac{1}{\nu} \int_{t_{0}}^{t} \! ds \; \caM_{s}$ 
      $\times(\sigma \theta_{s}+\caI\ddot{\theta}_{s})$  
      \end{center}\end{minipage} \\
\hline 
\end{tabular}
\end{center}
\vspfigB
\caption{
      Expressions for the work $\caW_{\pm}$ in   
   the NESS models described by 
   Langevin equations in Secs. \ref{ClassA}, \ref{ClassB} 
   and \ref{ComovModel} as Classes A and B. 
      We also show the external nonequilibrium parameter $\bfmu$, 
   the internal energy $E$, the relevant time-reversal 
   operator $\hat{I}_{\pm}$, the forces $F_{1\pm}^{(e)}$, 
   $F_{1\pm}^{(o)}$ and $f_{1\pm}$ to be used to obtain the 
   \NESSw, and the (strictly positive) average \NESSW rate. 
%      The corresponding heat is given by $\caQ_{\pm}  
%   = \caW_{\pm} - \Delta E_{\pm}$ with the energy difference 
%   $\Delta E_{\pm} \equiv E_{\pm}(\dot{x}_{t},x_{t},t;\bfeta)
%   -E_{\pm}(\dot{x}_{t_{0}},x_{t_{0}},t_{0};\bfeta)$. 
%      For the torsion pendulum model in Class B the external 
%   torque is given by $\caM_{s}=\xi s$ with a constant $\xi$, 
%   and for the electric field model in Class A  
%   the average particle velocity is given by 
%   $\overline{v} =q\caE/\alpha$. 
%      Note that in these models  
%   the average \NESSW rate, corresponding to the \NESSW 
%   ($\caW_{+}$ for Class A and $\caW_{-}$ for Class B),  
%   is strictly positive in the NESS.  
      The energy, the external nonequilibrium parameter and the work 
   in this Table are for Brownian particle (and pendulum) models, 
   and the corresponding quantities for electric circuit models 
   can be obtained using the correspondences 
   in Table \ref{CorreModel1}.  
}
\label{EnergWorkModel1}
\vspfigC
\end{table}%
%---------------------------------------------------------------------

%%%%%%%%%%%%%%%%%%%%%%%%%%%%%%%%%%%%%%%%%%%%%%%%%%%%%%%%%%%%%%%%%%%%%%
\subsection{Class B}
\label{WorkModelB}

   We now discuss the work and the heat for Class B. 
   For the dragged Brownian particle model of this class  
described by Eq. (\ref{LangeEquatB1}), 
the mechanical force is given by 
$F_{1}=-\kappa (x_{s}-vs)$ [cf. Eqs. (\ref{LangeEquatB1}) 
and (\ref{LangeEquat1})] and then  
the Lagrangian (\ref{LagraOM1}) by 
$
   \caL =  - (\alpha\beta/4)
   [\dot{x}_{s} + \kappa (x_{s}-vs)/\alpha 
         + (m/\alpha)\ddot{x}_{s} ]^{2}  
$. 
%%
%\begin{eqnarray}
%   \caL\!\left(\ddot{\bfx}_{s},\dot{\bfx}_{s},
%   \bfx_{s},s;\bfmu \right) 
%   =  - \frac{\alpha\beta}{4}
%         \left[\dot{x}_{s} + \frac{\kappa (x_{s}-vs)}{\alpha} 
%         + \frac{m}{\alpha}\ddot{x}_{s} \right]^{2} . 
%\label{LagraOMB1}
%\end{eqnarray}
%   
   For the driven torsion pendulum model 
[cf. Eq. (\ref{LangeEquatB3})], 
%(as well as for the corresponding electric circuit models)
the force $F_{1}$ and the Lagrangian $\caL$ can simply be obtained using Table \ref{CorreModel1} from the corresponding $F_{1}$ 
and $\caL$ for the dragged Brownian particle. 
%to Eq. (\ref{LagraOMB1}). 
   Then, in a similar fashion as   
in the previous section for Class A, we can obtain   
%from the Lagrangian or the force, %$\caL$ %(\ref{LagraOMB1}) 
expressions for the heat $\caQ_{\pm}$ and the work 
$\caW_{\pm}$, etc. for these models, using the NESS 
Onsager-Machlup theory.  
   These results are also 
summarized in Table \ref{EnergWorkModel1}. 
   
   From the works $\caW_{+}$ and $\caW_{-}$ 
in Table \ref{EnergWorkModel1}, the work $\caW_{-}$ gives 
the \NESSW to sustain 
a NESS for these models. 
%i.e. the work done by the harmonic potential 
%dragged by a constant velocity $v$ in the dragged Brownian 
%particle model, and the work done by the external torque 
%$\caM_{s}=\xi s$ in the driven torsion pendulum model. 
   Actually, the average work rates for these \NESSWs 
in the NESS are given by 
$\alpha v^{2}$ and $\xi^{2}t/\sigma$ for the 
dragged Brownian particle and the driven torsion pendulum, 
respectively, which are strictly positive and even functions 
of the external nonequilibrium parameter, as required 
for the \NESSw. 

   It is important to note that the \NESSW 
in Class B is $\caW_{-}$, 
different from Class A where the \NESSW is 
given by $\caW_{+}$. 
   Moreover, even within the same Class B with 
the same form of the Langevin equation, the \NESSW
is different for the dragged Brownian particle model,  
where $\caW_{-}=-\int_{t_{0}}^{t} ds \; \kappa (x_{s}- vs) v$, 
and the driven torsion pendulum model, where  
$\caW_{-}=\int_{t_{0}}^{t} ds \; \caM_{s} \dot{\theta}_{s}$. 
%   For, unlike in the dragged Brownian particle model,  
%where $f_{1-}=0$, there is an external driving force 
%$f_{1-}=\caM_{s}$ for the driven torsion pendulum model as 
%a force, which contributes to the \NESSW $\caW_{-}$. 
   This difference in the \NESSW is due to 
a difference of the external driving force $f_{1-}$, 
%in other words, due to themultiple possibilities to separate heat 
%into work and an energy difference, 
in other words, the difference between Eqs. (\ref{EnergB1}) 
and (\ref{EnergB2}) for the internal energy of these two models. 
  
   It may also be noted that there is no difference  
between the internal energies $E_{+}$ and $E_{-}$ 
[cf. Eq. (\ref{Energ1}) and Table \ref{EnergWorkModel1}] for 
the driven torsion pendulum model 
(as in the electric field driven model in Class A),  
so that we can obtain both the works $\caW_{+}$ and $\caW_{-}$ 
for this model.   
   On the other hand, there is no $E_{+}$ for the dragged 
Brownian particle model because that energy does not 
satisfy the time-symmetric condition   
(\ref{ReverEnerg1}) for the energy using $\hat{I}_{+}$, leaving 
$\hat{I}_{-}$ as the only possible 
time-reversal operator, giving the correct physical results. 
   Therefore, there is no $\caW_{+}$ in this model.

   Finally, as shown in Table \ref{EnergWorkModel1}, 
the \NESSWs for the models of Classes A and B in Secs. \ref{ClassA} 
and \ref{ClassB} are of two types: 
the type (\ref{Work3a}) 
for the dragged Brownian particle model (Class B), and 
the type (\ref{Work3b})                 
for the electric field model (Class A) and 
the driven torsion pendulum model (Class B). 
   In the next subsection \ref{ComovModel} 
we discuss another case in which 
the \NESSW is of the third 
type (\ref{Work3c}) of work.

%%%%%%%%%%%%%%%%%%%%%%%%%%%%%%%%%%%%%%%%%%%%%%%%%%%%%%%%%%%%%%%%%%%%%%
%\pagebreak
\subsection{Dragged Brownian Particle in a Comoving Frame} 
\label{ComovModel}

   The Langevin equation (\ref{LangeEquatB1}) describes 
the dynamics of a dragged Brownian particle 
in the laboratory frame, where $x_{s}$ is the particle position 
in the laboratory frame. 
   In this subsection we discuss this dragged Brownian particle  
in the comoving frame \cite{TC07a,TC07b}.  
   In that case, the \NESSW is given by 
a qualitatively different type of expression than 
discussed in the proceeding sections \ref{WorkModelA} 
and \ref{WorkModelB}. 
   This provides an interesting illustration of 
the general theory discussed in Sec. \ref{OMtheory}. 

   The spatial coordinate $y_{s}$ 
of the Brownian particle in the comoving frame for 
this system is given by 
\begin{eqnarray}
   y_{s}\equiv x_{s}- v s,
\label{ComovFrame1}
\end{eqnarray}
and its dynamics is expressed by the Langevin equation 
\begin{eqnarray}
   m\ddot{y}_{s} 
   = -\kappa y_{s} -\alpha v 
   -\alpha \dot{y}_{s} 
   + \zeta_{s} .
\label{LangeEquatB5}
\end{eqnarray}
   Note that different from the Langevin equation 
(\ref{LangeEquatB1}) for the laboratory frame, 
Eq. (\ref{LangeEquatB5}) does \emph{not} have an explicit 
time-dependence and the nonequilibrium effect appears 
just as a constant term $-\alpha v$. 
   In this frame the internal energy $E$ 
of the particle is given by  
\begin{eqnarray}
   E(\dot{y}_{s},y_{s}) = 
      \frac{1}{2}m \dot{y}_{s}^{2} 
      +\frac{1}{2}\kappa y_{s}^{2} 
\label{EnergB5}
\end{eqnarray}
which is independent of the external nonequilibrium parameter 
$\bfmu=v$, different from the laboratory case. 
   Note that the internal energy (\ref{EnergB5}) 
for the comoving frame is different from the internal energy 
(\ref{EnergB1}) for the laboratory frame 
because of a frame-dependence of the kinetic energy \cite{TC07b}. 

   In this model, the mechanical force and the Lagrangian 
are given by $F_{1} =  -\kappa y_{s} -\alpha v$ and 
$ \caL = - (\alpha\beta/4)[\dot{y}_{s} 
   +(\kappa /\alpha)y_{s} +v+ (m/\alpha)\ddot{y}_{s} ]^{2} $, 
respectively \cite{TC07a}.    
%
%\begin{eqnarray}
%   \caL\!\left(\ddot{y}_{s},\dot{y}_{s},
%   y_{s},s;v \right) = 
%   - \frac{\alpha\beta}{4}
%   \left(\dot{y}_{s} 
%   +\frac{y_{s}}{\tau_{r}} + v
%   +\frac{m}{\alpha}\ddot{y}_{s} \right)^{2} 
%\label{LagraOMB5}
%\end{eqnarray}
%
%with $\tau_{r}\equiv \alpha/\kappa$ \cite{TC07a}.  
%   Note again that there is no explicit time-dependence 
%in the Lagrangian \ref{LagraOMB5} for the comoving frame, 
%different from the Lagrangian (\ref{LagraB1}) for 
%the laboratory frame. 
   Using this force or Lagrangian, we can calculate 
the quantities like $\dot{S}_{\pm}$, $\caQ_{\pm}$, $\caW_{\pm}$, 
etc., in a similar way as in the previous subsections 
\ref{WorkModelA} and \ref{WorkModelB} 
for the models of Classes A and B in the laboratory frame.  
   We summarize the results in the 4-th column 
of Table \ref{EnergWorkModel1}. 

   Different from the models in Secs. \ref{ClassA} 
and \ref{ClassB}, the \NESSW $\caW_{-}$ for this 
model in the comoving frame is of the type (\ref{Work3c}) involving 
the odd part force $F_{1-}^{(o)}=-\alpha v$ of the force $F_{1}$.  
   Note that this \NESSW is obtained  
by using the same time-reversal operator $\hat{I}_{-}$ as  
in the laboratory frame, but that it   
includes an additional effect due to the d'Alembert type of force 
$-m\ddot{y}_{s}$, absent in the laboratory frame. 
   This d'Alembert type force has no effect 
on the \emph{average} \NESSW  
nor on the average work rate $\alpha v^{2}$ in the NESS, 
which are therefore frame-independent. 
   However, as discussed in Ref. \cite{TC07b}, 
fluctuation properties of the work are influenced 
by this d'Alembert type force.  

   Another difference between the comoving and the laboratory 
frames in the dragged Brownian particle model is that 
in the comoving frame, the energy is time-reversal invariant 
satisfying the condition (\ref{ReverEnerg1})  
under both time-reversal procedures $\hat{I}_{+}$ and 
$\hat{I}_{-}$.  
   Therefore, different from the laboratory frame,  
we can obtain the other work $\caW_{+}$, which is the work 
to overcome the friction, %force, 
i.e. $\caW_{+}=-\int_{-t}^{t} ds \; \alpha\dot{y}_{s} v$.%
\footnote{In Ref. \cite{TC07a} the work  
$-\int_{-t}^{t} ds \; \alpha\dot{y}_{s} v$ 
was called the ``energy loss by friction''.}   
   The average of this work $\caW_{+}$ is zero in the NESS, 
so that it cannot be the \NESSw.

%%%%%%%%%%%%%%%%%%%%%%%%%%%%%%%%%%%%%%%%%%%%%%%%%%%%%%%%%%%%%%%%%%%%%%
%\pagebreak
\subsection{Class C} 
\label{WorkModelC}

   As the last example for nonequilibrium work, 
we consider stochastic models with two random noises 
as in Class C introduced in Sec. \ref{ClassC}. 
   One of these models is an example in which the work 
is given by the type (\ref{Work3d}). 

   For Class C, whose Langevin equation is expressed 
by Eq. (\ref{LangeEquatC3}), the mechanical force $F_{j}$ 
is given by 
\begin{eqnarray}
   F_{j}(\bfx_{s},s;\bfeta)
   = \Gamma \delta_{j2} -\kappa \left(x_{js} - x_{ks}\right)  
\label{ForceC3}
\end{eqnarray}
for $j\neq k$, $j=1,2$ and $k=1,2$. 
   Inserting Eqs. (\ref{ForceC3}) and $N=2$ 
into Eq.  (\ref{LagraOM1}), 
the Lagrangian $\caL$ is given explicitly by 
\begin{eqnarray}
   \caL\!\left(\ddot{\bfx}_{s},\dot{\bfx}_{s},
   \bfx_{s},s;\bfmu\right) &=& 
      -\frac{\alpha_{1}\beta_{1}}{4}
      \left[\dot{x}_{1s} 
      +\frac{\kappa \left(x_{1s} - x_{2s}\right)}{\alpha_{1}} 
   +\frac{m_{1}}{\alpha_{1}}\ddot{x}_{1s} \right]^{2} 
   \nonumber \\
   && \spaEq -\frac{\alpha_{2}\beta_{2}}{4}
      \left[\dot{x}_{2s} 
      +\frac{\kappa \left(x_{2s} - x_{1s}\right) 
         - \Gamma}{\alpha_{2}}
      +\frac{m_{2}}{\alpha_{2}}\ddot{x}_{2s} \right]^{2}  
\label{LagraOM2}
\end{eqnarray}
as the sum of two terms due to the presence of two 
independent random noises. 
   Applying our general theory given in Sec. \ref{OMtheory} 
to this model expressed by the Lagrangian (\ref{LagraOM2}), 
we obtain the quantities like $\dot{S}_{\pm}$, 
$\caQ_{\pm}$, $\caW_{\pm}$, etc. 
   Especially, the work $\caW_{\pm}(\{\bfx_{s}\},\bfmu)$ is given by 
\begin{eqnarray}
   \caW_{+}(\{\bfx_{s}\},\bfmu) 
   &=&  \frac{\Delta T}{2T} 
      \int_{-t}^{t}ds\;\left[\kappa \left(x_{1s} - x_{2s}\right) 
      \left(\dot{x}_{1s} +\dot{x}_{2s}\right) 
      + m_{1}\dot{x}_{1s}\ddot{x}_{1s}
      - m_{2}\dot{x}_{2s}\ddot{x}_{2s} \right] 
      \nonumber \\
   && \spaEq
     +\left(1+\frac{\Delta T}{2T} \right) 
      \Gamma \int_{-t}^{t}ds\; \dot{x}_{2s} 
      +\mathcal{O}\left(\left|\frac{\Delta T}{T}\right|^{2}\right) 
%      \nonumber \\
%   &&\spaEq=  \frac{\Delta T}{2T} 
%      \left[\int_{-t}^{t}ds\;\kappa \left(x_{1s} - x_{2s}\right) 
%      \left(\dot{x}_{1s} +\dot{x}_{2s}\right) 
%      +\frac{1}{2}m_{1}\left(\dot{x}_{1t}^{2}
%         -\dot{x}_{1t_{0}}^{2}\right)  
%      - \frac{1}{2}m_{2}\left(\dot{x}_{2t}^{2}
%         -\dot{x}_{2t_{0}}^{2}\right)  \right] 
%      \nonumber \\
%   && \spaEq\spaEq
%     +\left(1+\frac{\Delta T}{2T} \right) 
%      \Gamma\left(x_{2t}-x_{2t_{0}}\right) 
%      +\mathcal{O}\left(\left|\frac{\Delta T}{T}\right|^{2}\right) 
      \label{WorkC3p} 
\end{eqnarray}
and
\begin{eqnarray}
   \caW_{-}(\{\bfx_{s}\},\bfmu) 
   &=&  
      \frac{\Delta T}{4T} \int_{-t}^{t}ds\;  
      \left\{\frac{\left[m_{1}\ddot{x}_{1s} 
         +\kappa \left(x_{1s} - x_{2s}\right)\right]^{2} 
         + (\alpha_{1}\dot{x}_{1s})^{2}}{\alpha_{1}} \right.
      \nonumber \\
   && \spaEq\spaEq\spaEq\spaEq\left.
      -\frac{\left[m_{2}\ddot{x}_{2s} 
         +\kappa \left(x_{2s} - x_{1s}\right) \right]^{2}
         + (\alpha_{2}\dot{x}_{2s}-\Gamma)^{2}}{\alpha_{2}} 
         \right\}
      \nonumber \\
   && \spaEq
      + \Gamma \int_{-t}^{t}ds\; \frac{m_{2}\ddot{x}_{2s} +\kappa 
         \left(x_{2s} - x_{1s}\right)}{\alpha_{2}} 
      +\mathcal{O}\left(\left|\frac{\Delta T}{T}\right|^{2}\right) ,
      \label{WorkC3m}
\end{eqnarray}
up to first order in $|\Delta T/\overline{T}|$, respectively, 
with $\Delta T =T_{1}-T_{2}= 2\Delta T_{1} 
= - 2\Delta T_{2}$ and $T=\overline{T}$. %
%\footnote{
%\cor{   
%   We note that to obtain Eq. (\ref{WorkC3m}) for $\caW_{-}$ 
%in the case of $\Delta T\neq 0$ we cannot use Eq. (\ref{Work2}), 
%which is for (i) both $\caW_{+}$ and $\caW_{-}$ 
%for $\Delta T_{j}=0$ or (ii) $\caW_{+}$ 
%for $\Delta T_{j}\neq 0$. 
%   [On the other hand, Eq. (\ref{WorkC3p}) for  
%$\caW_{+}$ can be derived from  Eq. (\ref{Work2}).] 
%   Eq. (\ref{WorkC3m}) for $\caW_{-}$ is derived from  
%$\caW_{-} = \overline{\beta}^{-1}\int_{-t}^{t}ds\;
%   \left[ \caL\!\left(\ddot{\bfx}_{s},\dot{\bfx}_{s},
%         \bfx_{s},s;\bfmu\right) 
%      - \caL\!\left(\ddot{\bfx}_{s},-\dot{\bfx}_{s},
%         \bfx_{s},-s;-\bfmu\right)\right] 
%      +  E_{\pm}(\dot{\bfx}_{t},\bfx_{t},t;\bfeta) 
%   - E_{\pm}(\dot{\bfx}_{-t},\bfx_{-t},-t;\bfeta) $ 
%[cf Eqs. (\ref{Heat3}), (\ref{EnergDiffe1}) and (\ref{Work1})] 
%with the energy (\ref{EnergC3}) and the Lagrangian (\ref{LagraOM2}). 
%   The quantity $\caW_{-}$ calculated in this way 
%for $\Delta T_{j}\neq 0$ still satisfies 
%the nonequilibrium detailed balance relation 
%(\ref{DetaiBalan1}) and the transient fluctuation theorem 
%(\ref{FluctTrans2}).
%}
%} 
   The works (\ref{WorkC3p}) 
and (\ref{WorkC3m}) are zero at equilibrium where 
$\Delta T=\Gamma=0$.  
%   For the models in Class C it is $\caW_{+}$ that gives the 
%\NESSW for Class C, as we will show below. 
   
%   So far we have discussed Class C using the  
%Langevin equation (\ref{LangeEquatC3}) which includes  
%both the energy transfer model driven by a temperature difference 
%and the electric circuit with two resistors. 
   
   Eqs. (\ref{WorkC3p}) and (\ref{WorkC3m}) gives the works in a unified form for the energy transfer model driven 
by a temperature difference and the electric circuit with 
two resistors in Class C. 
   In the remainder of this section, we discuss 
their physical meanings in these two models separately.

%---------------------------------------------------------------------
%\def\arraystretch{1.5}
%\def\Topspc{\rule{0pt}{12pt}}
%\def\Btmspc{\rule[-10pt]{0pt}{0pt}}
\newcommand{\widthTableP}{0.1\textwidth}
\newcommand{\widthTableQ}{0.45\textwidth}
\newcommand{\widthTableR}{0.34\textwidth}
\newcommand{\widthHeightP}{1cm}
\newcommand{\widthHeightQ}{1.7cm}
\newcommand{\widthHeightR}{1cm}
\newcommand{\widthHeightS}{1cm}
\newcommand{\widthHeightT}{3.2cm}
\newcommand{\widthHeightU}{2.65cm}
\newcommand{\widthHeightV}{3.8cm}
\begin{table}[t!]
\vspfigA
\begin{center}
\renewcommand{\arraystretch}{2}
\begin{tabular}{|c||c|c|}
\hline 
   Class &  \multicolumn{2}{|c|}{Class C} \\
\hline 
   \parbox[c][\widthHeightP]{\widthTableP}{
   \begin{minipage}{\widthTableP} \begin{center}
      \baselineskip 2 ex
      Model  
      \end{center}\end{minipage} }
   & \begin{minipage}{\widthTableQ} \begin{center}
      \baselineskip 0pt
      {\small Energy transfer model }
      \\ {\small driven by a temperature difference} 
      \end{center}\end{minipage} 
   & \begin{minipage}{\widthTableR} \begin{center}
      \baselineskip 0pt
      {\small Electric circuit}\\ {\small with two resisters}
      \end{center}\end{minipage}  \\
\hline 
\hline 
   \parbox[c][\widthHeightQ]{\widthTableA}{
    \begin{minipage}{\widthTableA} \begin{center}
      \baselineskip 2 ex
      Langevin \\ equation, \\ Figure, \\ Section
      \end{center}\end{minipage} } 
   & \begin{minipage}{\widthTableR} \begin{center}
      \baselineskip 3 ex Eq. (\ref{LangeEquatC1}),  
      Fig. \ref{fig3modelC}(a), 
      \\ Secs. \ref{ClassC} and \ref{WorkModelC1}
      \end{center}\end{minipage}
   & \begin{minipage}{\widthTableR} \begin{center}
      \baselineskip 3 ex 
      Eq. (\ref{LangeEquatC2}), Fig. \ref{fig3modelC}(b), 
      \\ Secs. \ref{ClassC} and \ref{WorkModelC2}  
      \end{center}\end{minipage} \\
\hline 
   \parbox[c][\widthHeightC]{\widthTableA}{
   \begin{minipage}{\widthTableA} \begin{center}
      \baselineskip 0 ex
      Nonequi- \\ librium \\ parameter
      \end{center}\end{minipage}} 
   & $\Delta T$ 
   & V \\
\hline 
   \parbox[c][\widthHeightR]{\widthTableA}{
   \begin{minipage}{\widthTableA} \begin{center}
      \baselineskip 2 ex
      Internal \\ energy 
      \end{center}\end{minipage}}
   & $ \sum_{j=1}^{2} \frac{1}{2}m \dot{x}_{js}^{2}
      + \frac{1}{2} \kappa \left(x_{1s} - x_{2s} 
      \right)^{2}$ 
   & $\frac{1}{2}L \dot{q}_{1s}^{2}
      + \frac{\left(q_{1s} - q_{2s} 
      \right)^{2}}{2C} $ \\
\hline 
   \parbox[c][\widthHeightC]{\widthTableA}{
   \begin{minipage}{\widthTableA} \begin{center}
      \baselineskip 0 ex
      Time-reversal \\ operator
      \end{center}\end{minipage}} 
   & \multicolumn{2}{|c|}{$\hat{I}_{+}$} \\
\hline 
   \parbox[c][\widthHeightU]{\widthTableA}{
      $\;$ Forces} 
   & \begin{minipage}{\widthTableQ} \begin{center}
      $F_{1+}^{(e)} = - F_{2+}^{(e)}= -\kappa (x_{1s}-x_{2s})$, \\
      $F_{1+}^{(o)}=F_{2+}^{(o)} = 0$,  \\ 
      $f_{1+} = f_{2+} = 0$
      \end{center}\end{minipage}
   &  \begin{minipage}{\widthTableR} \begin{center}
      $F_{1+}^{(e)}  = -\frac{1}{C} (q_{1s}-q_{2s})$, \\
      $F_{2+}^{(e)} = V -\frac{1}{C} (q_{2s}-q_{1s})$,  \\
      $F_{1+}^{(o)} = F_{2+}^{(o)} = 0$,  \\ 
      $f_{1+}= 0 $, $f_{2+}=V$
      \end{center}\end{minipage}\\
\hline 
   \parbox[c][\widthHeightV]{\widthTableA}{
      \begin{minipage}{\widthTableA} \begin{center}
      \baselineskip 0 ex
      NESS \\ work
      \end{center}\end{minipage}}
   & \begin{minipage}{\widthTableQ} \begin{center}
      $\caW_{+}=\caW_{+}^{(t)} $\\$ =\frac{\Delta T}{2T} 
      \Bigg[\int_{t_{0}}^{t}ds\;\kappa \left(x_{1s} - x_{2s}\right) 
      \left(\dot{x}_{1s} +\dot{x}_{2s}\right) $\\$
      +\frac{1}{2} m \left(\dot{x}_{1t}^{2}
         -\dot{x}_{1t_{0}}^{2}\right)  
      - \frac{1}{2} m \left(\dot{x}_{2t}^{2}
         -\dot{x}_{2t_{0}}^{2}\right) \Bigg] $\\$
      +\mathcal{O}\left(\left|\frac{\Delta T}{T}\right|^{2}\right)$
      \end{center}\end{minipage} 
   & $\caW_{+}=\caW_{+}^{(f)} = \int_{t_{0}}^{t} ds\;  V I_{2s}$\\ 
\hline 
   \parbox[c][\widthHeightE]{\widthTableA}{
   \begin{minipage}{\widthTableA} \begin{center}
      \baselineskip 0 ex
      Average \\ NESS  %\\ thermo-\\dynamic 
      \\ work rate  
      \end{center}\end{minipage}} 
   &$\frac{\alpha \kappa k_{B} \Delta T^{2}}
      {2 T (\alpha^{2}+m\kappa)} > 0$
   & $\frac{V^{2}}{R_{1}+R_{2}} > 0$\\
\hline 
   \parbox[c][\widthHeightT]{\widthTableA}{
   \begin{minipage}{\widthTableA} \begin{center}
      \baselineskip 2 ex
      Non-NESS \\ work 
      \end{center}\end{minipage}}
   & \begin{minipage}{\widthTableQ} \begin{center}
      $\caW_{-} = \caW_{-}^{(t)} 
      =\frac{\Delta T}{4 T} \int_{t_{0}}^{t}ds\;  
      \Bigg\{\alpha(\dot{x}_{1s}^{2}-\dot{x}_{2s}^{2}) $\\$
      + \frac{m}{\alpha} (\ddot{x}_{1s}+\ddot{x}_{2s}) 
         \bigg[m(\ddot{x}_{1s}-\ddot{x}_{2s}) $\\$
         +2\kappa\left(x_{1s} - x_{2s}\right)\bigg]  \Bigg\}
      +\mathcal{O}\left(\left|\frac{\Delta T}{T}\right|^{2}\right)$
      \end{center}\end{minipage} 
   & \begin{minipage}{\widthTableR} \begin{center}
      $ \caW_{-}=\caW_{-}^{(o)} $\\$=\frac{V}{R_{2}C} 
      \int_{t_{0}}^{t}ds\; (q_{2s} - q_{1s})$
      \end{center}\end{minipage} \\
\hline 
\end{tabular}
\end{center}
\vspfigB
\caption{
      Expressions for the work $\caW_{\pm}$ in   
   the NESS models of Class C.  
      We also show the external nonequilibrium parameter $\bfmu$, 
   the internal energy $E$, the relevant time-reversal 
   operator $\hat{I}_{\pm}$ and the forces $F_{j\pm}^{(e)}$, 
   $F_{j\pm}^{(o)}$ and $f_{j\pm}$ to be used to obtain the 
   \NESSw, and the (strictly positive) 
   average physical NESS work rate. 
%      For the work $\caW_{+}$ in the electric circuit model, 
%   we introduced $I_{2s}$ as the electric current $\dot{q}_{2s}$ 
%   generated by the battery. 
   }
   \label{EnergWorkModel2}
\vspfigC
\end{table}%
%---------------------------------------------------------------------

%%%%%%%%%%%%%%%%%%%%%%%%%%%%%%%%%%%%%%%%%%%%%%%%%%%%%%%%%%%%%%%%%%%%%%   
\subsubsection{Energy Transfer by a Temperature Difference}
\label{WorkModelC1}

   We first discuss the energy transfer driven 
by a temperature difference, 
which is described by the Langevin equation (\ref{LangeEquatC1}), 
i.e. Eq. (\ref{LangeEquatC3}) in the case of $m_{1}=m_{2}=m$, 
$\alpha_{1}=\alpha_{2}=\alpha$ and $\Gamma=0$. 
   In Table. \ref{EnergWorkModel2} we show 
the work $\caW_{\pm}$ obtained 
from Eqs. (\ref{WorkC3p}) and (\ref{WorkC3m}) in this case.
   These works are typical examples for the work $\caW_{+}^{(t)}$ 
as given by Eq. (\ref{Work3d}) for $N=2$.

   To choose the \NESSW from $\caW_{+}$ 
and $\caW_{-}$ in Table. \ref{EnergWorkModel2} 
it is enough to note that in the NESS 
the \NESSW should be zero 
in the case of $\kappa=0$, i.e. no coupling between 
the two reservoirs. 
   The work $\caW_{-}$  
does not vanish in the case of $\kappa=0$, 
while the work $\caW_{+}$ 
vanishes in such a case 
apart from a boundary term, which disappears   
in the NESS on average.   
   Therefore, we conclude that the work $\caW_{+}$ 
obtained by the time-reversal operator $\hat{I}_{+}$ gives 
the \NESSW in the energy transfer driven  
by a temperature difference. 
   As further evidence for the appropriateness 
of the \NESSW $\caW_{+}$, we note that 
the average work rate $\dot{\overline{\caW}}_{+}$ 
corresponding to this work $\caW_{+}$ in the NESS 
is given by 
$\dot{\overline{\caW}}_{+}=\alpha \kappa k_{B} \Delta T^{2}
/[2 T (\alpha^{2}+m\kappa)]$
%%
%\begin{eqnarray}
%   \dot{\overline{\caW}}_{+}
%   = \frac{\alpha \kappa k_{B} \Delta T^{2}}{2 T (\alpha^{2}+m\kappa)}
%\label{AveraWorkRateC2}
%\end{eqnarray}
%
up to the second order in $\Delta T$, 
as shown in  Appendix \ref{AveWorkAppC1}. 
   Thus, $\dot{\overline{\caW}}_{+}$ is strictly positive 
and an even function of the temperature difference $\Delta T$, 
which are necessary conditions 
for the average work rate to keep the system in a NESS. %
%\footnote{
%The average work rate $\dot{\overline{\caW}}_{+}$ 
%is also represented by  
%$k_{B} \Delta T^{2}/[2 T (\tau_{r}+\tau_{m})]$, 
%which is inversely proportional to the sum of two time-scales 
%$\tau_{r}\equiv \alpha/\kappa$ and $\tau_{m}\equiv m/\alpha$.
%}   
   This average work rate $\dot{\overline{\caW}}_{+}$ 
has some interesting features due to the term $m\kappa$, 
which includes inertia.  
   Its value is zero at $\alpha =0$ (as well as at $\kappa=0$),  
has a finite maximum value 
[$\dot{\overline{\caW}}_{+}
=k_{B}\Delta T^{2}\sqrt{\kappa}/(4T\sqrt{m})$ 
at $\alpha = \sqrt{m\kappa}$ as a function of $\alpha$, 
and $\dot{\overline{\caW}}_{+}
\rightarrow \alpha k_{B}\Delta T^{2}/(2mT)$  
for $\kappa\rightarrow +\infty$ as a function 
of $\kappa$], and is close to its over-damped value (at $m=0$) 
for large $\alpha$ or small $\kappa$.

%%%%%%%%%%%%%%%%%%%%%%%%%%%%%%%%%%%%%%%%%%%%%%%%%%%%%%%%%%%%%%%%%%%%%%   
\subsubsection{Electric Circuit with Two Resistors}
\label{WorkModelC2}
  
   We categorized the electric circuit with two resistors, 
described by the Langevin equation (\ref{LangeEquatC2}), 
in the same Class C as well as the above energy transfer model, 
although they look very different at first sight. 
   The common feature for these models are that both systems 
are coupled to two independent random noises. 
   However, different from the electric circuit models 
in Classes A and B, results for this electric circuit model 
in Class C cannot be derived from the ones for the corresponding 
Brownian model simply by using the correspondences  
in Table \ref{CorreModel1}, since the Langevin equations 
(\ref{LangeEquatC1}) and (\ref{LangeEquatC2}) for the models 
in Class C have different forms. 
   Therefore, we have to discuss physical quantities 
of this electric circuit model in Class C separately. 
  
  The electric circuit model with two resistors is described 
by the Langevin equation (\ref{LangeEquatC3}) 
in the case of $m_{1}=L$, $m_{2}=0$, $\Gamma =V$, $\kappa=1/C$, 
$\alpha_{j}=R_{j}$, $x_{js}=q_{js}$ and $\Delta T=0$. 
   Therefore, from Eqs. (\ref{WorkC3p}) and (\ref{WorkC3m}) 
the works $\caW_{\pm}$ for this system can be obtained  
(cf. Table \ref{EnergWorkModel2}).    
   It is clear that the work $\caW_{+}$ is the \NESSW done 
by the battery with the voltage $V$ to produce 
electric current $I_{2s} \equiv \dot{q}_{2s}$. 
   We note that the \NESSW $\caW_{+}$ for this model 
is of the type (\ref{Work3b}), different from the energy transfer 
model in which the \NESSW is of the type (\ref{Work3d}). 
   We also show in Appendix \ref{AveWorkAppC2} that in this model 
the average work rate $\dot{\overline{\caW}}_{+}$ in the 
NESS is given by 
$\dot{\overline{\caW}}_{+} = V^{2}/(R_{1}+R_{2})$. 
%%
%\begin{eqnarray}
%   \dot{\overline{\caW}}_{+} = \frac{V^{2}}{R_{1}+R_{2}} .
%\label{AveraWorkRateC3}
%\end{eqnarray}
%
   This average work rate is strictly positive  
and is an even function of the external nonequilibrium parameter $V$.

%%%%%%%%%%%%%%%%%%%%%%%%%%%%%%%%%%%%%%%%%%%%%%%%%%%%%%%%%%%%%%%%%%%%%%
\section{Summary and Remarks}
\label{SummaRemar}

   In this paper we have discussed a method to calculate the work 
done on and the heat removed from a system to maintain it in a 
NESS. 
   This was based on a NESS Onsager-Machlup theory 
for stochastic systems with Gaussian-white random noises.  
   The work and the heat to maintain a NESS appear only for NESSs, 
and not for equilibrium states. 
   In our approach we obtained the heat as the 
time-irreversible part of the probability functional 
for paths in a functional space 
via a Lagrangian, and from it the work, 
using the energy conservation law.   
   We incorporated multiple possibilities for the time-reversal 
procedure for NESSs, due to the external 
nonequilibrium parameters to specify the NESS. 
   We also indicated that the separation of the heat 
into work and an energy differences is not unique, so that 
we can get different expressions for the work for systems 
described by the same dynamical equation. 
   We showed that the work can consist of four parts: 
one coming from a partial derivative of the energy with respect 
to time, the second one due to an external driving force, 
the third one caused by a time-irreversible force  
and the last one due to temperature differences of reservoirs. 
   We also derived nonequilibrium generalizations of the 
detailed balance condition, leading to transient fluctuation 
theorems for the work distribution functions. 
   Our theory was illustrated by various NESS 
models, for example, dragged Brownian models, electric current 
models, an energy transfer model driven by a temperature difference, 
demonstrating the above four kinds of components for the work.  

   Finally, we make some remarks 
on the contents of this paper. 

%Entropy production <---> Transport coefficient

%   The NESS thermodynamics discussed here 
%is far from as completely developed as for equilibrium states. 
%   (For example, different from equilibrium thermodynamics 
%we still do not have a satisfactory method 
%to calculate some thermodynamics quantities like 
%specific heat and pressure, etc. in NESSs.)
%   But at least the two basic laws of equilibrium thermodynamics can be generalized to NESS, with unique and appropriate definitions of 
%heat and work. 
% steady current dependence in pressure, etc. 

   [1] We first make some remarks about the ambiguities to 
define the \NESSW by the NESS Onsager-Machlup theory.
   
   1a) The first ambiguity is a non-uniqueness 
of time-reversal procedures due to the presence of external 
parameters, defining the NESS.  
   This can be treated by the introduction of two 
time-reversal operators $\hat{I}_{+}$ and $\hat{I}_{-}$, 
leading to two candidates for the \NESSw, 
i.e., $\caW_{+}$ and $\caW_{-}$. 
   To chose the \NESSw, i.e. the actual work 
to maintain the system in a NESS, 
from $\caW_{+}$ and $\caW_{-}$ we had to use 
model-dependent physical arguments, for example, 
(i) The strict positivity of the average work in the NESS 
[cf. the electric field model (Class A) and 
the dragged Brownian particle model in the comoving frame (Class B)], 
since positive work has to be done to sustain the system in a NESS,   
(ii) The time-reversal symmetry (\ref{ReverEnerg1}) 
for the internal energy% 
\footnote{ 
   Note that if the internal energy were time-irreversible, then 
the energy of the final state of the forward path 
would not equal the initial energy of the backward path. 
}
[cf. the dragged Brownian particle model 
in the laboratory frame (Class B)].
   However, a general criterion to choose the \NESSW 
from $\caW_{+}$ and $\caW_{-}$ is an open problem. 
%   One should notice that both of $\caW_{+}$ and $\caW_{-}$ 
%satisfy the transient fluctuation theorems (\ref{FluctTrans2}), 
%so the transient fluctuation theorem cannot be an 
%criteria to choose the \NESSw. 

   1b) The second ambiguity to choose the \NESSW 
is due to multiple possibility to separate heat 
into work and an energy difference. 
   In our theory, this ambiguity appears in Eq. (\ref{Poten1}), 
i.e. when the even part $F_{j\pm}^{(e)}$ of the external force 
is separated into the two terms 
$- \partial U_{\pm} /\partial x_{js}$ and $f_{j\pm}$ 
as an essential step to define the energy $E_{\pm}$ 
by Eq. (\ref{Energ1}).  
   Note that like the first ambiguity discussed in the previous 
paragraph, this ambiguity also does not appear at equilibrium 
because there is then no external driving force $f_{j\pm}$. 
   We demonstrated this second ambiguity concretely 
using the dragged Brownian particle model and 
the driven torsion pendulum model, which are described 
by the same Langevin equation but have 
a different form for their internal energies. 
%   This ambiguity can be regarded as 
%the one to decide what is the ``system'' to be interested.   
%   In this viewpoint, for example, 
   In fact, in the driven torsion pendulum model, the torque 
$M_{s} (=\xi s)$ can be regarded as an external force 
not part of the system and not contributing to its internal 
energy, while in the dragged Brownian particle model 
the force $\kappa v s$, which corresponds to $M_{s}$, 
does contribute to the internal energy and 
is therefore regarded to be as part of the system.   
   In general, this ambiguity can be resolved only 
on physical grounds, 
i.e. by considering what is physically an internal energy   
for each nonequilibrium model.%
\footnote{
   This difficulty does not occur if one consider only 
the work (\ref{Work3a}) given by a partial 
time-derivative of the energy, as done in Refs. \cite{J97,C98}.  
   However, as shown in this paper, this relation 
connecting the work and the energy 
is not valid in general since the work can have other contributions  
(\ref{Work3b}), (\ref{Work3c}) and (\ref{Work3d}).  
}  %
   In another example,  
in the electric field model in Sec. \ref{ClassA},  
we took the internal energy as not to include  
the force $q\caE$, since this force was regarded as 
an external  force $f_{j+}$, leading to the \NESSW 
$\caW_{+} = \caW_{+}^{(f)}$ shown in Table \ref{EnergWorkModel1}.  
   However, purely mathematically, 
we could have chosen the internal energy as 
$(1/2)m\dot{x}_{s}^{2}+ q\caE x_{s}$ [cf. Eq. (\ref{EnergA1})]. 
   This choice of energy leads to the work $\caW_{+} = 0$, 
which is unphysical, since positive work must 
be done to keep the system in a NESS. 

  1c) An additional remark related to the above points is that 
the energy conservation law and the second law of thermodynamics  
are \emph{not} sufficient to determine 
the \NESSW $\caW$ and heat $\caQ$ uniquely, 
because one can always add 
the same functional $X$ of a path with a positive functional average 
($\pathaveA{X}>0$) to both the work and the heat so that these new ``work'' $\caW+X$ 
and ``heat'' $\caQ + X$ satisfy the energy conservation law 
$\caQ + X = (\caW+X) - \Delta E$ 
and the second law of thermodynamics $\pathaveA{\caQ+X}\geq 0$. 
   However, in our NESS Onsager-Machlup theory 
this ambiguity does not occur, because the heat 
%is introduced to have the anti-time-symmetry and 
is fixed by Eq. (\ref{Heat4}) via the probability functional 
$\caP_{x}(\{\bfx_{s}\};\bfmu)$ of paths.

[2] We now make some remarks on the time-reversal operator 
$\hat{I}_{\pm}$ for our NESS Onsager-Machlup theory. 
   In particular, we comment on the relation of these operators 
with other possible time-reversal operators which have been 
used in the literature. 

   2a) In Ref.  \cite{CCJ06} 
a time-reversal operator $\hat{I}^{\prime}$ is introduced by   
\begin{eqnarray}
   \hat{I}^{\prime}\int_{-t}^{t} ds\; Y
      \!\left(\ddot{\bfx}_{s},\dot{\bfx}_{s},\bfx_{s},s\right) 
      = \int_{-t}^{t} ds\; 
      Y\!\left(\ddot{\bfx}_{s},-\dot{\bfx}_{s},\bfx_{s},s\right), 
\label{TimeRever3} 
\end{eqnarray}
for any function $Y(\ddot{\bfx}_{s},\dot{\bfx}_{s},
\bfx_{s},s)$ of $\ddot{\bfx}_{s}$, $\dot{\bfx}_{s}$, 
$\bfx_{s}$, $s$, 
under which the explicit time-dependence $s$ 
in $Y(\ddot{\bfx}_{s},\dot{\bfx}_{s},
\bfx_{s},s)$ does not change its sign. 
   By the definition (\ref{TimeRever3}),  
the time-reversal operator $\hat{I}^{\prime}$ is independent 
of the external nonequilibrium parameter $\bfmu$. 
   This time-reversal operator $\hat{I}^{\prime}$ is different 
from the operator $\hat{I}_{\pm}$ used in this paper, 
but can yet be considered as a special case of the time-reversal 
operator $\hat{I}_{-}$. 
   This, because Eq. (\ref{TimeRever2}) 
for the operator $\hat{I}_{-}$ becomes  Eq. (\ref{TimeRever3}) 
if the explicit $\bfmu$- and $s$-dependences of 
$Y(\ddot{\bfx}_{s},\dot{\bfx}_{s},\bfx_{s},s;\bfmu)$ appear  
only via $\bfmu s$, 
i.e. if $Y(\ddot{\bfx}_{s},\dot{\bfx}_{s},\bfx_{s},s;\bfmu) 
= \tilde{Y}(\ddot{\bfx}_{s},\dot{\bfx}_{s},\bfx_{s},\bfmu s)$   
with a function 
$\tilde{Y}(\ddot{\bfx}_{s},\dot{\bfx}_{s},\bfx_{s},\bfmu s)$.   
   However, the operator $\hat{I}^{\prime}$ can not be 
used as an operator on a general functional $X(\{\bfx_{s}\})$, 
contrary to the operator $\hat{I}_{\pm}$ defined 
by Eq. (\ref{TimeRever1}). 
   Moreover, the time-reversal operator $\hat{I}^{\prime}$ 
fails to produce the correct \NESSW 
for some nonequilibrium models discussed in Sec. \ref{ComovModel}. 
   For example, if we were to apply $\hat{I}^{\prime}$ 
%for a Onsager-Machlup theory and apply it 
to the dragged Brownian particle model in the comoving frame, 
then we would obtain the work to overcome the friction only, % force, 
i.e. $\caW_{+}=-\int_{-t}^{t} ds \; \alpha\dot{y}_{s} v$, 
instead of the \NESSW 
$-\int_{-t}^{t} ds \; (\kappa y_{s}+m\ddot{y}_{s}) v$ 
for this model. 
   For these reasons we did not use 
the time-reversal operator (\ref{TimeRever3}) in this paper. 

  2b) As another example, Ref. \cite{LS99} considered 
the case in which the sign of the external nonequilibrium parameter 
does not change in a time-reversal procedure. 
   This case corresponds to the time-reversal operator 
$\hat{I}_{+}$ in this paper. 
   However, %as we showed in this paper 
this operator does not always produce the correct 
\NESSW for some models, for example, for the models
of Class B in Secs. \ref{WorkModelB} and 
\ref{ComovModel}. % (cf. Table \ref{EnergWorkModel1}).  
   Therefore, this operator is not general enough to construct 
the \NESSW and heat based on 
the NESS Onsager-Machlup theory for  
a sufficiently general class of nonequilibrium systems  
which include all NESS models discussed in this paper.   

%   2c) We introduced the time-reversal operator 
%$\hat{I}_{-}$ so that sign of the nonequilibrium parameter 
%$\bfmu$ is changed in the corresponding time-reversedmotion. 
%   In general, the nonequilibrium parameter can be a vector 
%$\bfmu=(\mu_{1}, \mu_{2}, \cdots)$ 
%containing multiple components $\mu_{1}$, $\mu_{2}$, $\cdots$.  
%   Mathematically speaking, we could introduce 
%a ``time-reversal'' operator $\hat{I}''$ in which only some of these 
%components of the nonequilibrium parameter change signs 
%in the corresponding time-reversal motion.  
%   Using this operator $\hat{I}''$ we could construct 
%a NESS Onsager-Machlup theory, and 
%derive, for example,  transient fluctuation theorems.  
%   However, the physical meaning of such 
%a time-reversal operator $\hat{I}''$ is not clear. 

\noindent [3] In this point we make some remarks on the 
transient and the asymptotic fluctuation theorems 
in the context of the NESS Onsager-Machlup theory.

3a) We first consider the transient fluctuation theorem \cite{ES94}.  
%as it appears in the literature.  
   We will argue that the transient fluctuation theorem 
can be derived purely formally, as a mathematical identity, 
without any specifications 
of the dynamics of the system.  
%[cf.also Ref. \cite{TC07a} Eqs. (106)-(110)]. 
   To show this, we first define, purely formally, 
without any physical interpretation, the functional 
$\caQ(\{\bfx_{s}\})$ by:
\begin{eqnarray}
   \caQ(\{\bfx_{s}\}) = \frac{1}{\beta}
   \ln\frac{\caP(\{\bfx_{s}\})}{\hat{\caJ}\caP(\{\bfx_{s}\})}
\label{Heat0}
\end{eqnarray}
where $\caP(\{\bfx_{s}\})$ is the probability functional 
of the path $\{x_{s}\}_{s\in [t_{0},t]}$.
   Here, the operator $\hat{\caJ}$ is a general 
time-reversal operator satisfying the condition
$\hat{\caJ}^{2}=1$.  
   Next we introduce again purely formally another functional 
$\caW(\{\bfx_{s}\})$ defined by:
\begin{eqnarray}
    \caW(\{\bfx_{s}\}) = \caQ(\{\bfx_{s}\}) + \Delta E 
\label{EnergConse0}
\end{eqnarray}
where $   \Delta E \equiv E(\dot{\bfx}_{t},\bfx_{t},t) 
   -  E(\dot{\bfx}_{t_{0}},\bfx_{t_{0}},t_{0})
$  is a boundary term as the energy difference 
between the energies of the system under consideration 
at the final time $t$ and the initial time $t_0$. 
   Inserting then (\ref{EnergConse0}) into Eq. (\ref{Heat0}), 
we obtain identically: 
\begin{eqnarray}
   e^{-\beta[\caW(\{\bfx_{s}\})-\Delta \caF]} \caP(\{\bfx_{s}\}) 
   \varrho(\dot{\bfx}_{t_{0}},\bfx_{t_{0}},t_{0}) 
   =  \varrho(\dot{\bfx}_{t},\bfx_{t},t) 
   \hat{\caJ} \caP(\{\bfx_{s}\}) ,
\label{NEDetaiBalan1}
\end{eqnarray}
where $\varrho(\dot{\bfx}_{s},\bfx_{s},s) 
\equiv Z_{s}^{-1} \exp[-\beta $ $E(\dot{\bfx}_{s},\bfx_{s},s) ]$ 
is a canonical-like distribution,
with the formal partition function 
$Z_{s}\equiv \int dx_{s}\int d\dot{x}_{s}\; 
\exp[-\beta E(\dot{\bfx}_{s},\bfx_{s},s) ]$. 
   Here $\Delta \caF \equiv \caF_{t}-\caF_{t_{0}}$ 
with $\caF_{s}\equiv -\beta^{-1}\ln Z_{s}$, 
a formal free energy-like quantity.
   If then  $\hat{\caJ}\caW = -\caW$ 
[cf. Eq. (\ref{ReverWork1}) for $\hat{\caJ}=\hat{I}_{\pm}$]
%(as in the NESS Onsager-Machlup theory with 
%$\hat{\caJ}=\hat{I}_{\pm}$) 
then the work distribution, 
$P_{w}(W,t) = \pathaveA{\delta \left(W-\caW(\{\bfx_{s}\})\right)}$  satisfies formally a generalized transient
fluctuation theorem similar in the form Eq. (\ref{FluctTrans2}):  
\begin{eqnarray}
   \frac{P_{w}(W,t)}{\hat{\caJ}P_{w}(-W,t)} 
   = e^{\beta (W-\Delta \caF)}
\label{FluctTheor2}
\end{eqnarray}
for the canonical-like initial condition 
$\varrho(\dot{\bfx}_{t_{0}},\bfx_{t_{0}},t_{0})$. 
%
%   As mentioned above, in this purely formal derivation 
%of the transient fluctuation theorem the specific dynamics 
%of the system does not enter in any way.  
   However, the derivation of Eq. (\ref{FluctTheor2}) 
shows that if one introduces formally 
\emph{any} quantities Q and W defined by the equations (\ref{Heat0}) and 
(\ref{EnergConse0}), respectively, then the Eq. (\ref{FluctTheor2}) 
follows as an identity.  
%where its particular form, involving $\Delta\caF$, was only 
%introduced to make contact with similar expressions 
%in the literature; any initial condition would suffice 
%[cf. Eqs. (106)-(110) in  Ref. \cite{TC07a}].
%
   To the contrary as shown in this paper the NESS Onsager-Machlup 
theory does define physical quantities $\caQ(\{\bfx_{s}\})$ 
and $\caW(\{\bfx_{s}\})$ [cf. Eqs. (\ref{Heat4}), (\ref{Work1}) 
and (\ref{ReverWork1})], which lead to a transient fluctuation 
theorem of the form (\ref{FluctTrans2}).  
%[with the proviso that the $\caW(\{\bfx_{s}\})$ cannot be 
%specified uniquely - so that there are two transient
%fluctuation theorems in the NESS Onsager-Machlup theory 
%- because of the ambiguity of the time reversal operator 
%$\hat{I}_{\pm}$ in this theory - being $\hat{I}_{+}$ or
%$\hat{I}_{-}$ and ``the internal energy difference'']. 

3b) In this paper we derived transient fluctuation theorems 
using the NESS Onsager-Machlup theory for the work 
distribution $P_{w\pm}(W,t;\bfmu)$. 
   The transient fluctuation theorems \cite{ES94}
(as well as those in Refs. \cite{J97,C99}) 
hold for any time for an equilibrium initial distribution function.
   For a general, i.e. any, initial condition 
(including equilibrium), an asymptotic fluctuation theorem  
\cite{GC95} can be derived, in the form:   
\begin{eqnarray}
   \lim_{t\rightarrow+\infty} 
   \frac{P_{w}(W,t)}{P_{w}(-W,t)} 
   = e^{\beta W} .
\label{FluctAsymp1}
\end{eqnarray}
for a work distribution function $P_{w}(W,t)$. 
   In spite of its analogy with Eq. (\ref{FluctTheor2}),
it is of an entirely different nature \cite{CG99,G06}. 
   To be sure, Eq. (\ref{FluctAsymp1}) can be formally derived 
in the long time limit from Eq. (\ref{FluctTheor2}), if
$P_{w}(W,t) = \hat{\caJ}P_{w}(W,t)$, $\Delta \caF=0$ 
and the initial condition is given by 
$\varrho(\dot{\bfx}_{t_{0}},\bfx_{t_{0}},t_{0})$. 
   However, Eq.(\ref{FluctAsymp1}) makes a much stronger statement, because it not only holds for the equilibrium initial condition,  
%[which the $\varrho(\dot{\bfx}_{t_{0}},\bfx_{t_{0}},t_{0})$ 
%is not, since it is a canonical-like condition, not necessarily
%equilibrium ?], 
but for any initial condition, which would require some dynamical
stability condition, e.g. a condition for the system to 
approach a unique 
steady state for $t\rightarrow +\infty$. 

   In this connection, one can ask whether both work 
distribution functions
$P_{w+}(W,t;\bfmu)$ and $P_{w-}(W,t;\bfmu)$,  
given by Eq. (\ref{WorkDistr1}),  
obey the asymptotic fluctuation theorem for any initial 
condition.
   The answer is \emph{no}, as was shown in Ref. \cite{TC07a} 
for the distribution function $P_{w+}(W,t;\bfmu)$ for the dragged Brownian particle. 
%[Is the occurrence of this special frame-dependence understandable
%for an asymptotic FT?].
%
%\footnote{
%In Ref. \cite{TC07a} we used the notations $P_{w}(W,t)$ and 
%$P_{r}(W,t)$ instead of $P_{w-}(W,t;\bfmu)$ and $P_{w+}(W,t;\bfmu)$,  
%respectively, for work distributions  
%in the dragged Brownian particle model. 
%}
   In fact, this could occur when the work distribution 
function $P_{w}(W,t)$ appears 
as a boundary term of the form $X(\bfx_{t})-X(\bfx_{t_{0}})$,  
for a function $X(\bfx_{s})$ of $\bfx_{s}$, 
rather than as a functional along the full path 
$\{\bfx_{s}\}_{s\in[t_{0},t]}$.

\section*{Acknowledgements}

   We gratefully acknowledge financial support 
of the National Science Foundation, under award PHY-0501315.

%%%%%%%%%%%%%%%%%%%%%%%%%%%%%%%%%%%%%%%%%%%%%%%%%%%%%%%%%%%%%%%%%%%%%%
%%%%%%%%%%%%%%%%%%%%%%%%%%%%%%%%%%%%%%%%%%%%%%%%%%%%%%%%%%%%%%%%%%%%%%
%\pagebreak
\appendix
\setcounter{section}{0} 
\makeatletter 
   \@addtoreset{equation}{section} 
   \makeatother 
   \def\theequation{\Alph{section}.% 
   \arabic{equation}}

%---------------------------------------------------------------------
\section{Work based on the NESS Onsager-Machlup theory}
\label{WorkOMtheory}

In this Appendix we give a derivation of Eq. (\ref{Work2}). 

   Using Eqs. (\ref{Heat2}), (\ref{Poten1}), (\ref{Energ1}),  
(\ref{EnergDiffe1}) and (\ref{Work1}) we obtain 
\begin{eqnarray}
   && \caW_{\pm}(\{\bfx_{s}\};\bfmu) 
      + \sum_{j=1}^{N} \frac{1}{\alpha_{j}} \int_{-t}^{t}ds\; 
      \left[F_{j\pm}^{(e)}(\bfx_{s},s;\bfeta)-m\ddot{x}_{js} \right]
      F_{j\pm}^{(o)}(\bfx_{s},s;\bfeta) 
      \nonumber \\
   &&\spaEq\spaEq 
      +\sum_{j=1}^{N} \frac{\Delta T_{j}}{\overline{T}} 
      \int_{-t}^{t}ds\; 
      \left[F_{j\pm}^{(e)}(\bfx_{s},s;\bfeta)-m_{j}\ddot{x}_{js}
      \right]\left[\dot{x}_{js} 
      - \frac{1}{\alpha}F_{j\pm}^{(o)}(\bfx_{s},s;\bfeta)\right] 
      \nonumber \\
   && \spaEq\spaEq - \left[1+(-1)^{\pm 1}\right]
      \sum_{j=1}^{N}  \frac{\alpha_{j}}{2} 
      \frac{\Delta T_{j}}{\overline{T}} 
      \nonumber \\
   && \spaEq\spaEq\spaEq \times
      \int_{-t}^{t}ds\;  
      \left[\dot{x}_{js} 
      + \frac{1}{\alpha_{j}} F_{j}(\bfx_{s},-s;\pm\bfeta)
      -\frac{m_{j}}{\alpha_{j}}\ddot{x}_{js} \right]^{2}
%      \nonumber \\
%   &&\spaEq\spaEq 
      +\mathcal{O}\left(\left|\frac{\Delta T_{j}}{\overline{T}}
      \right|^{2}\right) 
      \nonumber \\
   &&\spaEq 
      =  \sum_{j=1}^{N} \int_{-t}^{t}ds\; \left[-\frac{\partial 
         U_{\pm}(\bfx_{s},s;\bfeta)}
         {\partial x_{js}} + f_{j\pm}(\bfx_{s},s;\bfeta)
         -m\ddot{x}_{js} \right]\dot{x}_{js} 
      + \Delta E_{\pm} 
      \nonumber \\ 
   &&\spaEq 
      =  \int_{-t}^{t}ds\; \left[
         \frac{\partial U_{\pm}(\bfx_{s},s;\bfeta)}{\partial s}
         +\sum_{j=1}^{N}f_{j\pm}(\bfx_{s},s;\bfeta) 
         \dot{x}_{js}\right]
      + \Delta E_{\pm}      
      \nonumber \\
   &&\spaEq\spaEq
      - \int_{-t}^{t}ds\; \left[
         \sum_{j=1}^{N} 
         \frac{d}{ds}\left(\frac{1}{2} m \dot{x}_{js}^{2}\right)
         + \sum_{j=1}^{N}\frac{\partial U_{\pm}(\bfx_{s},s;\bfeta)}
         {\partial \bfx_{s}}\dot{x}_{js}    
         + \frac{\partial U_{\pm}(\bfx_{s},s;\bfeta)}{\partial s}
        \right] 
      \nonumber \\ 
   &&\spaEq 
      = \int_{-t}^{t}ds\;\left[ \frac{\partial 
         E_{\pm}(\dot{\bfx}_{s},\bfx_{s},s;\bfeta)}{\partial s}
         +\sum_{j=1}^{N} f_{j\pm}(\bfx_{s},s;\bfeta)
         \dot{x}_{js}\right]
         \nonumber \\
   &&\spaEq\spaEq
      + \Delta E_{\pm} -\int_{-t}^{t}ds\; 
         \frac{d E_{\pm}(\dot{\bfx}_{s},\bfx_{s},s;\bfeta)}{ds}
      \nonumber \\ 
   &&\spaEq 
      = \int_{-t}^{t}ds\;\left[
         \frac{\partial E_{\pm}(\dot{\bfx}_{s},\bfx_{s},s;\bfeta)}
         {\partial s}
         + \sum_{j=1}^{N} 
         f_{j\pm}(\bfx_{s},s;\bfeta)\dot{x}_{js}\right].
%\label{}
\end{eqnarray}
Therefore, we obtain Eq. (\ref{Work2}) with Eqs. 
(\ref{Work3a}), (\ref{Work3b}), (\ref{Work3c}) and (\ref{Work3d}).  
%

%%%%%%%%%%%%%%%%%%%%%%%%%%%%%%%%%%%%%%%%%%%%%%%%%%%%%%%%%%%%%%%%%%%%%%
\section{Transient Fluctuation Theorems} 
\label{TransFluctAp}

   In this Appendix we derive Eqs. (\ref{FluctTrans1}) 
and (\ref{FluctTrans2}) 
in the case of the initial condition (\ref{InitiCondi1}). 

   Using the definition (\ref{Efunct1}) of 
$\caE_{w\pm}(\sigma,t;\bfmu)$ with the functional average 
(\ref{PathAvera1}) under the initial distribution function  
(\ref{InitiCondi1}) we obtain 
\begin{eqnarray}
   \caE_{w\pm}(\overline{\beta}-\sigma,t;\bfmu) 
%      \nonumber\\
%   &=&  \pathaveA{e^{-(\overline{\beta}-\sigma)  
%      \caW_{\pm}(\{\bfx_{s}\};\bfmu)}} 
%      \nonumber \\
%   &=& \int d\bfx_{f}\int d\dot{\bfx}_{f}
%      \int_{(\bfx_{-t},\dot{\bfx}_{-t})
%         =(\bfx_{i},\dot{\bfx}_{i})}%
%      ^{(\bfx_{t},\dot{\bfx}_{t})=(\bfx_{f},\dot{\bfx}_{f})} 
%      \mathcal{D}\bfx_{s} \int d\bfx_{i}\int d\dot{\bfx}_{i}\; 
%      \nonumber \\
%   &&\spaEq \times 
%      e^{-(\overline{\beta}-\sigma)\caW_{\pm}(\{\bfx_{s}\};\bfmu)}
%      \caP_{x}(\{\bfx_{s}\}) f(\bfx_{i},\bfp_{i},-t) 
%      \nonumber \\
   &=& \int d\bfx_{f}\int d\dot{\bfx}_{f}
      \int_{(\bfx_{-t},\dot{\bfx}_{-t})
         =(\bfx_{i},\dot{\bfx}_{i})}%
      ^{(\bfx_{t},\dot{\bfx}_{t})=(\bfx_{f},\dot{\bfx}_{f})} 
      \mathcal{D}\bfx_{s} \int d\bfx_{i}\int d\dot{\bfx}_{i}\; 
      \nonumber \\
   &&\spaEq \times 
      e^{-(\overline{\beta}-\sigma)\caW_{\pm}(\{\bfx_{s}\};\bfmu)}
      \caP_{x}(\{\bfx_{s}\}) 
      \varrho_{\pm}(\dot{\bfx}_{i},\bfx_{i},-t;\bfeta)  
      \nonumber \\
%   &=& \int d\bfx_{f}\int d\dot{\bfx}_{f}
%      \int_{(\bfx_{-t},\dot{\bfx}_{-t})
%         =(\bfx_{i},\dot{\bfx}_{i})}%
%      ^{(\bfx_{t},\dot{\bfx}_{t})=(\bfx_{f},\dot{\bfx}_{f})} 
%      \mathcal{D}\bfx_{s} \int d\bfx_{i}\int d\dot{\bfx}_{i}\; 
%      \nonumber \\
%   &&\spaEq \times 
%      e^{-(\overline{\beta}-\sigma)\caW_{\pm}(\{\bfx_{s}\};\bfmu)}
%      \varrho_{\pm}(\dot{\bfx}_{f},\bfx_{f},t;\bfeta) 
%    \left[\hat{I}_{\pm}\caP_{x}(\{\bfx_{s}\};\bfmu)\right] 
%    e^{\overline{\beta}\left[\caW_{\pm}(\{\bfx_{s}\};\bfmu) 
%    - \Delta\caF_{\pm} \right]}
%      \nonumber \\
   &=&  e^{-\overline{\beta} \Delta\caF_{\pm}}
      \int d\bfx_{f}\int d\dot{\bfx}_{f}
      \int_{(\bfx_{-t},\dot{\bfx}_{-t})
         =(\bfx_{i},\dot{\bfx}_{i})}%
      ^{(\bfx_{t},\dot{\bfx}_{t})=(\bfx_{f},\dot{\bfx}_{f})} 
      \mathcal{D}\bfx_{s} \int d\bfx_{i}\int d\dot{\bfx}_{i}\; 
      \nonumber \\
   &&\spaEq \times 
      e^{\sigma\caW_{\pm}(\{\bfx_{s}\};\bfmu)}
      \varrho_{\pm}(\dot{\bfx}_{f},\bfx_{f},t;\bfeta)
      \hat{I}_{\pm}\caP_{x}(\{\bfx_{s}\};\bfmu) 
      \nonumber \\
%   &=&  e^{-\overline{\beta} \Delta\caF_{\pm}}
%      \int d\bfx_{f}\int d\dot{\bfx}_{f}
%      \int_{(\bfx_{-t},\dot{\bfx}_{-t})
%         =(\bfx_{i},\dot{\bfx}_{i})}%
%      ^{(\bfx_{t},\dot{\bfx}_{t})=(\bfx_{f},\dot{\bfx}_{f})} 
%      \mathcal{D}\bfx_{s} \int d\bfx_{i}\int d\dot{\bfx}_{i}\; 
%      \nonumber \\
%   &&\spaEq \times 
%      \varrho_{\pm}(\dot{\bfx}_{f},\bfx_{f},t;\bfeta)
%      \hat{I}_{\pm} e^{-\sigma\caW_{\pm}(\{\bfx_{s}\};\bfmu)}
%      \caP_{x}(\{\bfx_{s}\};\bfmu) 
%      \nonumber \\
   &=&  e^{-\overline{\beta} \Delta\caF_{\pm}}
      \int d\bfx_{f}\int d\dot{\bfx}_{f}
      \int_{(\bfx_{-t},\dot{\bfx}_{-t})
         =(\bfx_{i},\dot{\bfx}_{i})}%
      ^{(\bfx_{t},\dot{\bfx}_{t})=(\bfx_{f},\dot{\bfx}_{f})} 
      \mathcal{D}\bfx_{s} \int d\bfx_{i}\int d\dot{\bfx}_{i}\; 
      \nonumber \\
   &&\spaEq \times 
      \varrho_{\pm}(-\dot{\bfx}_{f},\bfx_{f},-t;\pm\bfeta)
      \hat{I}_{\pm} e^{-\sigma\caW_{\pm}(\{\bfx_{s}\};\bfmu)}
      \caP_{x}(\{\bfx_{s}\};\bfmu) 
      \nonumber \\
   &=&  e^{-\overline{\beta} \Delta\caF_{\pm}} \hat{I}_{\pm}
      \int d\bfx_{f}\int d\dot{\bfx}_{f}
      \int_{(\bfx_{t},\dot{\bfx}_{t})
         =(\bfx_{i},\dot{\bfx}_{i})}%
      ^{(\bfx_{-t},\dot{\bfx}_{-t})=(\bfx_{f},\dot{\bfx}_{f})} 
      \mathcal{D}\bfx_{s} \int d\bfx_{i}\int d\dot{\bfx}_{i}\; 
      \nonumber \\
   &&\spaEq \times 
      \varrho_{\pm}(-\dot{\bfx}_{f},\bfx_{f},-t;\bfeta)
      e^{-\sigma\caW_{\pm}(\{\bfx_{s}\};\bfmu)}
      \caP_{x}(\{\bfx_{s}\};\bfmu) 
      \label{TFTAppen1} \\
   &=&  e^{-\overline{\beta} \Delta\caF_{\pm}}
      \hat{I}_{\pm} \int d\bfx_{f}\int d\dot{\bfx}_{f}
      \int_{(\bfx_{-t},\dot{\bfx}_{-t})
         =(\bfx_{i},\dot{\bfx}_{i})}%
      ^{(\bfx_{t},\dot{\bfx}_{t})=(\bfx_{f},\dot{\bfx}_{f})} 
      \mathcal{D}\bfx_{s} \int d\bfx_{i}\int d\dot{\bfx}_{i}\; 
      \nonumber \\
   &&\spaEq \times 
      e^{-\sigma\caW_{\pm}(\{\bfx_{s}\};\bfmu)}
      \caP_{x}(\{\bfx_{s}\};\bfmu)
      \varrho_{\pm}(\dot{\bfx}_{i},\bfx_{i},-t;\bfeta) 
      \label{TFTAppen2} \\
%   &=&  e^{-\overline{\beta} \Delta\caF_{\pm}}
%      \hat{I}_{\pm} \pathaveA{e^{-\sigma  
%      \caW_{\pm}(\{\bfx_{s}\};\bfmu)}} 
%      \nonumber \\
   &=&  e^{-\overline{\beta} \Delta\caF_{\pm}}
      \hat{I}_{\pm} 
      \caE_{w\pm}(\sigma,t;\bfmu) 
%\label{}
\end{eqnarray}
where we used Eqs. (\ref{TimeRever1}), 
(\ref{ReverWork1}), (\ref{DetaiBalan1}) and 
$
   \varrho_{\pm}(\dot{\bfx}_{f},\bfx_{f},t;\bfeta)
   = \varrho_{\pm}(-\dot{\bfx}_{f},\bfx_{f},-t;\pm\bfeta)
$
noting Eq. (\ref{ReverEnerg1}). 
   Here, in the transformation from Eq. (\ref{TFTAppen1}) to 
Eq. (\ref{TFTAppen2}) we exchanged the integral variables 
$\bfx_{i},\dot{\bfx}_{i}, \bfx_{f}$ and $\dot{\bfx}_{f}$ 
with 
$\bfx_{f},\dot{\bfx}_{f}, \bfx_{i}$ and $\dot{\bfx}_{i}$, 
respectively, and used the relation 
$\varrho_{\pm}(\dot{\bfx}_{i},\bfx_{i},-t;\bfeta)
=\varrho_{\pm}(-\dot{\bfx}_{i},\bfx_{i},-t;\bfeta)$. 
   Therefore, we obtain Eq. (\ref{FluctTrans1}). 
   
   Using Eq. (\ref{FluctTrans1}) we obtain  
\begin{eqnarray}
   \caE_{w\pm}(\sigma,t;\bfmu) 
   &=&  e^{-\overline{\beta} \Delta\caF_{\pm}}
      \hat{I}_{\pm} 
   \caE_{w\pm}(\overline{\beta}-\sigma,t;\bfmu)  .
\label{FluctTrans3}
\end{eqnarray}
   From Eqs. (\ref{WorkDistr2})  
and (\ref{FluctTrans3}) we can derive 
\begin{eqnarray}
   P_{w\pm}(W,t;\bfmu) 
%      &=& \frac{1}{2\pi}\int_{-\infty}^{+\infty} 
%      d\sigma e^{i\sigma W} \caE_{w\pm}(i\sigma,t;\bfmu) 
%      \nonumber \\
   &=& \frac{1}{2\pi}\int_{-\infty}^{+\infty} 
      d\sigma \; e^{i\sigma W} e^{-\overline{\beta} \Delta\caF_{\pm}}
      \hat{I}_{\pm} 
      \caE_{w\pm}(\overline{\beta}-i\sigma,t;\bfmu) 
      \nonumber \\
%   &=& e^{-\overline{\beta} \Delta\caF_{\pm}} \hat{I}_{\pm}
%      \frac{1}{2\pi}\int_{+\infty+i\overline{\beta}}^{-\infty+i\overline{\beta}} 
%      d(-\sigma^{\prime}) e^{(\overline{\beta} -i\sigma^{\prime}) W} 
%      \caE_{w\pm}(i\sigma^{\prime},t;\bfmu) 
%      \nonumber \\
   &=& e^{\overline{\beta} (W-\Delta\caF_{\pm})} \hat{I}_{\pm}
      \frac{1}{2\pi}
      \int_{-\infty-i\overline{\beta}}^{+\infty-i\overline{\beta}} 
      d\sigma^{\prime} \; e^{i\sigma^{\prime} (-W)} 
      \caE_{w\pm}(i\sigma^{\prime},t;\bfmu) 
      \nonumber \\
%   &=& e^{\overline{\beta} (W-\Delta\caF_{\pm})} \hat{I}_{\pm}
%      \frac{1}{2\pi}\int_{-\infty}^{+\infty} 
%      d\sigma  e^{i\sigma (-W)} 
%      \caE_{w\pm}(i\sigma^{\prime},t;\bfmu) 
%      \nonumber \\
   &=& e^{\overline{\beta} (W-\Delta\caF_{\pm})} \hat{I}_{\pm}
      P_{w\pm}(-W,t;\bfmu)
\label{FluctTrans2b}
\end{eqnarray}
where we used 
%$\hat{I}_{\pm}^{2}=1$, 
$\sigma^{\prime}\equiv -i\overline{\beta} -\sigma$ and  
the fact that noting Eq. (\ref{Efunct1}) 
the function $\caE_{w\pm}(i\sigma^{\prime},t;\bfmu) $ $
e^{-i\sigma^{\prime} W}$ 
%appearing in Eq. (\ref{WorkRate3a}) 
does not have any pole in the complex plane for 
$\mbox{Im}\{\sigma^{\prime}\}\in [0,-\overline{\beta}]$ with   
the imaginary part $\mbox{Im}\{\sigma^{\prime}\}$ 
of $\sigma^{\prime}$.  
   Eq. (\ref{FluctTrans2b}) leads to Eq. (\ref{FluctTrans2}).

%%%%%%%%%%%%%%%%%%%%%%%%%%%%%%%%%%%%%%%%%%%%%%%%%%%%%%%%%%%%%%%%%%%%%%
\section{Average Work Rates in Class C}

%   In this Appendix, we give a derivation of the average 
%work rate for nonequilibrium models in Class C 
%in a NESS. 

%---------------------------------------------------------------------
\subsection{Energy Transfer Model by a Temperature Difference}
\label{AveWorkAppC1} 

   In this Appendix, we calculate the average 
work rate $\dot{\overline{\caW}}_{+}$ %(\ref{AveraWorkRateC2}) 
for an energy transfer model 
driven by a temperature difference in a NESS. 

   First we introduce $\bar{x}_{t}$ and $\Delta x_{t}$ as 
\begin{eqnarray}
   \bar{x}_{t} &\equiv& \frac{x_{1t}+x_{2t}}{2} ,
      \label{BarX1} \\
   \Delta x_{t} &\equiv&  x_{1t}-x_{2t} .
      \label{DeltaX1}
\end{eqnarray}
   From Eqs. (\ref{LangeEquatC1}), (\ref{BarX1}) and 
(\ref{DeltaX1}) we can derive the Langevin equations for 
$\bar{x}_{t}$ and $\Delta x_{t}$ separately as 
\begin{eqnarray} 
    &&m\ddot{\bar{x}}_{t}  + \alpha \dot{\bar{x}}_{t} 
       = e^{- \omega_{m} t} \frac{d}{dt} 
       e^{\omega_{m}t}\frac{d \bar{x}_{t}}{dt} 
       =\frac{\zeta_{1t}+\zeta_{2t}}{2} , 
      \label{LangeEquatC4}\\
    &&m\Delta\ddot{x}_{t} + \alpha \Delta\dot{x}_{t} 
          + 2\kappa \Delta x_{t}  
       = e^{-(\omega_{a}+\omega_{b})t}
          \frac{d}{dt} e^{\omega_{a}t}\frac{d}{dt} 
          e^{\omega_{b}t} \Delta x_{t} 
       = \zeta_{1t}-\zeta_{2t} 
      \label{LangeEquatC5}
\end{eqnarray}
with $\ddot{\bar{x}}_{t}\equiv d^{2}\bar{x}_{t}/dt^{2}$ 
$\dot{\bar{x}}_{t}\equiv d\bar{x}_{t}/dt$, 
$\Delta\ddot{x}_{t}\equiv d^{2}\Delta x_{t}/dt^{2}$ 
and $\Delta\dot{x}_{t}\equiv d\Delta x_{t}/dt$. 
   Here, $\omega_{m}$, $\omega_{a}$ and $\omega_{b}$ are defined by 
$\omega_{m} \equiv \alpha/m$, 
$\omega_{a} \equiv \sqrt{\omega_{m}^{2} - (8\kappa/m)}$ 
and $\omega_{b} \equiv (\omega_{m} -\omega_{a})/2$, 
respectively. 
   Solving Eqs. (\ref{LangeEquatC4}) and (\ref{LangeEquatC5}) 
we obtain 
\begin{eqnarray}
   \bar{x}_{t} &=& \bar{x}_{t_{0}} 
      + \frac{1}{\omega_{m}} 
      \left[1-e^{-\omega_{m}(t-t_{0})}\right]
      \dot{\bar{x}}_{t_{0}}  
      \nonumber \\
   &&\spaEq 
      + \frac{1}{2m} 
      \int_{t_{0}}^{t}du_{1} 
      \int_{t_{0}}^{u_{1}}du_{2} \; 
      e^{-\omega_{m}(u_{1}-u_{2})} 
      \left(\zeta_{1u_{2}}+\zeta_{2u_{2}}\right) , 
      \label{BarX2} \\
   \Delta x_{t} &=&  \left\{ \Delta x_{t_{0}} 
      +\frac{1}{\omega_{a}}
      \left[1-e^{-\omega_{a}(t-t_{0})}\right] 
      \left(\omega_{b} \Delta x_{t_{0}} 
      + \Delta\dot{x}_{t_{0}}\right) \right\} 
      e^{-\omega_{b}(t-t_{0})}
      \nonumber \\
   &&\spaEq 
      + \frac{1}{m}\int_{t_{0}}^{t} du_{1}\int_{t_{0}}^{u_{1}}du_{2}\; 
      e^{-\omega_{a}(u_{1}-u_{2})-\omega_{b}(t-u_{2})}  
      \left(\zeta_{1u_{2}}-\zeta_{2u_{2}}\right) .
      \label{DeltaX2}
\end{eqnarray}
   Using Eq. (\ref{BarX2}) we obtain 
\begin{eqnarray}
   \dot{\bar{x}}_{t} &=& 
   e^{-\omega_{m}(t-t_{0})}  \dot{\bar{x}}_{t_{0}}  
   + \frac{1}{2m} \int_{t_{0}}^{t}du\; e^{-\omega_{m}(t-u)} 
   \left(\zeta_{1u}+\zeta_{2u}\right) . 
   \label{BarDotX1} 
\end{eqnarray}
for the time-derivative of the position $\bar{x}_{t}$. 

   By the expression of the work $\caW_{+}$ 
in Table. \ref{EnergWorkModel2} %Eqs. (\ref{WorkC1p}), 
and using Eqs. (\ref{BarX1}) and (\ref{DeltaX1})
the average of the work rate can be expressed as 
\begin{eqnarray} 
   \left\langle \frac{d\caW_{+}(\{\bfx_{s}\},\Delta T)}{dt}\right\rangle 
%   &=&  \frac{\Delta T}{2T} 
%      \left\langle \left[\kappa \left(x_{1t} - x_{2t}\right) 
%      \left(\dot{x}_{1t} +\dot{x}_{2t}\right) 
%      +m\dot{x}_{1t}\ddot{x}_{1t}
%      - m\dot{x}_{2t}\ddot{x}_{2t} \right] \right\rangle
%      +\mathcal{O}\left(\left|\frac{\Delta T}{T}\right|^{2}\right) \nonumber \\
%   &=&  \frac{\Delta T}{2T} 
%      \left\langle \left[2\kappa \Delta x_{t} \dot{\bar{x}}_{t} 
%      +m(\dot{x}_{1t}\ddot{x}_{1t}
%      - \dot{x}_{2t}\ddot{x}_{2t}) \right] \right\rangle
%      +\mathcal{O}\left(\left|\frac{\Delta T}{T}\right|^{2}\right) \nonumber \\
%   &=&  \frac{\Delta T}{T} \left[\kappa 
%      \left\langle \Delta x_{t} \dot{\bar{x}}_{t} \right\rangle
%      +\frac{m}{4} 
%      \frac{d}{dt}\left\langle (\dot{x}_{1t}^{2}
%      - \dot{x}_{2t}^{2})  \right\rangle\right]
%      +\mathcal{O}\left(\left|\frac{\Delta T}{T}\right|^{2}\right) \nonumber \\
   &=&  \frac{\Delta T}{T} \left[\kappa 
      \left\langle \Delta x_{t} \dot{\bar{x}}_{t} \right\rangle
      +\frac{m}{2} \frac{d\left\langle \Delta \dot{x}_{t} 
      \dot{\bar{x}}_{t}  \right\rangle}{dt}\right]
      +\mathcal{O}\left(\left|\frac{\Delta T}{T}\right|^{2}\right) 
\label{WorkC2Appen1}
\end{eqnarray} 
using $\Delta \dot{x}_{t}$ and $\bar{x}_{t}$
for the energy transfer model driven by a temperature difference. 
   In the NESS, the quantity 
$\left\langle \Delta \dot{x}_{t} \dot{\bar{x}}_{t}  \right\rangle$ 
should be independent of time $t$, so we can neglect 
the term $(m/2) $ $d\langle \Delta \dot{x}_{t} 
\dot{\bar{x}}_{t}\rangle/dt$ in the right-hand side of 
Eq. (\ref{WorkC2Appen1}) in such a state. 
   Moreover, the NESS should be realized 
in the long-time limit, so we can neglect the first terms 
of the right-hand side of Eqs. (\ref{DeltaX2}) and (\ref{BarDotX1}) 
to calculate a quantity in the NESS. 
   Using these points, the average work rate 
$\dot{\overline{\caW}}_{+}$ in the NESS 
up to the second order of $\Delta T^{2}$ 
can be calculated by 
\begin{eqnarray}
   \dot{\overline{\caW}}_{+}
   &=& \lim_{t\rightarrow+\infty}\frac{\kappa\Delta T}{T} 
      \left\langle \Delta x_{t} \dot{\bar{x}}_{t} \right\rangle 
      \nonumber\\
   &=& \lim_{t\rightarrow+\infty} 
      \frac{\kappa \Delta T}{2 m^{2}T}
      \int_{t_{0}}^{t} du_{1}\int_{t_{0}}^{u_{1}}du_{2}
      \int_{t_{0}}^{t} du_{3}\; 
      \langle (\zeta_{1u_{2}}+\zeta_{2u_{2}})
      (\zeta_{1u_{3}}+\zeta_{2u_{3}})\rangle 
      \nonumber \\
   &&\spaEq \times 
      e^{-\omega_{a}(u_{1}-u_{2})-\omega_{b}(s-u_{2}) 
      -\omega_{m}(s-u_{3})} 
      \nonumber\\
   &=&  \frac{\alpha \kappa k_{B} \Delta T^{2}}
      {m^{2}T (\omega_{a}+\omega_{b}+\omega_{m}) 
       (\omega_{b}+\omega_{m})}
      \nonumber\\
   &=& \frac{\alpha \kappa k_{B} \Delta T^{2}}
      {2 T (\alpha^{2}+m\kappa)}
\label{WorkC2Appen2}
\end{eqnarray}
where we used the relation 
$\langle (\zeta_{1s}+\zeta_{2s})(\zeta_{1u}-\zeta_{2u})\rangle 
= 2\alpha k_{B} \Delta T \delta(s-u)$. 
%   Therefore, we obtain Eq. (\ref{AveraWorkRateC2}). 

%%%%%%%%%%%%%%%%%%%%%%%%%%%%%%%%%%%%%%%%%%%%%%%%%%%%%%%%%%%%%%%%%%%%%%
\subsection{Electric Circuit with Two Resistors}
\label{AveWorkAppC2}

   In this Appendix, we calculate the average work rate 
$\dot{\overline{\caW}}_{+}$ %(\ref{AveraWorkRateC3}) 
for the electric circuit model with two resistors. 
   By the expression shown in Table 
\ref{EnergWorkModel2} %(\ref{WorkC2p})
for the work $\caW_{+}$ in the electric circuit with two Resistors, 
the average work rate is given by 
$\langle V I_{2t}\rangle = V \langle \dot{q}_{2t} \rangle$, 
so that calculation of  
$\langle \dot{q}_{2t} \rangle$ in the long time limit 
is sufficient to calculate the average work rate 
$\dot{\overline{\caW}}_{+}$ in the NESS. 

   To calculate $\langle \dot{q}_{2t} \rangle$ we first note that 
\begin{eqnarray}
   C \left(L \langle\ddot{q}_{1t}\rangle 
   + R_{1} \langle\dot{q}_{1t} \rangle\rangle\right) 
      &=&\langle q_{2t}\rangle - \langle q_{1t}\rangle
      \label{LangeAveraC2a} \\
   C \left(V -  R_{2} \langle\dot{q}_{2t} \rangle\right)
      &=& \langle q_{2t}\rangle - \langle q_{1t}\rangle
      \label{LangeAveraC2b}
\end{eqnarray}
by taking the ensemble average of the Langevin equation 
(\ref{LangeEquatC2}). 
   Inserting $ \langle q_{1t}\rangle 
= \langle q_{2t}\rangle + C (R_{2} \langle\dot{q}_{2t} \rangle - V) $ 
from Eq. (\ref{LangeAveraC2b}) into Eq. (\ref{LangeAveraC2a}) 
we obtain 
\begin{eqnarray}
   LCR_{2}\frac{d^{2}\phi_{t}}{dt^{2}} 
   +\left(L+CR_{1}R_{2}\right) \frac{d\phi_{t}}{dt} 
   +\left(R_{1}+R_{2}\right)\phi_{t} = 0
\label{LangeAveraC2c}
\end{eqnarray}
for $\phi_{t}\equiv \langle \dot{q}_{2t} \rangle - V/(R_{1}+R_{2})$. 
   Note that the differential equation (\ref{LangeAveraC2c}) 
for $\phi_{t}$ has the same form as the one for 
$\langle \Delta x_{t} \rangle$ obtained by taking the ensemble 
average of the Langevin equation (\ref{LangeEquatC5}), 
whose solution satisfies 
$\lim_{t\rightarrow+\infty}\langle \Delta x_{t} \rangle = 0$ 
by the average of Eq. (\ref{DeltaX2}). 
   In a similar way for $\langle \Delta x_{t} \rangle$ 
we can also show that $\lim_{t\rightarrow+\infty}\phi_{t} = 0$, i.e. 
\begin{eqnarray}
   \lim_{t\rightarrow+\infty}\langle \dot{q}_{2t} \rangle 
   = \frac{V}{R_{1}+R_{2}}
%\label{}
\end{eqnarray}
which is the value $\langle \dot{q}_{2t} \rangle$ 
in the NESS. 
   Using this and $\dot{\overline{\caW}}_{+} 
=  \lim_{t\rightarrow+\infty}V \langle \dot{q}_{2t} \rangle$, 
we obtain $\dot{\overline{\caW}}_{+} = V^{2}/(R_{1}+R_{2})$.
%Eq. (\ref{AveraWorkRateC3}). 

%%%%%%%%%%%%%%%%%%%%%%%%%%%%%%%%%%%%%%%%%%%%%%%%%%%%%%%%%%%%%%%%%%%%%%

%%%%%%%%%%%%%%%%%%%%%%%%%%%%%%%%%%%%%%%%%%%%%%%%%%%%%%%%%%%%%%%%%%%%%%

\end{document}